\documentclass[preprint,10pt,authoryear]{elsarticle}
\usepackage{amssymb,amsmath,latexsym,mathrsfs,wasysym,graphicx,subfigure,epsfig,varioref,xr-hyper,hyperref,color,multirow,array,float}
\usepackage[utf8]{inputenc}
\usepackage[T1]{fontenc}
\newcolumntype{?}{!{\vrule width 3pt}}

\hypersetup{colorlinks=true,urlcolor=blue,citecolor=blue,linkcolor=blue,menucolor=blue,anchorcolor=blue,filecolor=blue,pdfauthor=author}

\journal{Journal of High Energy Astrophysics}

\begin{document}

\begin{frontmatter}

\title{Dark energy, spatial curvature, and star formation efficiency from JWST photometric and spectroscopic high-redshift galaxies}

\author[unitn]{Leonardo Comini}
\ead{leo1800.com@gmail.com}

\author[unitn,tifpa]{Sunny Vagnozzi}\corref{cor1}
\ead{sunny.vagnozzi@unitn.it}\cortext[cor1]{Corresponding author}

\author[harv]{Abraham Loeb}
\ead{aloeb@cfa.harvard.edu}

\address[unitn]{Department of Physics, University of Trento, Via Sommarive 14, 38123 Povo (TN), Italy}
\address[tifpa]{\mbox{Trento Institute for Fundamental Physics and Applications, Via Sommarive 14, 38123 Povo (TN), Italy}}
\address[harv]{\mbox{Department of Astronomy, Harvard University, 60 Garden Street, Cambridge, MA 02138, USA}}

\begin{abstract}
\noindent Early observations from the James Webb Space Telescope (JWST) have revealed an overabundance of massive high-redshift galaxies, raising the question of whether this points to new physics beyond $\Lambda$CDM, or an enhanced formation efficiency of massive stars. We revisit this issue going beyond earlier analyses based on direct comparisons to theoretical bounds at a fixed cosmology, by performing a full Bayesian analysis of the most extreme galaxies in the CEERS imaging and FRESCO spectroscopic samples, jointly constraining cosmological parameters and the baryon-to-star conversion efficiency $\epsilon$. We do so not only within the spatially flat $\Lambda$CDM model, but also in models where the dark energy equation of state $w$ and/or the spatial curvature parameter $\Omega_K$ are allowed to vary, carefully discussing the impact of both $w$ and $\Omega_K$ on the cumulative comoving stellar mass density. Within the flat $\Lambda$CDM model, once cosmological parameters are marginalized over, the CEERS sample provides a weak $2\sigma$ lower limit of $\epsilon \gtrsim 0.07$, compatible with astrophysical expectations. In contrast, the FRESCO sample requires $\epsilon \gtrsim 0.5$ at $2\sigma$, with values $\epsilon \lesssim 0.2$ disfavored at $>5.0\sigma$. These results do not qualitatively change when we allow $w$ and/or $\Omega_K$ to vary, with no evidence for deviations from $w=-1$ or $\Omega_K=0$. Our results therefore suggest that the origin of the ``JWST tension'' is unlikely to be cosmological, but lies in the astrophysics of galaxy formation.
\end{abstract}


\end{frontmatter}

\section{Introduction}
\label{sec:introduction}

The \textit{James Webb Space Telescope} (JWST), a space-based infrared observatory designed to study the formation of the first galaxies and stars~\citep{Gardner:2006ky}, is providing a unique view of sources at redshifts up to $z \sim 15$, corresponding to the first few hundred million years of cosmic history. This regime, previously only indirectly constrained by low-redshift observations, is one where both the efficiency of early star formation and the growth of structure are still very uncertain. The most massive high-redshift galaxies are particularly interesting in this sense, as they populate the most massive dark matter (DM) halos, whose abundance depends very strongly on the underlying cosmology. Such systems therefore provide a very powerful probe of both cosmology and the physics of galaxy formation.

Early JWST imaging obtained with the Near Infrared Camera (NIRCam) as part of the Cosmic Evolution Early Release Science (CEERS) program led to the identification of a surprisingly high abundance of extremely massive galaxies (with stellar masses exceeding $\sim 10^{10}M_{\odot}$) at redshifts $6 \lesssim z \lesssim 10$~\citep{Labbe:2022ahb}. As demonstrated by~\citet{Boylan-Kolchin:2022kae}, the abundance of such systems can be used to stress test cosmological models in a conservative way with minimal assumptions about galaxy formation physics. In particular, considering objects of a given stellar mass, an upper limit on their cumulative comoving stellar mass density can be obtained once the cosmic baryon fraction $f_b \equiv \Omega_b/\Omega_m$ and DM halo mass function (HMF) are known, since the total stellar mass cannot exceed the available baryonic mass. Applying this ``baryon availability'' argument, \citet{Boylan-Kolchin:2022kae} argued that within the $\Lambda$CDM cosmological model the population of high-$z$ galaxies of~\citet{Labbe:2022ahb} can only be accommodated if the baryon-to-star conversion efficiency $\epsilon$ is implausibly high, close to the theoretical upper bound $\epsilon=1$. Stated differently, within $\Lambda$CDM the population of galaxies identified by~\citet{Labbe:2022ahb} appears too abundant, too massive, and too early.

The above results were driven by early JWST imaging photometry, and are therefore subject to the intrinsic limitations of photometric redshift estimation and spectral energy distribution (SED)-based stellar mass determinations, motivating spectroscopic follow-up. Spectroscopic observations from the JWST First Reionization Epoch Spectroscopically Complete Observations (FRESCO) NIRCam/grism survey have now provided secure redshifts for a sample of massive galaxies at $5 \lesssim z \lesssim 9$~\citep{Xiao:2023ghw}. Under conservative assumptions, the properties of three ultra-massive galaxies in the FRESCO sample imply unusually high baryon-to-star conversion efficiencies of order $\epsilon \sim 0.5$~\citep{Xiao:2023ghw}. This value is significantly larger than typical expectations, which typically indicate efficiencies $\epsilon \lesssim 0.2$ level or lower, even at higher redshifts~\citep{Conroy:2008dx,Bigiel:2008jw,Leroy:2008kh,Kennicutt:2012ea,Moster:2012fv,Behroozi:2012iw,Behroozi:2012sp,Wechsler:2018pic,Tacchella:2018qny,Shuntov:2022qwu}.~\footnote{See however~\citet{Dekel:2023ddd,Boylan-Kolchin:2024roq,Shen:2025isu} for studies suggesting mechanisms for increasing the efficiency of star formation at high redshifts.} FRESCO spectroscopy therefore not only qualitatively confirms the trend suggested by CEERS photometry, but also places it on a firmer footing by providing secure redshifts and demonstrating that the efficiency problem is already present at lower redshifts~\citep{Xiao:2023ghw}.

The above conclusions of~\citet{Boylan-Kolchin:2022kae} and~\cite{Xiao:2023ghw} were obtained from a ``baryon availability'' consistency argument based on direct comparison to theoretical bounds, checking whether individual extreme systems lay above or below theoretical predictions (assuming a $\Lambda$CDM cosmology with fixed parameters), rather than via a likelihood-based analysis. This approach, while simple and conservative, only really tests whether individual systems are consistent with theoretical expectations: it does not provide a probabilistic constraint on $\epsilon$, nor does it account for uncertainties in the underlying cosmological model and/or parameters. Given the potentially very important (astrophysical and cosmological) implications of these results, it is important to revisit the problem within a framework allowing for consistent comparison between data and theory, while accounting for uncertainties on the cosmological side. In this work we therefore move beyond the earlier fixed cosmological assumptions and per-object minimum-efficiency arguments, and perform a full Bayesian analysis of most extreme galaxies within the CEERS and FRESCO samples. We do so not only within the spatially flat $\Lambda$CDM model, but also within some of its simplest extensions, for instance those with a free dark energy equation of state and/or spatial curvature. We stress that our goal is not that of ruling out $\Lambda$CDM, but rather to quantify, in a conservative and fully probabilistic manner, the range of baryon-to-star conversion efficiencies implied by current JWST observations once cosmological uncertainties are consistently accounted for. In short, we find that while CEERS provides only weak constraints and actually still allows relatively low efficiencies, FRESCO robustly favors relatively large efficiencies $\epsilon \gtrsim 0.3-0.5$ within all the cosmological extensions considered, suggesting that the origin of the ``JWST tension'' lies primarily in the efficiency of early star formation rather than in simple modifications of the background cosmology.

The rest of this paper is then organized as follows. In Sec.~\ref{sec:highredshift} we introduce the main observable used in this work (the cumulative comoving stellar mass density), and describe how theoretical predictions for this quantity depend on the underlying cosmology; we also present the four cosmological models considered in our analysis. In Sec.~\ref{sec:datamethods} we present the CEERS and FRESCO datasets and discuss the statistical framework adopted in our analysis. We present our results in Sec.~\ref{sec:results}, and critically discuss them in Sec.~\ref{sec:discussion}. We close in Sec.~\ref{sec:conclusions} by drawing concluding remarks.

\section{From cosmological models to high-redshift stellar mass density}
\label{sec:highredshift}

We now describe the procedure for obtaining theoretical predictions for the main observable used here, i.e.\ the cumulative comoving stellar mass density above a stellar mass threshold, which we denote by $\rho_{\star}(>M_{\star},z)$. This observable, which we will refer to simply as ``cumulative stellar mass density'', measures the total stellar mass per unit comoving volume contained in galaxies with stellar mass larger than a given value $M_{\star}$ at a given redshift.

To compute the \textit{observed} cumulative stellar mass density, we specify a stellar mass threshold $M_{\star}$ within a redshift bin $z \in [z_{\min},z_{\max}]$, and centered at an effective redshift $z_{\text{eff}}$. We then sum over the stellar masses of all galaxies which fall above the threshold, and divide by the survey comoving volume $V$ (i.e.\ the comoving volume contained between $z_{\min}$ and $z_{\max}$), as follows:
\begin{equation}
\rho_{\star}^{\text{obs}}(>M_{\star},z_{\text{eff}})=\frac{1}{V}\sum_{M_i>M_{\star}}M_i\,,
\label{eq:rhostarobs}
\end{equation}
with $M_i$ being the stellar masses of the individual galaxies above the threshold within the redshift bin. This definition agrees with the estimator adopted by~\citet{Boylan-Kolchin:2022kae}. Since the comoving volume depends on the assumed cosmology, the observable defined in Eq.~(\ref{eq:rhostarobs}) carries an implicit cosmological dependence which must be treated consistently when comparing observations against theoretical predictions, as we discuss in more detail later. For the purposes of our work, this observable presents several advantages. Firstly, it is by construction relatively insensitive to individual outliers or the properties of any single object, and instead captures the global abundance of massive systems. Additionally, $\rho_{\star}(>M_{\star},z)$ can be directly compared to theoretical cosmological upper bounds once the dark matter HMF and cosmic baryon fraction are known, without requiring detailed knowledge of galaxy formation physics or galaxy-halo matching prescriptions, as highlighted by~\citet{Boylan-Kolchin:2022kae}. Last but not least, $\rho_{\star}(>M_{\star},z)$ has already been used in earlier JWST-based ``stress test'' works~\citep[see for instance][]{Boylan-Kolchin:2022kae,Xiao:2023ghw}, therefore facilitating direct comparisons.

In the standard picture of hierarchical structure formation, the abundance of DM halos is determined by the linear matter power spectrum and the growth history, and is measured by the DM halo mass function $dn(M,z)/dM$, i.e.\ the number of DM halos per unit mass per unit comoving volume at a given redshift. We follow the Sheth-Tormen prescription and compute the DM HMF as follows~\citep{Sheth:1999su}:
\begin{equation}
\frac{dn(M,z)}{dM}=\frac{\bar{\rho}_m}{M}A\sqrt{\frac{2a}{\pi}} \left [ 1+(a\nu^2)^{-p} \right ] \nu e^{-\frac{a\nu^2}{2}} \left \vert \frac{d\nu}{dM} \right \vert \,.
\label{eq:shethtormen}
\end{equation}
In Eq.~(\ref{eq:shethtormen}) above, $\bar{\rho}_m=\Omega_m\rho_{\rm crit,0}$ is the mean comoving matter density, whereas $\Omega_m$ and $\rho_{\rm crit,0}$ are respectively the present-day matter density parameter and critical density. In our calculation we fix the Sheth-Tormen parameters to $A=0.3222$, $a=0.707$, and $p=0.3$. The peak height parameter is instead given by $\nu=\delta_c/\sigma(M,z)$, where $\delta_c \simeq 1.686$ is the critical (linear) overdensity for collapse. Finally, we denote by $\sigma(M,z)$ the root mean square linear matter fluctuation at redshift $z$, smoothed on the scale $R=(3M/4\pi\bar{\rho}_m)^{1/3}$:
\begin{equation}
\sigma^2(M,z)=D^2(z)\int\frac{dk}{2\pi^2}\,k^2P(k)W^2(kR)\,.
\label{eq:sigmamz}
\end{equation}
In Eq.~(\ref{eq:sigmamz}) above, $P(k)$ is the linear matter power spectrum at $z=0$, $D(z)$ is the linear growth factor normalized such that $D(0)=1$, and $W(kR)$ is the Fourier transform of a real-space top-hat filter of radius $R$.

We then compute the cumulative comoving DM halo mass density contained in halos more massive than a threshold $M_{\text{halo}}$ from the DM HMF as follows:
\begin{equation}
\rho_m(>M_{\text{halo}},z)=\int_{M_{\text{halo}}}^{\infty}dM\,M\frac{dn(M,z)}{dM}\,.
\label{eq:rhomgmhalo}
\end{equation}
In order to connect Eq.~(\ref{eq:rhomgmhalo}) to the statistics of galaxies, we need a map between $M_{\text{halo}}$ and the stellar content of a given halo, $M_{\star}$. It is obvious that the cosmic baryon fraction $f_b \equiv \Omega_b/\Omega_m$ (where $\Omega_b$ is the present-day baryon density parameter) controls the maximum possible value of $M_{\star}$. Following~\citet{Boylan-Kolchin:2022kae}, we parametrize the relation between $M_{\star}$ and $M_{\text{halo}}$ as $M_{\star}=\epsilon f_bM_{\text{halo}}$, where $\epsilon \leq 1$ is a dimensionless parameter which quantifies the efficiency for converting gas into stars. Stated differently, we are assuming that only a fraction $\epsilon$ of the available baryonic mass is converted into stars. We can then express the (theoretical) cumulative stellar mass density as follows:
\begin{align}
\rho_{\star}^{\text{th}}(>M_{\star},z)&=\epsilon f_b\rho_m \left ( >\frac{M_{\star}}{\epsilon f_b},z \right ) \nonumber \\
&= \epsilon f_b\int_{\frac{M_{\star}}{\epsilon f_b}}^{\infty}dM\,M\frac{dn(M,z)}{dM}\,.
\label{eq:rhostarth}
\end{align}
Given a certain value of $\epsilon \leq 1$, consistency between theory and observations requires that the following holds:
\begin{equation}
\rho_{\star}^{\text{obs}}(>M_{\star},z_{\text{eff}})\le \rho_{\star}^{\text{th}}(>M_{\star},z_{\text{eff}})\,.
\label{eq:consistency}
\end{equation}
The above condition is precisely the ``baryon availability'' argument introduced in~\citet{Boylan-Kolchin:2022kae}~\citep[and also used in][]{Xiao:2023ghw}. This provides a conservative consistency test between observed galaxy properties and theoretical predictions for the halo population within a given cosmological model. In~\citet{Boylan-Kolchin:2022kae} this stress test was performed in the limiting case $\epsilon=1$, and other two (discrete) values of $\epsilon$. The $\epsilon=1$ case corresponds to the (theoretically allowed but otherwise extreme and rather implausible) scenario of complete conversion of all available baryons into stars, which sets an absolute upper bound on the allowed cumulative stellar mass density. Here we instead choose to treat $\epsilon$ as a free parameter. As stressed above, the consistency condition set by Eq.~(\ref{eq:consistency}) conservatively compresses all the uncertainties associated to galaxy formation physics into a single parameter $\epsilon$. The underlying cosmological model instead enters our theoretical prediction through both the expansion history and the growth of structure, see Eq.~(\ref{eq:rhostarth}). In our analysis, we will compare different cosmological models while fixing the present-day amplitude of fluctuations, controlled by $\sigma_8$. As we shall see, it follows that what matters is the relative growth of perturbations with respect to today, a subtle point to which we return later.

We now introduce the four cosmological models considered in our later analysis. In order of increasing complexity, these models are:
\begin{itemize}
\item the concordance flat $\Lambda$CDM model, which represents our baseline model, and where dark energy (DE) takes the form of a cosmological constant with equation of state (EoS) $w=-1$;
\item the flat $w$CDM model, where the DE EoS takes values $w \neq -1$, while still remaining constant: this is one of the simplest extensions of our baseline model~\citep[see for instance][for recent examples of constraints on this model]{Vagnozzi:2019ezj,Escamilla:2023oce,Luongo:2024fww,Sammut:2025eik,Yadav:2025vgo};
\item the non-flat $\Lambda$CDM model, also referred to as $o\Lambda$CDM model, which allows for non-zero spatial curvature, characterized by a spatial curvature parameter $\Omega_K \neq 0$;~\footnote{See for instance~\citet{Handley:2019tkm,Wang:2019yob,DiValentino:2019qzk,Efstathiou:2020wem,DiValentino:2020hov,Chudaykin:2020ghx,Benisty:2020otr,Vagnozzi:2020rcz,Vagnozzi:2020dfn,DiValentino:2020kpf,Yang:2021hxg,Cao:2021ldv,Dhawan:2021mel,Gonzalez:2021ojp,Dinda:2021ffa,Zuckerman:2021kgm,Bargiacchi:2021hdp,Akarsu:2021max,Glanville:2022xes,Bel:2022iuf,Wu:2022fmr,Yang:2022kho,Stevens:2022evv,Favale:2023lnp,Qi:2023oxv,Giare:2023ejv,Wu:2024faw,Liu:2024yib,Forconi:2025zzu,Specogna:2025ufe} for recent examples of constraints on spatial curvature.};
\item finally, the non-flat $w$CDM model, also referred to as $ow$CDM model, where $w \neq -1$ and $\Omega_K \neq 0$ simultaneously.
\end{itemize}
The last three models are among the simplest, best studied extensions to the concordance $\Lambda$CDM model~\citep[see for instance][]{CosmoVerseNetwork:2025alb}. In the context of the present work, they provide a minimal set of controlled extensions which allow us to study how departures from flat $\Lambda$CDM affect the abundance of high-$z$ massive DM halos and galaxies. The expansion history within each model is characterized by the dimensionless expansion rate $E(z) \equiv H(z)/H_0$, where $H(z)$ and $H_0$ are the Hubble rate and Hubble constant respectively. For the flat $\Lambda$CDM model $E(z)$ is given as follows:
\begin{equation}
E^2_{\Lambda{\text{CDM}}}(z)=\Omega_m(1+z)^3+\Omega_r(1+z)^4+\Omega_{\Lambda}\,,
\label{eq:ezlcdm}
\end{equation}
where $\Omega_r$ is the radiation density parameter, and $\Omega_{\Lambda}$ is the present-day cosmological constant density parameter. For the flat $w$CDM model, $E(z)$ is instead given by the following:
\begin{equation}
E^2_{w{\text{CDM}}}(z)=\Omega_m(1+z)^3+\Omega_r(1+z)^4+\Omega_{\text{DE}}(1+z)^{3(1+w)}\,,
\label{eq:ezwcdm}
\end{equation}
where $\Omega_{\text{DE}}$ is the present-day DE density parameter. For the non-flat $\Lambda$CDM model, $E(z)$ is given as follows:
\begin{equation}
E^2_{o\Lambda{\text{CDM}}}(z)=E^2_{\Lambda{\text{CDM}}}(z)+\Omega_K(1+z)^2\,,
\label{eq:nonflatlcdm}
\end{equation}
where $\Omega_K$ is the present-day spatial curvature density parameter. Finally, in the non-flat $w$CDM model, $E(z)$ is given by the following:
\begin{equation}
E^2_{ow{\text{CDM}}}(z)=E^2_{w{\text{CDM}}}+\Omega_K(1+z)^2\,.
\label{eq:eznonflatwcdm}
\end{equation}
For all four models we neglect the radiation component, which is completely subdominant at the relevant epochs. The closure relation therefore implies that $\Omega_{\Lambda} \approx 1-\Omega_m$ within the flat $\Lambda$CDM model, $\Omega_{\text{DE}} \approx 1-\Omega_m$ within the flat $w$CDM model, $\Omega_{\Lambda} \approx 1-\Omega_m-\Omega_K$ within the non-flat $\Lambda$CDM model, and $\Omega_{\text{DE}} \approx 1-\Omega_m-\Omega_K$ within the non-flat $w$CDM model. When we introduce non-zero spatial curvature, we therefore implicitly compare different models at fixed $\Omega_m$, with the remaining energy budget redistributed between spatial curvature and the dark energy sector. Moreover, within all four models we can divide the matter sector into a cold DM and a baryon component, i.e.\ $\Omega_m=\Omega_c+\Omega_b$. We then define the cosmic baryon fraction as $f_b=\Omega_b/\Omega_m$.

The cosmological dependence of our theoretical predictions for the cumulative stellar mass density in Eq.~(\ref{eq:rhostarth}) enters primarily through the linear growth factor $D(z)$, which determines the amplitude of matter fluctuations at the redshifts probed by JWST, and is itself controlled by the background expansion rate. In our analysis, we will compare different cosmological models at fixed $\sigma_8$ and $H_0$. Stated differently, we fix both the present-day normalization of density fluctuations (and not their primordial amplitude), as well as the present-day expansion rate. The relevant quantity controlling the DM HMF is therefore the linear growth factor relative to its present-day value, given that the amplitude of matter fluctuations at redshift $z$ can be expressed in the following way:
\begin{equation}
\sigma(M,z)=\sigma(M,0)\frac{D(z)}{D(0)}=\sigma(M,0)D(z)\,,
\label{eq:growth}
\end{equation}
where we adopt the conventional normalization $D(0)=1$. Because models are compared \textit{at fixed $\sigma_8$}, it is the growth \textit{with respect to today} that matters, leading to a somewhat counterintuitive behaviour. Consider for instance a quintessence-like DE scenario with $w>-1$. If we fix $H_0$, the high-redshift expansion rate increases relative to $\Lambda$CDM. This increases the friction term in the growth equation for matter overdensities $\delta$, and therefore slows the growth of structure. This implies that the amplitude of fluctuations at higher redshifts had to be larger in order to reach the same value today, given that we have chosen to fix $\sigma_8$. To put it differently, imagine evolving the system backwards from $D(0)=1$. Then we would find that $D(z)$ is larger relative to $\Lambda$CDM, which directly implies a larger abundance of high-redshift DM halos. Analogous considerations hold for a spatially open Universe ($\Omega_K>0$). The converse is true for models which (at fixed $H_0$) reduce the expansion rate at high redshift relative to $\Lambda$CDM: in our case, this corresponds to phantom DE models (i.e.\ those with $w<-1$) and a spatially closed Universe ($\Omega_K<0$), which therefore accelerate the growth of structure, and correspondingly suppress the abundance of DM halos at high redshift.

\begin{figure*}[!htb]
\centering
\includegraphics[width=0.49\textwidth]{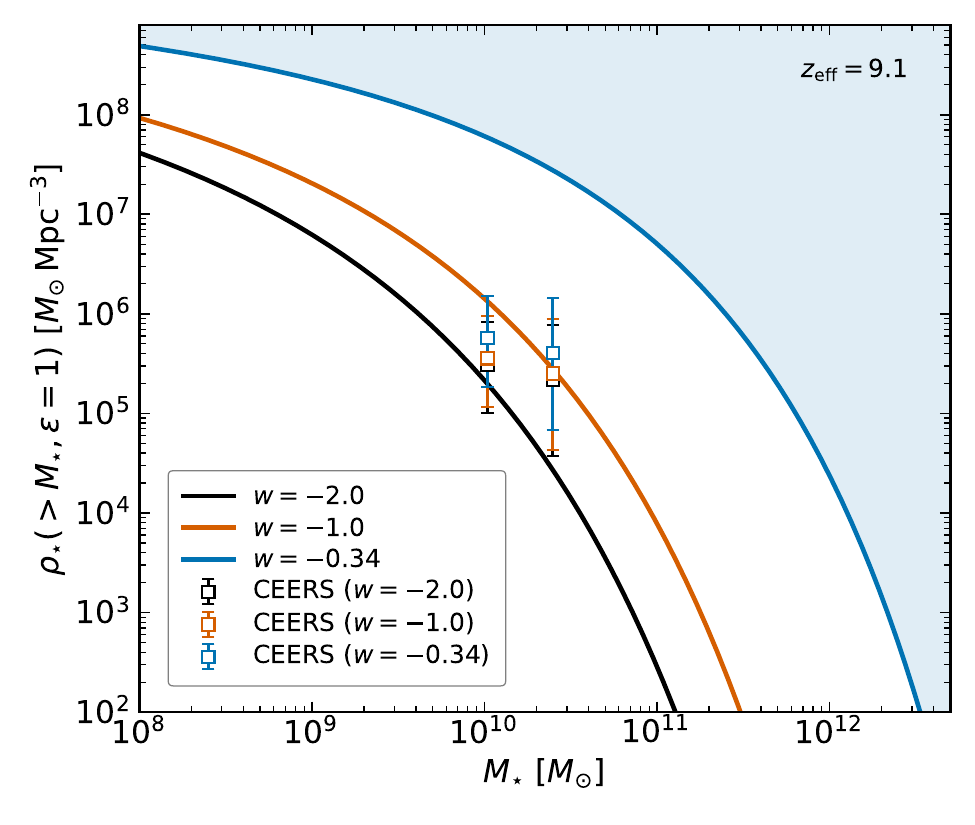}\hfill\includegraphics[width=0.49\textwidth]{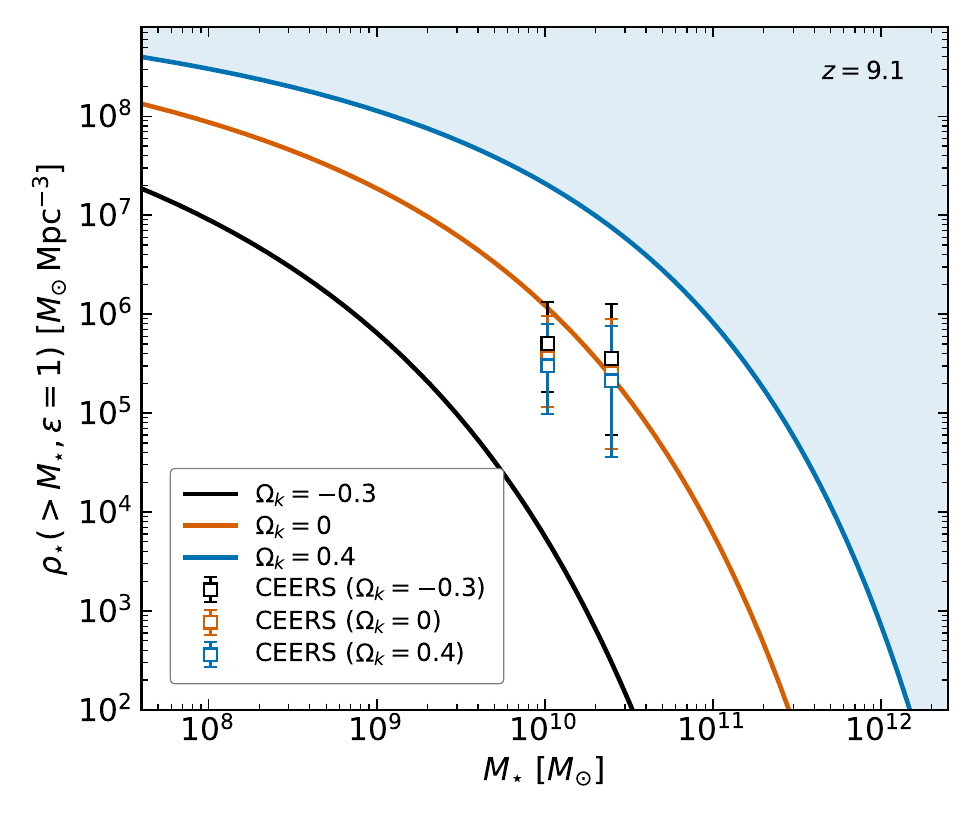}
\caption{Maximum allowed cumulative comoving stellar mass density as a function of stellar mass predicted within different cosmological models, compared at fixed $\Omega_m$, $\sigma_8$, and $H_0$, with baryon fraction fixed to $f_b \simeq 0.15$. The curves correspond to the limiting case $\epsilon=1$, and thus represent the upper envelope of the cumulative stellar mass density allowed by each cosmological model within the ``baryon availability'' argument (the blue shaded region is excluded by the ``baryon availability'' argument in the most generous case). The datapoints correspond to the CEERS measurements at $z_{\text{eff}} \simeq 9.1$, and are recomputed for each cosmology considered to consistently account for the fiducial cosmology dependence of the inferred cumulative stellar mass density and stellar masses. \textit{Left panel}: comparison between models with different values of the dark energy equation of state $w=-1$ ($\Lambda$CDM, orange curve), $-0.34$ (blue curve), and $-2.0$ (black curve). We clearly see that, at fixed $\sigma_8$, quintessence-like (phantom) models with $w>-1$ ($w<-1$) enhance (suppress) the abundance of high-redshift galaxies relative to $\Lambda$CDM. \textit{Right panel}: comparison between models with different values of the spatial curvature parameter $\Omega_K=0$ ($\Lambda$CDM, orange curve), $0.4$ (blue curve), and $-0.3$ (black curve). We clearly see that, at fixed $\Omega_m$ (with the difference in energy density implied by spatial curvature being accounted for by the dark energy sector), spatially open (closed) models with $\Omega_K>0$ ($\Omega_K<0$) enhance (suppress) the abundance of high-redshift galaxies relative to $\Lambda$CDM.}
\label{fig:varywomegak}
\end{figure*}

These points are illustrated in Fig.~\ref{fig:varywomegak}, where we plot the \textit{maximum} allowed cumulative stellar mass density as a function of stellar mass predicted within different cosmological models, showing the impact of changing the DE EoS $w$ (left panel) and the spatial curvature parameter $\Omega_K$ (right panel). To compute the underlying DM HMF we adopt the Python package \texttt{hmf}~\citep{Murray:2013qza}, relying on the \texttt{astropy} library~\citep{Astropy:2013muo}. The datapoints correspond to the CEERS cumulative stellar mass density measurements at $z_{\text{eff}} \simeq 9.1$~\citep[which have also been used in][]{Boylan-Kolchin:2022kae}, and are recomputed for each cosmology considered to consistently account for the fiducial cosmology dependence of the inferred cumulative stellar mass density and stellar masses, as discussed later in Sec.~\ref{subsec:data}. The curves correspond to the limiting case $\epsilon=1$, and thus represent the upper envelope of the cumulative stellar mass density allowed by each cosmological model within the ``baryon availability'' argument. In this limiting case, the quantity appearing on the $x$ axis is therefore given by $M_{\star}=f_bM_{\text{halo}}$. We immediately observe that models with $w>-1$ ($w<-1$) enhance (suppress) the abundance of high-redshift halos and galaxies relative to the reference $\Lambda$CDM model, and similarly for models with $\Omega_K>0$ ($\Omega_K<0$), consistently with our earlier discussion. The differences between the curves directly reflect how changes in the background expansion history $E(z)$ modify the relative growth factor $D(z)/D(0)$. We stress once more that the different models are being compared at fixed $\Omega_m$, $\sigma_8$, and $H_0$, and the baryon fraction $f_b$ is similarly fixed to $f_b \simeq 0.15$.

\section{Datasets and methodology}
\label{sec:datamethods}

Having described how our theoretical predictions for the cumulative stellar mass density are obtained, we now introduce the data adopted in our later analysis, as well as the statistical methods used to compare the data against theoretical predictions.

\subsection{Data}
\label{subsec:data}

The \textit{CEERS} dataset is based on early JWST multi-band imaging obtained with the NIRCam as part of the CEERS program, which covers an effective survey area of $\sim 40\,{\text{arcmin}}^2$. Within this dataset, \citet{Labbe:2022ahb} identified a population of extremely massive, intrinsically red galaxy candidates with photometric redshifts $6 \lesssim z \lesssim 10$. We focus in particular on the highest redshift bin considered by~\citet{Boylan-Kolchin:2022kae}, containing galaxies with photometric redshifts $8.5<z<10$, and centered at an effective redshift $z_{\text{eff}} \simeq 9.1$. The individual galaxy properties are given in the upper three rows of Tab.~\ref{tab:data}. We compute the cumulative stellar mass density following~\citet{Boylan-Kolchin:2022kae} and considering the three most extreme systems, with the stellar masses of the two most extreme ones exceeding $10^{10}\,M_{\odot}$. Following Eq.~(\ref{eq:rhostarobs}), we evaluate $\rho_{\star}^{\text{obs}}(>M_{\star},z_{\text{eff}}=9.1)$ at the two thresholds reported in \citet{Boylan-Kolchin:2022kae}. We obtain values of order $\rho_{\star} \lesssim 10^6\,M_{\odot}{\text{Mpc}}^{-3}$, in agreement with those reported by~\citet{Boylan-Kolchin:2022kae}. Only the most massive object contributes to the highest mass bin, whereas all three objects contribute to the lowest mass bin. These two measurements of $\rho_{\star}^{\text{obs}}$, which are also those reported in Fig.~\ref{fig:varywomegak}, define the CEERS datapoints which enter the likelihood adopted in our later statistical analysis. The uncertainties are computed following~\citet{Labbe:2022ahb}, and reflect both Poisson statistics and cosmic variance. We stress that, since the CEERS sample is based on photometric redshifts and stellar mass estimates, the resulting cumulative stellar mass density measurements carry substantial uncertainties. We expect this to translate into rather loose constraints on $\epsilon$, as our full Bayesian analysis will confirm.

The \textit{FRESCO} dataset is based on JWST NIRCam/grism spectroscopy obtained within the FRESCO program, covering a total survey area of $\sim 124\,{\text{arcmin}}^2$ across the GOODS-North and GOODS-South fields. Within the survey, \citet{Xiao:2023ghw} identified a population of spectroscopically confirmed extremely massive, red, emission-line galaxies, many of which had been missed by earlier UV-selected surveys due to heavy dust obscuration. The spectroscopic redshifts of these galaxies fall in the range $5 \lesssim z \lesssim 9$. Following~\citet{Xiao:2023ghw}, we focus on the lowest redshift bin $5<z<6$ centered at an effective redshift $z_{\text{eff}} \simeq 5.5$, since the stellar masses derived for galaxies in this bin are significantly more robust than those of galaxies at higher redshifts. The individual galaxy properties are given in the lower three rows of Tab.~\ref{tab:data}. We compute the cumulative stellar mass density by considering the three most extreme objects in this redshift bin, named $S1$, $S2$, and $S3$, whose masses exceed $10^{11}\,M_{\odot}$~\citep{Xiao:2023ghw}. We then evaluate $\rho_{\star}^{\text{obs}}(>M_{\star},z_{\text{eff}}=5.5)$ at three stellar mass thresholds: $M_{\star}=1.096 \times 10^{11}\,M_{\odot}$, $1.514 \times 10^{11}\,M_{\odot}$, and $2.344 \times 10^{11}\,M_{\odot}$. We note that all three galaxies contribute above the first threshold, whereas only S2 and S1 contribute above the second, and only S1 contributes above the third. This yields cumulative stellar mass densities of order $\rho_{\star} \lesssim 10^6\,M_{\odot}{\text{Mpc}}^{-3}$. These three measurements of $\rho_{\star}^{\text{obs}}$ define the FRESCO datapoints which enter the likelihood adopted in our later statistical analysis, and their uncertainties reflect both Poisson statistics and cosmic variance. Owing to the spectroscopic confirmation of the galaxies and, accordingly, the more robust redshift and stellar mass estimates, the FRESCO measurements have substantially smaller relative uncertainties compared to their CEERS counterparts. We therefore expect them to lead to significantly tighter limits on $\epsilon$, as our analysis will confirm.

\begin{table*}[!t]
\centering
\begin{tabular}{|c?c|c|c|}
\hline
\textbf{Dataset} & ID & $z$ & $\log_{10}(M_{\star}/M_{\odot})$ \\
\hline\hline
CEERS & 35300 & 9.08 & 10.40 \\ \hline
CEERS & 14924 & 8.83 & 10.02 \\ \hline
CEERS & 21834 & 8.54 & 9.61 \\ \hline
\hline
FRESCO & S-18258 (S1) & 5.58 & 11.37 \\ \hline
FRESCO & N-7496 (S2/GN10) & 5.31 & 11.18 \\ \hline
FRESCO & N-2663 (S3) & 5.18 & 11.04 \\ \hline
\end{tabular}
\caption{Properties of the most extreme galaxies used to compute the cumulative stellar mass densities for the CEERS and FRESCO datasets. For each galaxy, we report the catalog identifier, redshift (photometric redshifts for CEERS, spectroscopic redshifts for FRESCO), and (logarithm of the) stellar mass in units of solar masses. For the FRESCO dataset, we report both the catalog identifiers and the labels (S1, S2, S3) used in~\citet{Xiao:2023ghw}; note that S2 is also known as GN10 in the literature. Stellar masses correspond to values inferred from SED fitting assuming the fiducial cosmology.}
\label{tab:data}
\end{table*}

We stress that the observed value of $\rho_{\star}^{\text{obs}}(>M_{\star})$ depends on an assumed fiducial cosmology in two different ways~\citep[see][for discussions]{Menci:2022wia,Menci:2024rbq}. In first place, since stellar masses derived from photometric observations are inferred from observed fluxes, they scale with luminosity distance as $M_{\star} \propto d_L^2$. This is true even for galaxies with spectroscopic redshifts such as the FRESCO ones, given that their stellar masses are still derived from SED modelling and therefore follow the same scaling. Moreover, as Eq.~(\ref{eq:rhostarobs}) makes explicit, the cumulative stellar mass density depend on the comoving survey volume: this is a cosmology-dependent quantity, as computing requires assuming a specific expansion history to compute the distance-redshift relation, i.e.\ a fiducial cosmology. These issues were not relevant to the earlier works of~\citet{Boylan-Kolchin:2022kae} and~\citet{Xiao:2023ghw}, since they worked at fixed cosmology, but are essential when conducting a full Bayesian analysis scanning over different cosmologies, as we aim to do here. Therefore, at each step of our Monte Carlo Markov Chain (MCMC) analysis, we will rescale $\rho_{\star}^{\text{obs}}$ by the ratio of the comoving volume within the fiducial cosmology to that in the cosmology at the given MCMC step. The assumed fiducial cosmology is the best-fit \textit{Planck} 2018 $\Lambda$CDM cosmological model, with $H_0=67.4\,{\text{km}}/{\text{s}}/{\text{Mpc}}$, $\Omega_m=0.315$, and $\sigma_8=0.811$~\citep{Planck:2018vyg}. Similarly, the values of $M_{\star}$ at each MCMC step are rescaled by the square of the ratio of the luminosity distance within the cosmology at the given MCMC step to that in the fiducial cosmology.

\subsection{Statistical framework}
\label{subsec:statisticalframework}

For each of the four models studied, characterized by the unnormalized expansion rates given in Eqs.~(\ref{eq:ezlcdm}--\ref{eq:eznonflatwcdm}), the baseline set of parameters is given by the matter density parameter $\Omega_m$, the present-day amplitude of fluctuations $\sigma_8$, and the baryon-to-star conversion efficiency $\epsilon$. We have explicitly checked that the physical matter density $\omega_b$ and the Hubble constant $H_0$ do not have a strong impact on the predicted cumulative stellar mass density, and therefore opt to fix these parameters to $\omega_b=0.02233$ and $H_0=67.4\,{\text{km}}/{\text{s}}/{\text{Mpc}}$ respectively~\citep{Planck:2018vyg}. We note that both $\omega_b$ and $H_0$ are required to compute the cosmic baryon fraction $f_b \equiv \Omega_b/\Omega_m=\omega_b/\Omega_mh^2$, which enters the theoretical prediction for the cumulative stellar mass density as in Eq.~(\ref{eq:rhostarth}), and is consistently computed at each step of our MCMC analysis to account for the different values of $\Omega_m$ we sample. For our purposes the flat $\Lambda$CDM model parameter space is therefore 3-dimensional, and described by the parameter vector $\{\Omega_m,\sigma_8,\epsilon\}$. Similarly, the 4-dimensional flat $w$CDM and non-flat $\Lambda$CDM models are described by the parameter vectors $\{\Omega_m,\sigma_8,\epsilon,w\}$ and $\{\Omega_m,\sigma_8,\epsilon,\Omega_K\}$ respectively. Finally, the 5-dimensional non-flat $w$CDM model is described by the parameter vector $\{\Omega_m,\sigma_8,\epsilon,w,\Omega_K\}$.

We sample the posterior distributions of the four cosmological models running MCMC chains through the \texttt{emcee} sampler~\citep{Foreman-Mackey:2012any}. Predictions for the HMF within each model are obtained through the publicly available Python package \texttt{hmf}~\citep{Murray:2013qza}, which relies on the \texttt{astropy} library~\citep{Astropy:2013muo}. For each model, we evolve an ensemble of $N_w=8$ walkers for $N_{\text{step}}=10000$ steps (except for the non-flat $w$CDM model, for which we use 18000 steps), and conservatively discard the first 2000 points of each chain as burn-in. We assess the convergence of the chains by estimating the integrated autocorrelation time $\tau$ for each parameter, and verifying that the chain length satisfies $N_{\text{step}} \gg 50\tau$. We confirm that the total number of independent samples, $N_{\text{ind}} \simeq N_wN_{\text{step}}/\tau$, exceeds by far $\sim 10^3$ for all parameters. We then analyze the resulting MCMC chains using the \texttt{GetDist} package~\citep{Lewis:2019xzd}.

We set wide, flat priors on all three cosmological parameters, verifying a posteriori that our posteriors are not affected by the choice of prior bounds. In particular, we set uniform prior ranges on $\sigma_8 \in [0.76,0.84]$, $\epsilon \in [0.01,1.00]$, $w \in [-2.00,-0.33]$, and $\Omega_K \in [-0.2,0.4]$, all of which are very broad and consistent with the constraints imposed from independent cosmological probes. In particular, for what concerns $\epsilon$, we do not allow the prior range to go all the way down to $\epsilon=0$ since in that limit the cumulative stellar mass density is identically zero, and therefore in obvious disagreement with data. As for $w$, the upper limit of $w=-0.33$ is set by the requirement that the DE component should drive cosmic acceleration, which requires $w<-1/3$.

For the matter density parameter $\Omega_m$, following the reasoning of~\citet{Pedrotti:2025ccw} we set a Gaussian prior $\Omega_m=0.30 \pm 0.03$. This prior reflects the fact that $\Omega_m$ is very well constrained by a wide variety of late-time probes. At the same time, there remains a mild level of disagreement between different datasets~\citep{Sakr:2023hrl,Akarsu:2024qiq,Pedrotti:2024kpn,Colgain:2024mtg,Lynch:2025ine,Lee:2025hjw,Lee:2025kbn,Wang:2025znm,Weiner:2026sfm,Shlivko:2026jxa}, with \textit{DESI} BAO measurements indicating lower values $\Omega_m \sim 0.28$~\citep{DESI:2024mwx,DESI:2025zgx,Chudaykin:2025aux}, whereas several Type Ia Supernovae datasets indicate somewhat larger values $\Omega_m \gtrsim 0.33$~\citep{Rubin:2023jdq,DES:2024jxu,Baryakhtar:2024rky,Colgain:2024ksa}. Even accounting for this spread across different probes, it is clear that extreme values such as $\Omega_m \lesssim 0.25$ and $\Omega_m \gtrsim 0.35$ are strongly disfavored by virtually all datasets. Our choice of prior is therefore conservative, in the sense that its rather generous width more than reflects the current spread among different determinations of $\Omega_m$, while its central value is not tied to any single measurement~\citep[although it is consistent with the model-independent ``uncalibrated cosmic standards'' determinations of][]{Lin:2021sfs,Wang:2025mqz}.

We adopt a one-sided semi-Gaussian likelihood. Denoting by $\boldsymbol{\theta}$ the parameter vector sampled in our MCMC, the log-likelihood takes the following form:
\begin{equation}
-\ln {\cal L}(\boldsymbol{\theta})=\frac{1}{2}\sum_i
\begin{cases}
\dfrac{ \left [ \rho_{\star,i}^{\text{th}}(\boldsymbol{\theta})-\rho_{\star,i}^{\text{obs}} \right ] ^2}{\sigma_i^2} & (\rho_{\star,i}^{\text{th}}(\boldsymbol{\theta})<\rho_{\star,i}^{\text{obs}}) \\
0 & (\rho_{\star,i}^{\text{th}}(\boldsymbol{\theta}) \geq \rho_{\star,i}^{\text{obs}})
\end{cases}
\label{eq:likelihood}
\end{equation}
where the sum runs over the different cumulative stellar mass density measurements (two for CEERS and three for FRESCO), and $\sigma_i$ are the associated observational uncertainties. The rationale behind the semi-Gaussian likelihood mirrors that of~\citet{Vagnozzi:2021tjv,Wei:2022plg,Binici:2024smk}, and is as follows. Parameters predicting cumulative stellar mass densities smaller than the observed ones should be penalized according to a Gaussian likelihood, i.e.\ exponentially suppressed as the discrepancy increases. On the other hand, parameters predicting larger values of the cumulative stellar mass density compared to the observed ones should not be penalized~\citep[although see the discussion in][]{Costa:2023cmu}. The reason is that the observed cumulative stellar mass density should be interpreted conservatively, because incompleteness and selection effects may lead to an underestimation of the true abundance of massive galaxies. A symmetric Gaussian likelihood would instead artificially strengthen the obtained constraints unless such effects were modeled explicitly, which is beyond the scope of this work. Our semi-Gaussian likelihood can be viewed as a continuous generalization of the binary ``baryon availability'' stress test carried out by~\citet{Boylan-Kolchin:2022kae}, in which models either pass or fail depending on whether they can predict a cumulative stellar mass density at least as large as observed. Our treatment implements the same logic in the sense that parameters failing the consistency condition are penalized, whereas parameters satisfying it are not required to precisely match the observed values.

From our posterior distributions we derive one-sided lower limits on $\epsilon$, quoting $68\%$ and $95\%$ one-sided lower limits, which correspond to the $32^{\text{nd}}$ and $5^{\text{th}}$ percentiles of the posterior distributions.~\footnote{All the limits we compute are one-sided as our likelihood penalizes only parameters whose corresponding cumulative stellar mass densities are too low. As a result, only sufficiently low values of $\epsilon$ are statistically disfavored, and the posterior distribution of $\epsilon$ is of the ``limit-type'' frequently encountered in cosmology, where only one side of the parameter space is constrained. See for instance the case of the sum of neutrino masses $\sum m_{\nu}$, for which upper limits are typically obtained, at least when the physical prior $\sum m_{\nu} \geq 0\,{\text{eV}}$ is assumed~\citep{Vagnozzi:2017ovm,Vagnozzi:2018jhn,RoyChoudhury:2018vnm,RoyChoudhury:2019hls,Tanseri:2022zfe,Wang:2024hen,Green:2024xbb,Naredo-Tuero:2024sgf,Du:2024pai,Jiang:2024viw,RoyChoudhury:2024wri,Loverde:2024nfi,RoyChoudhury:2025dhe,RoyChoudhury:2025iis,Feng:2026pzs}.} We also quantify the statistical significance at which a given reference value of $\epsilon=\epsilon_0$ is excluded, computing the posterior probability $p=P(\epsilon<\epsilon_0)$ from the fraction of MCMC samples satisfying the condition $\epsilon<\epsilon_0$. We convert this to a Gaussian-equivalent one-sided significance $Z$ by computing the corresponding one-sided tail probability for a standard normal distribution as $Z=\Phi^{-1}(1-p)$, where $\Phi$ is the cumulative distribution function of the unit normal distribution. The benchmark reference value we choose is $\epsilon_0=0.2$: this is a relatively high conversion efficiency, yet one which is still consistent with upper estimates from empirical stellar-to-halo mass relations~\citep{Conroy:2008dx,Bigiel:2008jw,Leroy:2008kh,Kennicutt:2012ea,Moster:2012fv,Behroozi:2012iw,Behroozi:2012sp,Wechsler:2018pic,Tacchella:2018qny,Shuntov:2022qwu}. Values $\epsilon \gtrsim 0.2$ therefore already imply very efficient star formation, and the statistical significance at which $\epsilon=0.2$ is excluded provides a conservative but meaningful measure of the tension with standard galaxy formation scenarios. For completeness, we also consider values of $\epsilon_0=0.1$ and $0.15$, noting that the former value allows for a direct comparison with the results of~\citet{Boylan-Kolchin:2022kae}.

\section{Results}
\label{sec:results}

We begin with a brief anticipation of our main results before discussing them in detail. We find that the CEERS dataset mildly favors high values of $\epsilon$, but the comparatively large uncertainties imply that lower, astrophysically more plausible values of $\epsilon$ remain consistent. On the other hand, the FRESCO dataset requires significantly larger values of $\epsilon \gtrsim 0.5-0.7$ within $\Lambda$CDM. Freeing up $w$ or $\Omega_K$ somewhat relaxes this requirement but does not eliminate the need for high values of $\epsilon \gtrsim 0.3$ for extensions to $\Lambda$CDM. Overall, we find no evidence for deviations from $\Lambda$CDM, with $w=-1$ and $\Omega_K=0$ remaining fully consistent with the data: this suggests that the tension is of astrophysical rather than cosmological nature.

\subsection{Flat $\Lambda$CDM model}
\label{subsec:lcdm}

\begin{figure}[!htb]
\centering
\includegraphics[width=0.75\textwidth]{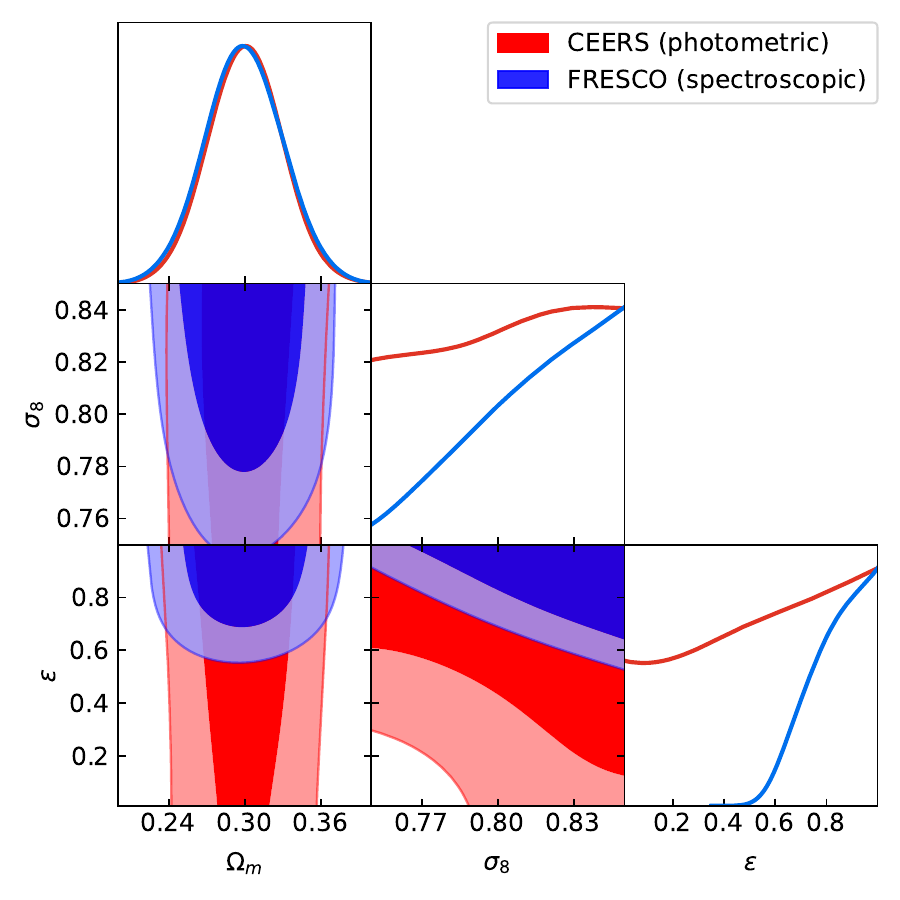}
\caption{Triangular plot showing 2D joint and 1D marginalized posterior probability distributions for the present-day matter density parameter $\Omega_m$, the present-day amplitude of fluctuations $\sigma_8$, and baryon-to-star conversion efficiency $\epsilon$, obtained within the flat $\Lambda$CDM model in light of the photometric CEERS (red contours) and spectroscopic FRESCO (blue contours) samples. We note that the FRESCO constraints on $\epsilon$ are significantly tighter than their CEERS counterparts, due to the comparatively larger uncertainties of the latter.}
\label{fig:comparisonlcdm}
\end{figure}

We begin by discussing the flat $\Lambda$CDM model, where we recall that the parameters being varied are $\Omega_m$, $\sigma_8$, and $\epsilon$. The posterior distributions for these parameters are shown in Fig.~\ref{fig:comparisonlcdm}, for both the CEERS (red) and FRESCO (blue) datasets.

For the CEERS dataset, we find the one-sided lower limits $\epsilon \gtrsim 0.39$ and $\epsilon \gtrsim 0.07$ at $68\%$ and $95\%$~C.L. respectively. For what concerns the baryon-to-star conversion efficiency, we find that our benchmark reference value $\epsilon_0=0.2$ is consistent with the data within $1.0\sigma$, confirming that the constraints on $\epsilon$ are relatively weak. Similarly, values of $\epsilon_0=0.15$ and $0.1$ are consistent with the data within $1.2\sigma$ and $1.4\sigma$ respectively. These results differ from those of~\citet{Boylan-Kolchin:2022kae}, where a stronger tension with $\Lambda$CDM was reported, prompting an ongoing discussion on whether high-redshift galaxies observed by JWST call for new physics.~\footnote{See for instance~\citet{Ferrara:2022dqw,Liu:2022bvr,Biagetti:2022ode,Haslbauer:2022vnq,Hutsi:2022fzw,Gandolfi:2022bcm,Maio:2022lzg,Wang:2022jvx,Yuan:2023bvh,Dayal:2023nwi,Ilie:2023zfv,Jiao:2023wcn,Parashari:2023cui,Lei:2023mke,Shen:2023cva,Qin:2023rtf,Padmanabhan:2023esp,Su:2023jno,Lin:2023ewc,Forconi:2023izg,Guo:2023hyp,Huang:2023chx,Pallottini:2023yqg,Bird:2023pkr,Wang:2023ros,Wang:2023xmm,Adil:2023ara,Sun:2023ocn,Casey:2023ghw,Pacucci:2023oci,Gupta:2023mgg,Wang:2023gla,Forconi:2023hsj,vanPutten:2023ths,Iocco:2024rez,Ellis:2024wdh,Hegde:2024kph,Colgain:2024clf,Lu:2024oli,Huang:2024aog,Jiang:2024tll,Menci:2024hop,Urrutia:2024hwc,Wang:2024hce,Lei:2025ooq,Ziegler:2025plz,Fakhry:2025tma,Lei:2025zjx,Zhou:2025nkb,Fakhry:2025yeu,Das:2025bnr,Menci:2026ajy} for examples of various works investigating the implications of the JWST observations for fundamental physics and/or our understanding of galaxy formation.} The origin of the discrepancy is entirely in the different choice of statistical methodology: a full Bayesian analysis marginalizing over cosmological parameters and properly accounting for observational uncertainties shows that the apparent tension is actually rather weak, since relatively low values of $\epsilon$ remain statistically allowed because of the large uncertainties of the CEERS dataset.

In contrast, for the FRESCO dataset we find significantly stronger constraints, i.e.\ $\epsilon \gtrsim 0.79$ and $\epsilon \gtrsim 0.63$ at $68\%$ and $95\%$~C.L. respectively. These results, which are qualitatively consistent with those of~\citet{Xiao:2023ghw} (while being obtained within a full Bayesian framework), point to a strong requirement for very efficient star formation in the early Universe within the flat $\Lambda$CDM model. We find that benchmark values of $\epsilon_0=0.2$, $0.15$, and $0.1$ are all disfavored by the data at a statistical significance of $>5.0\sigma$.

For what concerns the cosmological parameters, we find that $\Omega_m$ and $\sigma_8$ are mostly prior dominated, with no significant shifts between the CEERS and FRESCO datasets. Furthermore, as is clear from the posterior distributions shown in Fig.~\ref{fig:comparisonlcdm}, we observe mild degeneracies between $\epsilon$ and $\sigma_8$. In particular, the two parameters exhibit a negative correlation, reflecting the fact that a reduction in the growth of structure can be compensated by a more efficient star formation. Nevertheless, this degeneracy is not sufficient to alleviate the strong preference for high values of $\epsilon$ in the FRESCO case, as doing so would require implausibly high values of $\sigma_8$.

\subsection{Flat $w$CDM model}
\label{subsec:wcdm}

\begin{figure}[!htb]
\centering
\includegraphics[width=0.75\textwidth]{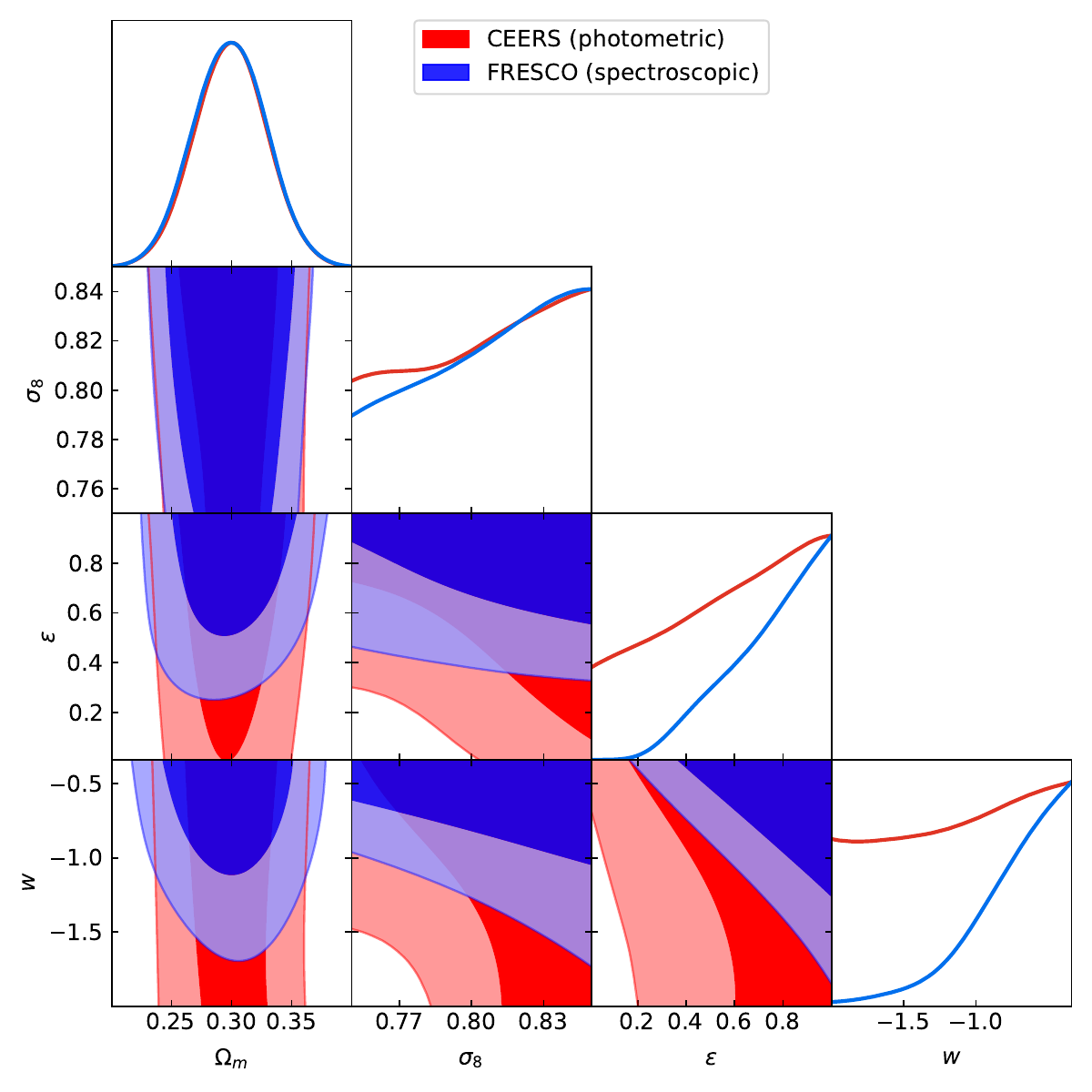}
\caption{As in Fig.~\ref{fig:comparisonlcdm}, but for the flat $w$CDM model, therefore including the dark energy equation of state $w$ among the parameters.}
\label{fig:comparisonwcdm}
\end{figure}

We move on to the flat $w$CDM model, where the parameters being varied are $\Omega_m$, $\sigma_8$, $\epsilon$, and $w$. Their posterior distributions in light of both the CEERS and FRESCO datasets are shown in Fig.~\ref{fig:comparisonwcdm}, with the same color coding as earlier. Using the CEERS dataset, we derive the one-sided lower limits $\epsilon \gtrsim 0.39$ and $\epsilon \gtrsim 0.07$ at $68\%$ and $95\%$~C.L. respectively. Both figures are in very good agreement with their flat $\Lambda$CDM counterparts. Similarly, we find that the data is consistent with benchmark values $\epsilon_0=0.2$, $0.15$, and $0.1$ at the $0.9\sigma$, $1.1\sigma$, and $1.3\sigma$  levels respectively. This confirms that the constraints on $\epsilon$ remain relatively weak. As we could have expected, the DE EoS remains poorly constrained, with $w \gtrsim -1.41$ and $w \gtrsim -1.91$ at $68\%$ and $95\%$~C.L. respectively, and a posterior probability for quintessence-like values of $p(w>-1)=0.45$, indicating no evidence for deviations from a cosmological constant in the DE sector.

When instead considering the FRESCO dataset, the trend observed earlier in the flat $\Lambda$CDM model is fully confirmed. We find significantly stronger constraints of $\epsilon \gtrsim 0.67$ and $\epsilon \gtrsim 0.38$ at $68\%$ and $95\%$~C.L. respectively, whereas values of $\epsilon_0=0.2$, $0.15$, and $0.1$ are disfavored at the $3.2\sigma$, $4.0\sigma$, and $>5.0\sigma$ levels respectively. This suggests that allowing $w$ to vary somewhat relaxes the requirement of a very strong baryon-to-star conversion efficiency relative to flat $\Lambda$CDM, but is unable to eliminate the need for high efficiencies. Finally, for the DE EoS we infer $w \gtrsim -0.85$ and $w \gtrsim -1.41$ at $68\%$ and $95\%$~C.L., whereas the posterior probability for quintessence-like values is $p(w>-1)=0.80$, corresponding to an extremely weak preference for a quintessence-like DE EoS, one not worthy of further discussion.

As one could have anticipated from our earlier discussion on the growth factor, there is a clear degeneracy between $w$ and $\epsilon$, which again is reflected by a negative correlation between the two. The reason is that more quintessence-like values of $w>-1$ increase the abundance of massive halos, therefore relaxing the requirement of high values of $\epsilon$, with the converse being true for phantom values of $w<-1$. Similar considerations hold for the degeneracy between $w$ and $\sigma_8$, since changes in the expansion history also affect the normalization of structure growth. For the same reason, the introduction of $w$ as a free parameter weakens the negative correlation between $\epsilon$ and $\sigma_8$ observed in flat $\Lambda$CDM or, more precisely, redistributes the earlier $\epsilon$-$\sigma_8$ degeneracy among $\epsilon$, $w$, and $\sigma_8$. Nevertheless, FRESCO data still require relatively high values of $\epsilon$, indicating that the simple modification of the expansion history we are considering is unable to resolve the tension, which is reflected in the absence of indications for deviations from the $\Lambda$CDM limit $w=-1$.

\subsection{Non-flat $\Lambda$CDM model}
\label{subsec:lcdmomegak}

\begin{figure}[!htb]
\centering
\includegraphics[width=0.75\textwidth]{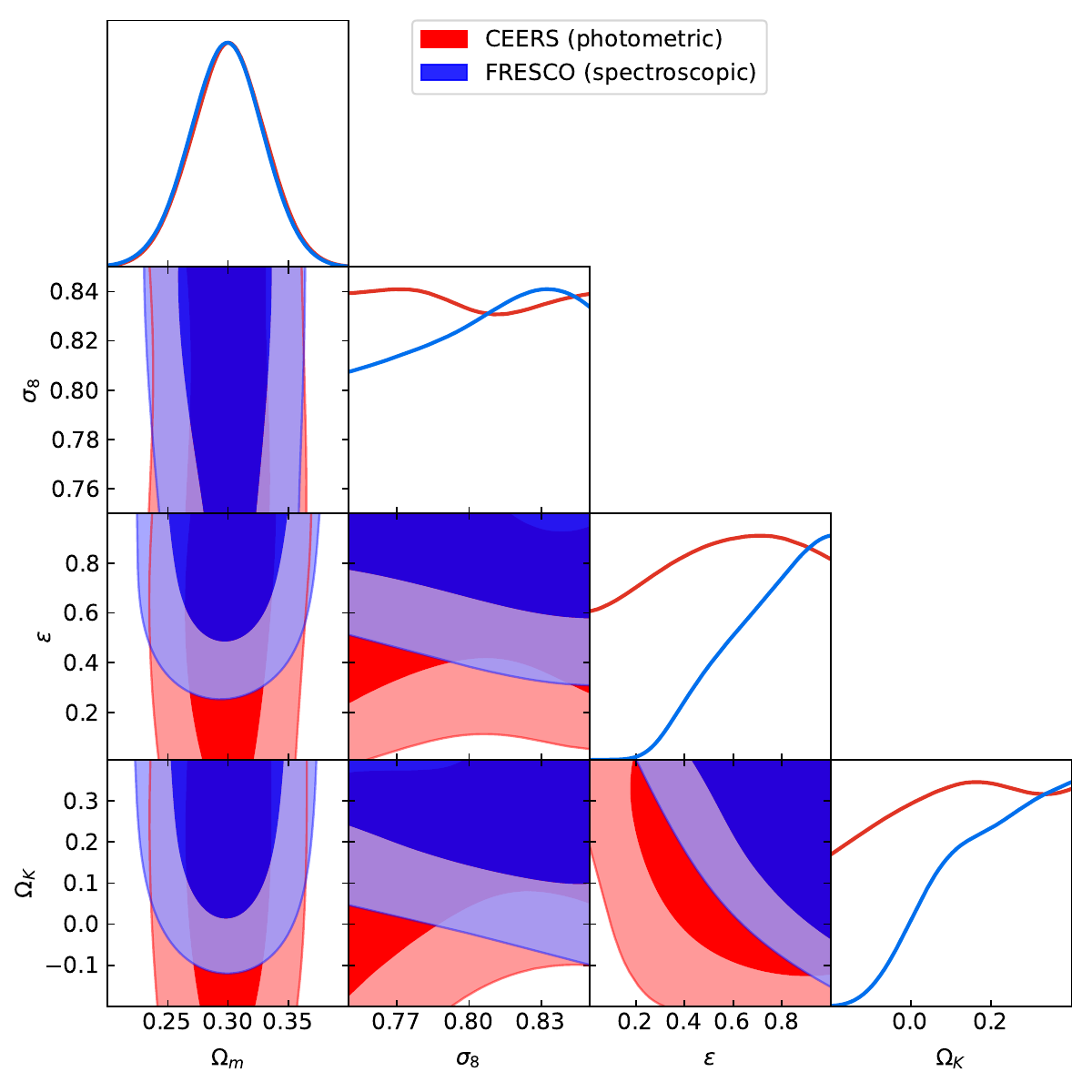}
\caption{As in Fig.~\ref{fig:comparisonlcdm}, but for the non-flat $\Lambda$CDM model, therefore including the spatial curvature parameter $\Omega_K$ among the parameters.}
\label{fig:comparisonlcdmomegak}
\end{figure}

We now study the non-flat $\Lambda$CDM model, where the parameters being varied are $\Omega_m$, $\sigma_8$, $\epsilon$, and $\Omega_K$. The corresponding posterior distributions are shown in Fig.~\ref{fig:comparisonlcdmomegak}.

For the CEERS dataset, we find one-sided lower limits $\epsilon \gtrsim 0.38$ and $\epsilon \gtrsim 0.07$ at $68\%$ and $95\%$~C.L. respectively, in very good agreement with the earlier flat $\Lambda$CDM results. The data is consistent with our benchmark values $\epsilon_0=0.2$, $0.15$, and $0.1$ at the $0.9\sigma$, $1.1\sigma$, and $1.3\sigma$ levels respectively. This again confirms that constraints on $\epsilon$ are weak due to the relatively large uncertainties of the data. Similarly, we obtain comparatively weak constraints on the spatial curvature parameter, with one-sided lower limits of $\Omega_K \gtrsim 0.02$ and $\Omega_K \gtrsim -0.16$ at $68\%$ and $95\%$~C.L. respectively. The posterior probability for negative spatial curvature (i.e.\ positive spatial curvature parameter, corresponding to a spatially open Universe) is $p(\Omega_K>0)=0.70$, with no statistically significant preference for deviations from spatial flatness.

For the FRESCO dataset, we confirm the trends observed earlier in the flat $\Lambda$CDM and $w$CDM models. In particular, we find stronger constraints on the baryon-to-star conversion efficiency compared to those obtained with the CEERS sample, with $\epsilon \gtrsim 0.65$ and $\epsilon \gtrsim 0.38$ at $68\%$ and $95\%$~C.L. respectively, whereas our benchmark values of $\epsilon_0=0.2$, $0.15$, and $0.1$ are disfavored at the $4.0\sigma$, $>5.0\sigma$, and $>5.0\sigma$ levels respectively. This indicates that, while allowing for non-zero spatial curvature somewhat relaxes the constraints relative to flat $\Lambda$CDM, the requirement for very high efficiencies persists, similar to the trend observed when the DE EoS was freed. For the spatial curvature parameters, we find $\Omega_K \gtrsim 0.13$ and $\Omega_K \gtrsim -0.03$ at $68\%$ and $95\%$~C.L. respectively, with a posterior probability for negative spatial curvature $p(\Omega_K>0)=0.92$, corresponding to a very weak preference for an open Universe.

Similarly to the flat $w$CDM model, we find that allowing for non-zero spatial curvature introduces a degeneracy between $\Omega_K$ and $\epsilon$. This is reflected by a negative correlation between the two, since negative spatial curvature ($\Omega_K>0$) modifies the expansion history (albeit through its geometric contribution to the Friedmann equation rather than through a true dynamical component), and hence the growth of structure, in such a way as to enhance the latter at early times, which can be compensated by smaller baryon-to-star conversion efficiencies. Moreover, the introduction of $\Omega_K$ as a free parameter redistributes the earlier $\epsilon$-$\sigma_8$ degeneracy among $\epsilon$, $\Omega_K$, and $\sigma_8$. Nevertheless, even allowing for non-zero spatial curvature, FRESCO data continue to require high values of $\epsilon$, and show no statistically significant indications for departures from spatial flatness.

\subsection{Non-flat $w$CDM model}
\label{subsec:wcdmomegak}

\begin{figure}[!htb]
\centering
\includegraphics[width=0.75\textwidth]{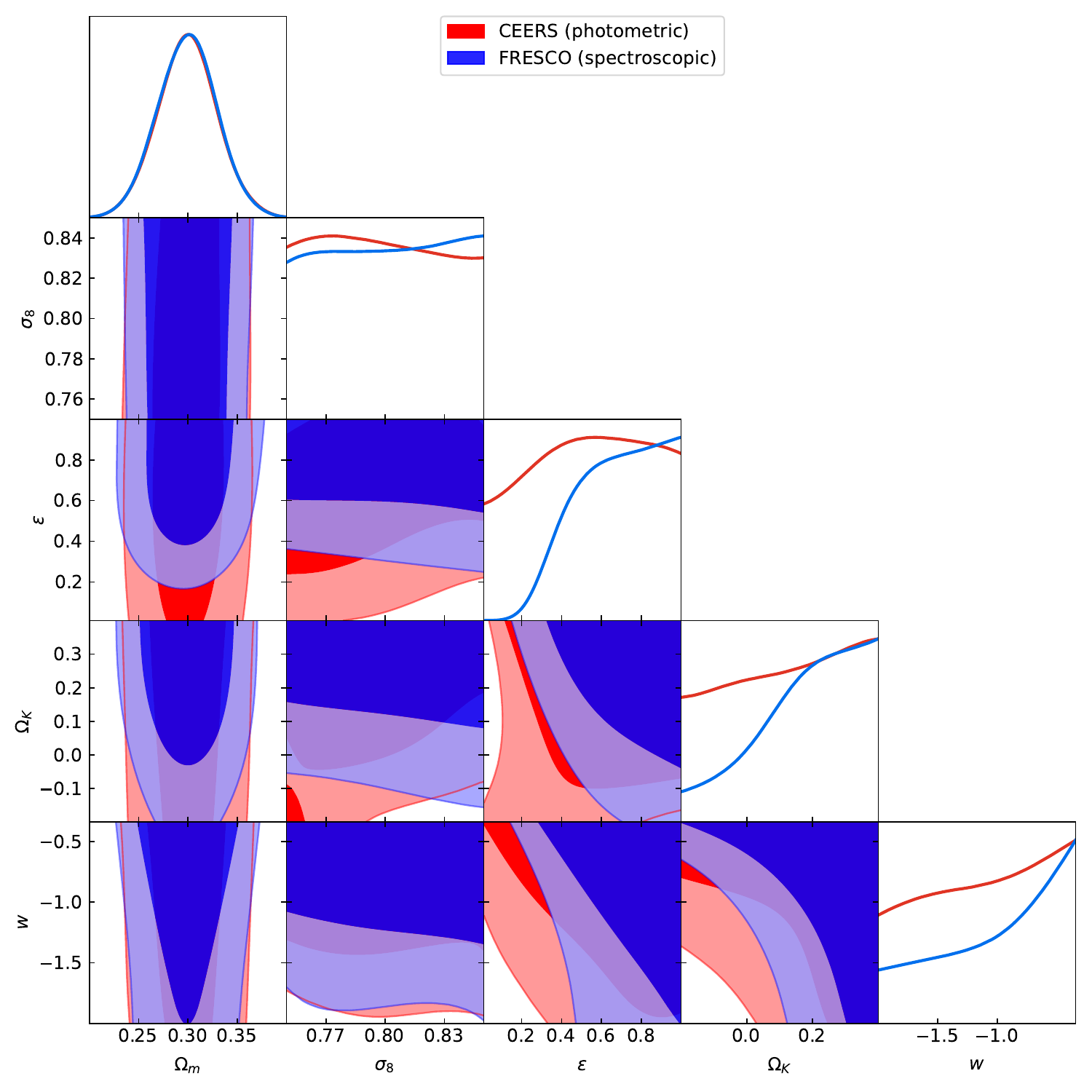}
\caption{As in Fig.~\ref{fig:comparisonlcdm}, but for the non-flat $w$CDM model, therefore including the dark energy equation of state $w$ and spatial curvature parameter $\Omega_K$ among the parameters.}
\label{fig:comparisonwcdmomegak}
\end{figure}

The most general model we consider is the non-flat $w$CDM model, where the parameters being varied are $\Omega_m$, $\sigma_8$, $\epsilon$, $w$, and $\Omega_K$. The corresponding posterior distributions are shown in Fig.~\ref{fig:comparisonwcdmomegak}.

For the CEERS dataset, we find one-sided lower limits $\epsilon \gtrsim 0.38$ and $\epsilon \gtrsim 0.07$ at $68\%$ and $95\%$~C.L. respectively, in very good agreement with the results obtained in all previous models. We find the data to be in agreement with our benchmark values $\epsilon_0=0.2$, $0.15$, and $0.1$ at the $0.9\sigma$, $1.1\sigma$, and $1.3\sigma$ levels respectively. This confirms once again that the constraints on $\epsilon$ are relatively weak, due to the large uncertainties of the data. We also find that both non-$\Lambda$CDM cosmological parameters are poorly constrained, with $w \gtrsim -1.39$ [$w \gtrsim -1.89$] and $\Omega_K \gtrsim 0.02$ [$\Omega_K \gtrsim -0.17$] at $68\%$~C.L. [$95\%$~C.L.], and posterior probabilities $p(w>-1)=0.44$ and $p(\Omega_K>0)=0.70$. All of these figures show no indication for deviations from $\Lambda$CDM.

For the FRESCO dataset, we once again find much stronger constraints on $\epsilon$, with $\epsilon \gtrsim 0.56$ and $\epsilon \gtrsim 0.32$ at $68\%$ and $95\%$~C.L. respectively. Similarly, the data disagrees with our benchmark values $\epsilon_0=0.2$, $0.15$, and $0.1$ at the $2.8\sigma$, $3.8\sigma$, and $>5.0\sigma$ levels respectively. Our results still indicate that even in this rather general scenario the requirement for high star formation efficiency persists. On the other hand, the non-$\Lambda$CDM cosmological parameters remain rather weakly constrained, with $w \gtrsim -1.23$ [$w \gtrsim -1.86$] and $\Omega_K \gtrsim 0.12$ [$\Omega_K \gtrsim -0.11$] at $68\%$~C.L. [$95\%$~C.L.]. We find no statistically significant indications for deviations from a cosmological constant and/or spatial flatness, with posterior probabilities $p(w>-1)=0.56$ and $p(\Omega_K>0)=0.86$, confirming all earlier results.

The degeneracies between $\epsilon$, $w$, and $\Omega_K$ observed and discussed earlier persist here, although at a weaker level, since they are now redistributed across the parameters: in particular, we still find that $\epsilon$ is negatively correlated with both $w$ and $\Omega_K$. However, even when moving away from the cosmological constant and allowing for non-zero spatial curvature, FRESCO data continues to require high values of $\epsilon$, and show no statistically significant indications for departures from the $\Lambda$CDM model.

\section{Discussion}
\label{sec:discussion}

\begin{figure*}[!htb]
\centering
\includegraphics[width=0.49\textwidth]{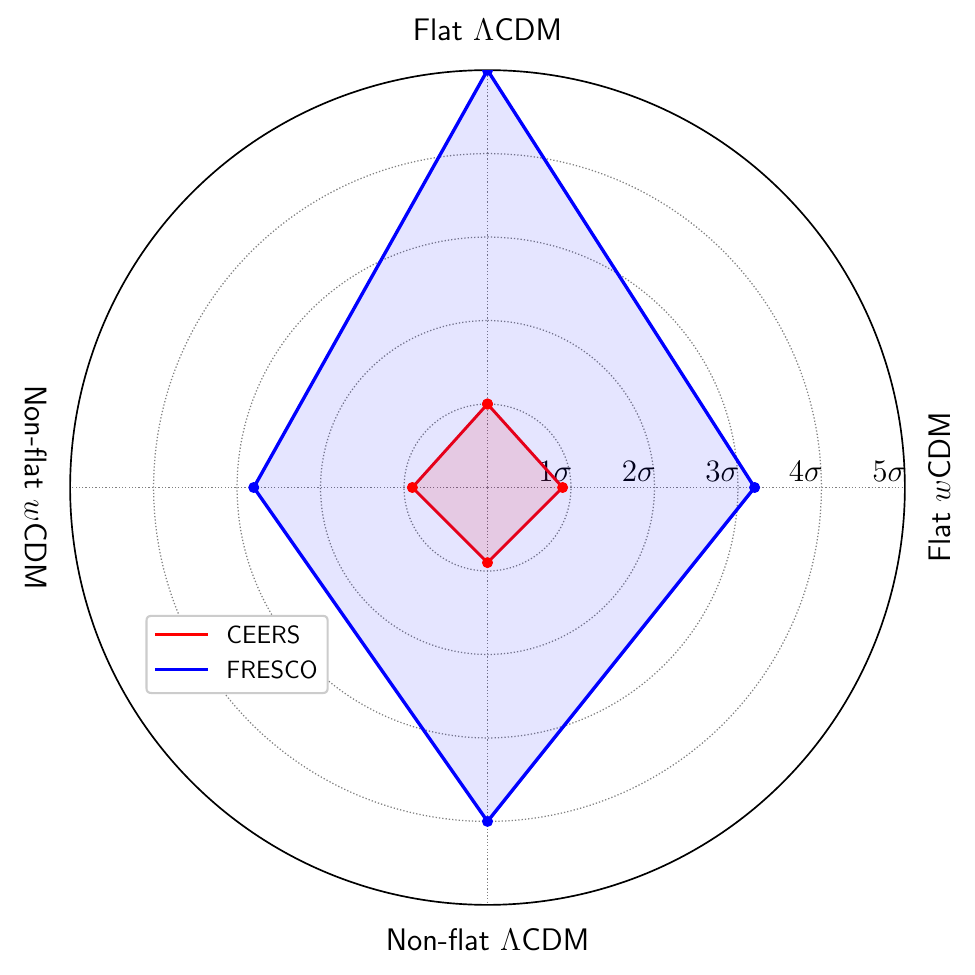}\hfill\includegraphics[width=0.49\textwidth]{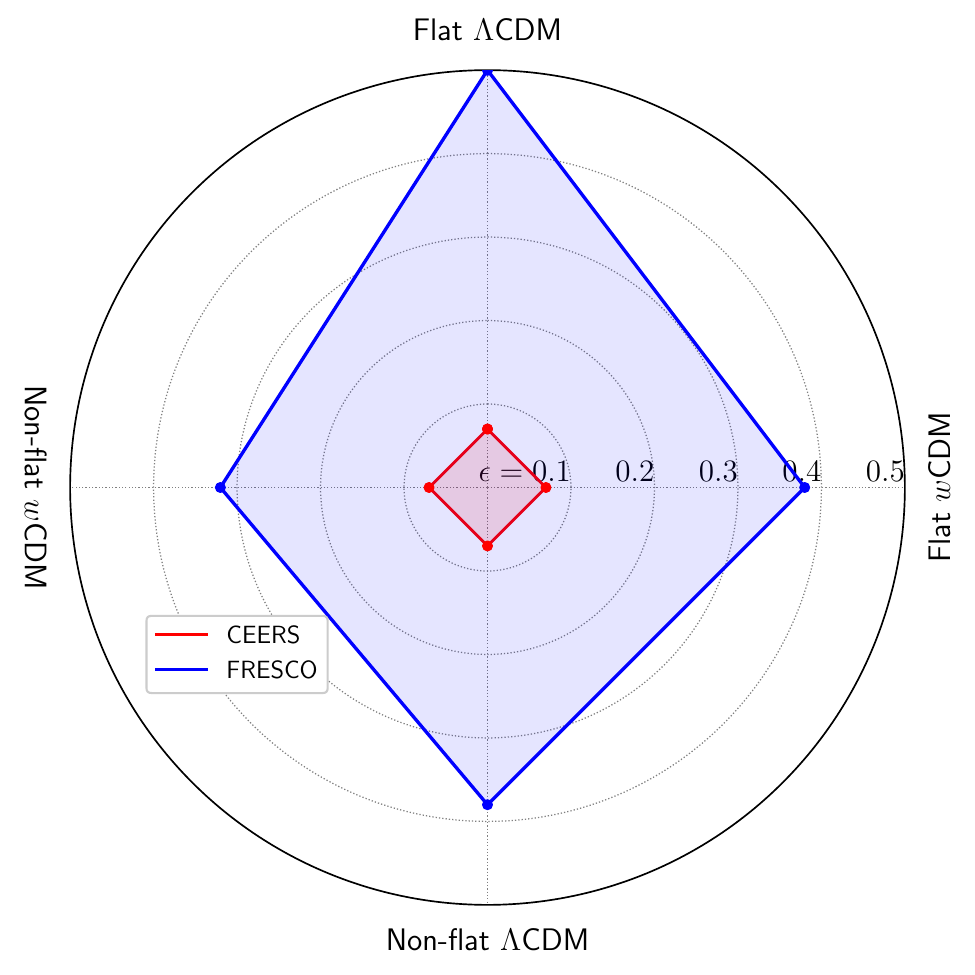}
\caption{Radar plots summarizing our main results, with each spoke corresponding to one of the four cosmological models studied. \textit{Left panel}: radar plot showing the Gaussian-equivalent one-sided significance at which values of the baryon-to-star conversion efficiency $\epsilon<0.2$ are excluded in light of the photometric CEERS (red) and spectroscopic FRESCO (blue) samples. For visualization purposes, the significance is capped at $5.0\sigma$. \textit{Right panel}: radar plot showing the $95\%$ one-sided lower limits on $\epsilon$ for the same samples. For visualization purposes, the values are capped at $0.5$.}
\label{fig:radar}
\end{figure*}

Our main results are visually summarized in the radar plots of Fig.~\ref{fig:radar}. There we show the Gaussian-equivalent one-sided significance at which values of the baryon-to-star conversion efficiency $\epsilon<0.2$ are excluded (left), and $95\%$ one-sided lower limits on $\epsilon$, for various different choices of galaxy sample and cosmological model (right). Across all cosmological models considered earlier, we find that photometric data from CEERS provides weak constraints on the baryon-to-star conversion efficiency, albeit hinting at values somewhat larger than those typically assumed in astrophysical contexts, consistently with the earlier results of~\citet{Boylan-Kolchin:2022kae}. This weak hint is dramatically confirmed by the FRESCO spectroscopic dataset, which robustly requires $\epsilon \gtrsim 0.3$-$0.5$ across all cosmological models, with the benchmark $\epsilon_0=0.1$ value always excluded at $>5.0\sigma$ significance within all four cosmological models. All the extensions to $\Lambda$CDM we consider relax our constraints but do not qualitatively alter them. Moreover, we find no evidence of deviations from a cosmological constant and spatial flatness. We therefore conclude that, once cosmological uncertainties are marginalized over, cumulative comoving stellar mass density measurements at high redshifts primarily constrain $\epsilon$, and show no indication for departures from the flat $\Lambda$CDM model. While we have explicitly tested this for a limited set of representative models, our expectation is that this conclusion should hold more generally for cosmological models that smoothly modify the expansion history relative to $\Lambda$CDM.

Let us return to one of the key questions raised by early JWST observations, i.e.\ whether the abundance of massive galaxies at $z \gtrsim 8$ signals a breakdown of $\Lambda$CDM or instead points to more efficient early galaxy and/or star formation~\citep{Labbe:2022ahb,Xiao:2023ghw}. Our results provide a clear answer: within the class of models we have considered, which are arguably among the simplest and best motivated extensions to $\Lambda$CDM, modifying the background expansion history, and with it the growth of structure, is not sufficient to account for the observed abundances. As discussed earlier, cosmology affects the cumulative stellar mass density primarily through the growth of structure, and in particular through the ratio $D(z)/D(0)$ when comparing models at fixed $\sigma_8$. However, as is clear from Fig.~\ref{fig:varywomegak}, even allowing for very generous variations in $w$ and $\Omega_K$, the resulting changes in the growth factor at the redshifts of interest are typically at the level of tens of percent at best. On the other hand, the values of $\epsilon$ required by the FRESCO dataset correspond to nearly order unity shifts relative to standard astrophysical expectations. This mismatch in scales implies that generous yet reasonable modifications of the expansion history cannot bridge this gap: \textit{the origin of the ``JWST tension'' is very unlikely to be cosmological in nature, and should instead be sought in the astrophysics of galaxy formation}.

Our results can be critically compared to earlier works based on the same galaxy samples, most notably those of~\citet{Boylan-Kolchin:2022kae} and~\citet{Xiao:2023ghw} for the CEERS and FRESCO samples respectively. These works already highlighted the potential tension with $\Lambda$CDM, but their analyses on the cosmological side were based on a direct comparison between observed and theoretically allowed cumulative stellar mass densities for fixed cosmological and astrophysical parameters, effectively implementing a binary consistency test. We have instead adopted a fully Bayesian approach, which allowed us to simultaneously vary cosmological and astrophysical parameters, while consistently accounting for and propagating uncertainties. As expected, this leads to overall weaker limits on $\epsilon$, which nevertheless remain consistent with previous findings. Importantly, we have explicitly demonstrated that the earlier results are largely robust against marginalization over cosmological parameters, and are therefore not driven by assumptions on the underlying cosmological model.

While our results clearly indicate that the observed tension is astrophysical rather than cosmological in nature, the astrophysical interpretation of our findings remains open. Given the cosmological focus of this work, we refrain from detailed speculation on the mechanisms which could lead to the large values of $\epsilon$ implied by our results. Nevertheless, it is useful to briefly discuss possible effects going in the required direction. The values $\epsilon \approx 0.7$ implied by the FRESCO dataset are significantly higher than typical expectations from empirical stellar-to-halo mass relations, and may reflect residual systematic uncertainties in the inferred stellar masses and/or number densities. On the observational side, systematic uncertainties in stellar mass estimates are known to arise, for instance, from assumptions about the initial mass function (IMF), dust attenuation, and star formation histories, all of which could plausibly bias the inferred masses high. As is clear from Fig.~\ref{fig:varywomegak}, our systems all lie on the steep, exponential decline of the high-mass end of the stellar mass function. Therefore, even rather modest systematic shifts can have a very significant impact on the required values of $\epsilon$. On the same token, lensing magnification may also play a role in enhancing the apparent abundance of massive systems. Moreover, we note that most of the UV luminosity is contributed by massive stars. Therefore, a top-heavy IMF of Population-III stars is expected to reduce the required star formation efficiency~\citep[see for instance][]{Wang:2023xmm,Hutter:2024cvr,Ziegler:2025plz}. Finally, the inferred cumulative stellar mass densities may also be affected by the limited size of the current galaxy samples. On the theoretical side, additional uncertainties arise from the calibration of the halo mass function at very high redshift. Existing fitting functions are calibrated on numerical simulations over a limited range of masses and redshifts, and are extrapolated in the regime relevant for JWST observations~\citep{Reed:2006rw,Tinker:2008ff,Watson:2012mt}. Moreover, additional uncertainties can arise from baryonic processes which are yet to be fully understood~\citep{Cui:2011xc,Bocquet:2015pva,McCarthy:2016mry}. Improved calibration of the high-redshift HMF, informed by dedicated simulations, will thus be a crucial step towards assessing the robustness of our results.

Our findings indicate that resolving the ``JWST tension'' will ultimately require improvements on both the data and modelling sides. On the data side, larger spectroscopic samples will help reduce statistical uncertainties and confirm the nature of the most massive high-redshift systems, thanks to their more robust redshift and stellar mass determinations. At the same time, deep JWST photometry can enable more reliable SED fits which, when combined with spectroscopic information, can help break degeneracies with stellar ages and star-formation history, reducing systematic uncertainties in stellar mass estimates~\citep{Luberto:2025bct}. On the theory side, a key role will be played by progress in the calibration of the high-redshift HMF. Advances on these and related points will help determine whether the origin of the ``JWST tension'' is astrophysical, or lies in new fundamental physics~\citep[see also][for a recent discussion on how to distinguish between these two scenarios]{Mehta:2025xwf}.

\section{Conclusions}
\label{sec:conclusions}

We have revisited the issue of whether the unexpectedly high abundance of massive, high-redshift galaxies observed by JWST requires new physics beyond $\Lambda$CDM, or calls for more astrophysical explanations. Our analysis goes beyond earlier works in three respects. Firstly, while building upon the same ``baryon availability'' argument adopted earlier~\citet{Boylan-Kolchin:2022kae,Xiao:2023ghw}, we recast the problem using a full likelihood-based analysis which consistently accounts for the fiducial cosmology dependence of the inferred cumulative stellar mass density and stellar masses: this improves over the (simple but conservative) direct comparison, binary consistency test approach adopted earlier, allowing us to consistently propagate uncertainties and ultimately derive a full probability distribution for the baryon-to-star conversion efficiency $\epsilon$. Next, to improve the robustness of our results we consider not only the most extreme galaxies within the (photometric) CEERS sample (which provided the first indications of a possible tension with $\Lambda$CDM), but also those within the spectroscopic FRESCO sample. Last but not least, we go beyond the $\Lambda$CDM model and consider some of its simplest extension which relax assumptions about the dark energy component and spatial curvature, to assess whether the requirement for high values of $\epsilon$ can be relaxed going beyond $\Lambda$CDM. In short, our work moves beyond earlier fixed cosmological assumptions and per-object minimum-efficiency arguments, performing a full Bayesian analysis of the most extreme galaxies within the CEERS and FRESCO samples. We stress that our goal is not to rule out $\Lambda$CDM, but rather to quantify in a fully probabilistic framework what is the range of baryon-to-star conversion efficiencies required once cosmological uncertainties are consistently accounted for.

Our key results are summarized in Fig.~\ref{fig:radar}, and are in qualitative agreement with the earlier findings of~\citet{Boylan-Kolchin:2022kae} and~\citet{Xiao:2023ghw}, while being placed on a more robust statistical footing. In a broad brush, we find that the CEERS photometric sample actually leads to comparatively weak constraints on $\epsilon$: while there is a mild preference for an enhanced formation efficiency of massive stars, lower values ($\epsilon \approx 0.1$) in agreement with astrophysical expectations are perfectly allowed at $95\%$~C.L., because of the comparatively large uncertainties. Importantly, this picture does not change significantly when moving beyond the flat $\Lambda$CDM model. On the other hand, the constraints we derive from FRESCO spectroscopic data are much tighter within all cosmological models, indicating $\epsilon \gtrsim 0.3-0.5$ at $95\%$~C.L., with values $\epsilon<0.2$ ruled out at no less than $2.0\sigma$, and values $\epsilon<0.1$ always ruled out at more than $5.0\sigma$ within all cosmological models. We obtain especially tight limits within $\Lambda$CDM, where we find $\epsilon \gtrsim 0.63$ at $95\%$~C.L., whereas values $\epsilon<0.2$ are ruled out at $>5.0\sigma$. While moving beyond $\Lambda$CDM somewhat relaxes these figures, the need for large values of $\epsilon$ is only softened but not removed. At the same time, there is no evidence for new physics, as both the dark energy equation of state $w$ and the spatial curvature parameter $\Omega_K$ remain consistent with $-1$ and $0$ respectively. Our main finding can therefore be summarized as follows: \textit{the origin of the ``JWST tension'' is very unlikely to be cosmological in nature, and should instead be sought in the astrophysics of galaxy formation}.

The astrophysical implications of our findings remain open, and we have refrained from speculating on possible effects which could lead to the large values of $\epsilon$ implied by our results (although we note that our limits assume that the IMF of high-redshift, massive stars is similar to the present-day IMF). Nevertheless, we can confidently state that determining the origin of the ``JWST tension'' will require improvements on both the data and modelling sides, with larger spectroscopic samples and deeper photometry playing an important role in the former case. On the theory side, we anticipate that it could be interesting to explore additional scenarios, including for instance a (astrophysically well-motivated) redshift- and/or mass-dependent efficiency $\epsilon(M_{\text{halo}},z)$, or looking at the impact of mass functions beyond the Sheth-Tormen one. It is in fact worth recalling that our framework extrapolates at very high redshifts a model which is only well tested at much lower redshifts. While on the cosmology side we have only considered models directly altering the expansion history, and with it the growth of structure, it could be interesting to go beyond this class of models and consider those where the properties of primordial fluctuations are altered, including e.g.\ models featuring non-Gaussianities or modifications to the primordial power spectrum, all of which could arise from non-standard models of cosmic inflation. At the same time, it is unlikely that models which only directly alter the expansion history, including e.g.\ those hinted to by the recent DESI observations~\citep[see for example][]{Colgain:2024xqj,Carloni:2024zpl,Giare:2024smz,Wang:2024dka,Yang:2024kdo,Li:2024qso,Giare:2024gpk,Jiang:2024xnu,Giare:2024oil,Giare:2025pzu,Scherer:2025esj,Chaudhary:2025pcc,Zhang:2025dwu,Li:2025vuh,Capozziello:2025qmh}, may loosen the high efficiency requirement while remaining consistent with high-redshift cosmological observations~\citep[for the difficulties in this sense, see for instance][]{Chakraborty:2025yuo}. Nevertheless, this point remains to be fully explored.

In closing we note that, for the sake of clarity, we have focused our analysis on the CEERS and FRESCO samples from JWST, which are among the best-studied high-redshift galaxy datasets currently available. It will of course be interesting to extend this work to data from additional JWST programs, especially spectroscopic ones, as well as individual objects that have attracted significant attention in the literature. An example in this sense could be the red, dust-rich galaxy EGS-Z11-R0~\citep{Rodighiero:2026ghw}, although a preliminary assessment shows that its properties are not sufficiently extreme to be in tension with $\Lambda$CDM. We leave a full study of these points, and those mentioned earlier, to future work.

\section*{Acknowledgements}

\noindent We thank Mike Boylan-Kolchin, Ivo Labb\'{e}, and Nicola Menci for various very useful discussions, and Steve Murray for promptly fixing an important bug in \texttt{hmf} which we identified while this work was being completed. S.V.\ acknowledges support from the University of Trento and the Provincia Autonoma di Trento (PAT, Autonomous Province of Trento) through the UniTrento Internal Call for Research 2023 grant ``Searching for Dark Energy off the beaten track'' (DARKTRACK, grant agreement no.\ E63C22000500003), and from the Istituto Nazionale di Fisica Nucleare (INFN) through the Commissione Scientifica Nazionale 4 (CSN4) Iniziativa Specifica ``Quantum Fields in Gravity, Cosmology and Black Holes'' (FLAG). A.L.\ is partially supported by the Black Hole Initiative at Harvard University, which is funded by grants from the John Templeton Foundation and the Gordon and Betty Moore Foundation. This publication is based upon work from the COST Action CA21136 ``Addressing observational tensions in cosmology with systematics and fundamental physics'' (CosmoVerse), supported by COST (European Cooperation in Science and Technology).

\footnotesize

\bibliography{JWST}

\begin{thebibliography}{178}
\expandafter\ifx\csname natexlab\endcsname\relax\def\natexlab#1{#1}\fi
\providecommand{\url}[1]{\texttt{#1}}
\providecommand{\href}[2]{#2}
\providecommand{\path}[1]{#1}
\providecommand{\DOIprefix}{doi:}
\providecommand{\ArXivprefix}{arXiv:}
\providecommand{\URLprefix}{URL: }
\providecommand{\Pubmedprefix}{pmid:}
\providecommand{\doi}[1]{\href{http://dx.doi.org/#1}{\path{#1}}}
\providecommand{\Pubmed}[1]{\href{pmid:#1}{\path{#1}}}
\providecommand{\bibinfo}[2]{#2}
\ifx\xfnm\relax \def\xfnm[#1]{\unskip,\space#1}\fi
\bibitem[{Abbott et~al.(2024)}]{DES:2024jxu}
\bibinfo{author}{Abbott, T.M.C.}, et~al. (\bibinfo{collaboration}{DES}),
  \bibinfo{year}{2024}.
\newblock \bibinfo{title}{{The Dark Energy Survey: Cosmology Results with
  {\ensuremath{\sim}}1500 New High-redshift Type Ia Supernovae Using the Full 5
  yr Data Set}}.
\newblock \bibinfo{journal}{Astrophys. J. Lett.} \bibinfo{volume}{973},
  \bibinfo{pages}{L14}.
\newblock \DOIprefix\doi{10.3847/2041-8213/ad6f9f},
  \href{http://arxiv.org/abs/2401.02929}{{\tt arXiv:2401.02929}}.
\bibitem[{Abdul~Karim et~al.(2025)}]{DESI:2025zgx}
\bibinfo{author}{Abdul~Karim, M.}, et~al. (\bibinfo{collaboration}{DESI}),
  \bibinfo{year}{2025}.
\newblock \bibinfo{title}{{DESI DR2 results. II. Measurements of baryon
  acoustic oscillations and cosmological constraints}}.
\newblock \bibinfo{journal}{Phys. Rev. D} \bibinfo{volume}{112},
  \bibinfo{pages}{083515}.
\newblock \DOIprefix\doi{10.1103/tr6y-kpc6},
  \href{http://arxiv.org/abs/2503.14738}{{\tt arXiv:2503.14738}}.
\bibitem[{Adame et~al.(2025)}]{DESI:2024mwx}
\bibinfo{author}{Adame, A.G.}, et~al. (\bibinfo{collaboration}{DESI}),
  \bibinfo{year}{2025}.
\newblock \bibinfo{title}{{DESI 2024 VI: cosmological constraints from the
  measurements of baryon acoustic oscillations}}.
\newblock \bibinfo{journal}{JCAP} \bibinfo{volume}{02}, \bibinfo{pages}{021}.
\newblock \DOIprefix\doi{10.1088/1475-7516/2025/02/021},
  \href{http://arxiv.org/abs/2404.03002}{{\tt arXiv:2404.03002}}.
\bibitem[{Adil et~al.(2023)Adil, Mukhopadhyay, Sen and Vagnozzi}]{Adil:2023ara}
\bibinfo{author}{Adil, S.A.}, \bibinfo{author}{Mukhopadhyay, U.},
  \bibinfo{author}{Sen, A.A.}, \bibinfo{author}{Vagnozzi, S.},
  \bibinfo{year}{2023}.
\newblock \bibinfo{title}{{Dark energy in light of the early JWST observations:
  case for a negative cosmological constant?}}
\newblock \bibinfo{journal}{JCAP} \bibinfo{volume}{10}, \bibinfo{pages}{072}.
\newblock \DOIprefix\doi{10.1088/1475-7516/2023/10/072},
  \href{http://arxiv.org/abs/2307.12763}{{\tt arXiv:2307.12763}}.
\bibitem[{Aghanim et~al.(2020)}]{Planck:2018vyg}
\bibinfo{author}{Aghanim, N.}, et~al. (\bibinfo{collaboration}{Planck}),
  \bibinfo{year}{2020}.
\newblock \bibinfo{title}{{Planck 2018 results. VI. Cosmological parameters}}.
\newblock \bibinfo{journal}{Astron. Astrophys.} \bibinfo{volume}{641},
  \bibinfo{pages}{A6}.
\newblock \DOIprefix\doi{10.1051/0004-6361/201833910},
  \href{http://arxiv.org/abs/1807.06209}{{\tt arXiv:1807.06209}}.
  \bibinfo{note}{[Erratum: Astron.Astrophys. 652, C4 (2021)]}.
\bibitem[{Akarsu et~al.(2024)Akarsu, Colg{\'a}in, Sen and
  Sheikh-Jabbari}]{Akarsu:2024qiq}
\bibinfo{author}{Akarsu, {\"O}.}, \bibinfo{author}{Colg{\'a}in, E.{\'O}.},
  \bibinfo{author}{Sen, A.A.}, \bibinfo{author}{Sheikh-Jabbari, M.M.},
  \bibinfo{year}{2024}.
\newblock \bibinfo{title}{{{\ensuremath{\Lambda}}CDM Tensions: Localising
  Missing Physics through Consistency Checks}}.
\newblock \bibinfo{journal}{Universe} \bibinfo{volume}{10},
  \bibinfo{pages}{305}.
\newblock \DOIprefix\doi{10.3390/universe10080305},
  \href{http://arxiv.org/abs/2402.04767}{{\tt arXiv:2402.04767}}.
\bibitem[{Akarsu et~al.(2023)Akarsu, Di~Valentino, Kumar, Ozyigit and
  Sharma}]{Akarsu:2021max}
\bibinfo{author}{Akarsu, O.}, \bibinfo{author}{Di~Valentino, E.},
  \bibinfo{author}{Kumar, S.}, \bibinfo{author}{Ozyigit, M.},
  \bibinfo{author}{Sharma, S.}, \bibinfo{year}{2023}.
\newblock \bibinfo{title}{{Testing spatial curvature and anisotropic expansion
  on top of the \ensuremath{\Lambda}CDM model}}.
\newblock \bibinfo{journal}{Phys. Dark Univ.} \bibinfo{volume}{39},
  \bibinfo{pages}{101162}.
\newblock \DOIprefix\doi{10.1016/j.dark.2022.101162},
  \href{http://arxiv.org/abs/2112.07807}{{\tt arXiv:2112.07807}}.
\bibitem[{Bargiacchi et~al.(2022)Bargiacchi, Benetti, Capozziello, Lusso,
  Risaliti and Signorini}]{Bargiacchi:2021hdp}
\bibinfo{author}{Bargiacchi, G.}, \bibinfo{author}{Benetti, M.},
  \bibinfo{author}{Capozziello, S.}, \bibinfo{author}{Lusso, E.},
  \bibinfo{author}{Risaliti, G.}, \bibinfo{author}{Signorini, M.},
  \bibinfo{year}{2022}.
\newblock \bibinfo{title}{{Quasar cosmology: dark energy evolution and spatial
  curvature}}.
\newblock \bibinfo{journal}{Mon. Not. Roy. Astron. Soc.} \bibinfo{volume}{515},
  \bibinfo{pages}{1795--1806}.
\newblock \DOIprefix\doi{10.1093/mnras/stac1941},
  \href{http://arxiv.org/abs/2111.02420}{{\tt arXiv:2111.02420}}.
\bibitem[{Baryakhtar et~al.(2024)Baryakhtar, Simon and
  Weiner}]{Baryakhtar:2024rky}
\bibinfo{author}{Baryakhtar, M.}, \bibinfo{author}{Simon, O.},
  \bibinfo{author}{Weiner, Z.J.}, \bibinfo{year}{2024}.
\newblock \bibinfo{title}{{Cosmology with varying fundamental constants from
  hyperlight, coupled scalars}}.
\newblock \bibinfo{journal}{Phys. Rev. D} \bibinfo{volume}{110},
  \bibinfo{pages}{083505}.
\newblock \DOIprefix\doi{10.1103/PhysRevD.110.083505},
  \href{http://arxiv.org/abs/2405.10358}{{\tt arXiv:2405.10358}}.
\bibitem[{Behroozi et~al.(2013a)Behroozi, Wechsler and
  Conroy}]{Behroozi:2012sp}
\bibinfo{author}{Behroozi, P.S.}, \bibinfo{author}{Wechsler, R.H.},
  \bibinfo{author}{Conroy, C.}, \bibinfo{year}{2013}a.
\newblock \bibinfo{title}{{On the Lack of Evolution in Galaxy Star Formation
  Efficiency}}.
\newblock \bibinfo{journal}{Astrophys. J. Lett.} \bibinfo{volume}{762},
  \bibinfo{pages}{L31}.
\newblock \DOIprefix\doi{10.1088/2041-8205/762/2/L31},
  \href{http://arxiv.org/abs/1209.3013}{{\tt arXiv:1209.3013}}.
\bibitem[{Behroozi et~al.(2013b)Behroozi, Wechsler and
  Conroy}]{Behroozi:2012iw}
\bibinfo{author}{Behroozi, P.S.}, \bibinfo{author}{Wechsler, R.H.},
  \bibinfo{author}{Conroy, C.}, \bibinfo{year}{2013}b.
\newblock \bibinfo{title}{{The Average Star Formation Histories of Galaxies in
  Dark Matter Halos from $z=$0-8}}.
\newblock \bibinfo{journal}{Astrophys. J.} \bibinfo{volume}{770},
  \bibinfo{pages}{57}.
\newblock \DOIprefix\doi{10.1088/0004-637X/770/1/57},
  \href{http://arxiv.org/abs/1207.6105}{{\tt arXiv:1207.6105}}.
\bibitem[{Bel et~al.(2022)Bel, Larena, Maartens, Marinoni and
  Perenon}]{Bel:2022iuf}
\bibinfo{author}{Bel, J.}, \bibinfo{author}{Larena, J.},
  \bibinfo{author}{Maartens, R.}, \bibinfo{author}{Marinoni, C.},
  \bibinfo{author}{Perenon, L.}, \bibinfo{year}{2022}.
\newblock \bibinfo{title}{{Constraining spatial curvature with large-scale
  structure}}.
\newblock \bibinfo{journal}{JCAP} \bibinfo{volume}{09}, \bibinfo{pages}{076}.
\newblock \DOIprefix\doi{10.1088/1475-7516/2022/09/076},
  \href{http://arxiv.org/abs/2206.03059}{{\tt arXiv:2206.03059}}.
\bibitem[{Benisty and Staicova(2021)}]{Benisty:2020otr}
\bibinfo{author}{Benisty, D.}, \bibinfo{author}{Staicova, D.},
  \bibinfo{year}{2021}.
\newblock \bibinfo{title}{{Testing late-time cosmic acceleration with
  uncorrelated baryon acoustic oscillation dataset}}.
\newblock \bibinfo{journal}{Astron. Astrophys.} \bibinfo{volume}{647},
  \bibinfo{pages}{A38}.
\newblock \DOIprefix\doi{10.1051/0004-6361/202039502},
  \href{http://arxiv.org/abs/2009.10701}{{\tt arXiv:2009.10701}}.
\bibitem[{Biagetti et~al.(2023)Biagetti, Franciolini and
  Riotto}]{Biagetti:2022ode}
\bibinfo{author}{Biagetti, M.}, \bibinfo{author}{Franciolini, G.},
  \bibinfo{author}{Riotto, A.}, \bibinfo{year}{2023}.
\newblock \bibinfo{title}{{High-redshift JWST Observations and Primordial
  Non-Gaussianity}}.
\newblock \bibinfo{journal}{Astrophys. J.} \bibinfo{volume}{944},
  \bibinfo{pages}{113}.
\newblock \DOIprefix\doi{10.3847/1538-4357/acb5ea},
  \href{http://arxiv.org/abs/2210.04812}{{\tt arXiv:2210.04812}}.
\bibitem[{Bigiel et~al.(2008)Bigiel, Leroy, Walter, Brinks, de~Blok, Madore and
  Thornley}]{Bigiel:2008jw}
\bibinfo{author}{Bigiel, F.}, \bibinfo{author}{Leroy, A.},
  \bibinfo{author}{Walter, F.}, \bibinfo{author}{Brinks, E.},
  \bibinfo{author}{de~Blok, W.J.G.}, \bibinfo{author}{Madore, B.},
  \bibinfo{author}{Thornley, M.D.}, \bibinfo{year}{2008}.
\newblock \bibinfo{title}{{The Star Formation Law in Nearby Galaxies on Sub-Kpc
  Scales}}.
\newblock \bibinfo{journal}{Astron. J.} \bibinfo{volume}{136},
  \bibinfo{pages}{2846--2871}.
\newblock \DOIprefix\doi{10.1088/0004-6256/136/6/2846},
  \href{http://arxiv.org/abs/0810.2541}{{\tt arXiv:0810.2541}}.
\bibitem[{Binici et~al.(2024)Binici, Deliduman and Dilsiz}]{Binici:2024smk}
\bibinfo{author}{Binici, S.S.}, \bibinfo{author}{Deliduman, C.},
  \bibinfo{author}{Dilsiz, F.{\c{S}}.}, \bibinfo{year}{2024}.
\newblock \bibinfo{title}{{The ages of the oldest astrophysical objects in an
  ellipsoidal universe}}.
\newblock \bibinfo{journal}{Phys. Dark Univ.} \bibinfo{volume}{46},
  \bibinfo{pages}{101600}.
\newblock \DOIprefix\doi{10.1016/j.dark.2024.101600},
  \href{http://arxiv.org/abs/2402.16646}{{\tt arXiv:2402.16646}}.
\bibitem[{Bird et~al.(2024)Bird, Chang, Cui and Yang}]{Bird:2023pkr}
\bibinfo{author}{Bird, S.}, \bibinfo{author}{Chang, C.F.},
  \bibinfo{author}{Cui, Y.}, \bibinfo{author}{Yang, D.}, \bibinfo{year}{2024}.
\newblock \bibinfo{title}{{Enhanced early galaxy formation in JWST from axion
  dark matter?}}
\newblock \bibinfo{journal}{Phys. Lett. B} \bibinfo{volume}{858},
  \bibinfo{pages}{139062}.
\newblock \DOIprefix\doi{10.1016/j.physletb.2024.139062},
  \href{http://arxiv.org/abs/2307.10302}{{\tt arXiv:2307.10302}}.
\bibitem[{Bocquet et~al.(2016)Bocquet, Saro, Dolag and Mohr}]{Bocquet:2015pva}
\bibinfo{author}{Bocquet, S.}, \bibinfo{author}{Saro, A.},
  \bibinfo{author}{Dolag, K.}, \bibinfo{author}{Mohr, J.J.},
  \bibinfo{year}{2016}.
\newblock \bibinfo{title}{{Halo mass function: Baryon impact, fitting formulae
  and implications for cluster cosmology}}.
\newblock \bibinfo{journal}{Mon. Not. Roy. Astron. Soc.} \bibinfo{volume}{456},
  \bibinfo{pages}{2361--2373}.
\newblock \DOIprefix\doi{10.1093/mnras/stv2657},
  \href{http://arxiv.org/abs/1502.07357}{{\tt arXiv:1502.07357}}.
\bibitem[{Boylan-Kolchin(2023)}]{Boylan-Kolchin:2022kae}
\bibinfo{author}{Boylan-Kolchin, M.}, \bibinfo{year}{2023}.
\newblock \bibinfo{title}{{Stress testing {\ensuremath{\Lambda}}CDM with
  high-redshift galaxy candidates}}.
\newblock \bibinfo{journal}{Nature Astron.} \bibinfo{volume}{7},
  \bibinfo{pages}{731--735}.
\newblock \DOIprefix\doi{10.1038/s41550-023-01937-7},
  \href{http://arxiv.org/abs/2208.01611}{{\tt arXiv:2208.01611}}.
\bibitem[{Boylan-Kolchin(2025)}]{Boylan-Kolchin:2024roq}
\bibinfo{author}{Boylan-Kolchin, M.}, \bibinfo{year}{2025}.
\newblock \bibinfo{title}{{Accelerated by dark matter: a high-redshift pathway
  to efficient galaxy-scale star formation}}.
\newblock \bibinfo{journal}{Mon. Not. Roy. Astron. Soc.} \bibinfo{volume}{538},
  \bibinfo{pages}{3210--3218}.
\newblock \DOIprefix\doi{10.1093/mnras/staf471},
  \href{http://arxiv.org/abs/2407.10900}{{\tt arXiv:2407.10900}}.
\bibitem[{Cao et~al.(2021)Cao, Ryan and Ratra}]{Cao:2021ldv}
\bibinfo{author}{Cao, S.}, \bibinfo{author}{Ryan, J.}, \bibinfo{author}{Ratra,
  B.}, \bibinfo{year}{2021}.
\newblock \bibinfo{title}{{Using Pantheon and DES supernova, baryon acoustic
  oscillation, and Hubble parameter data to constrain the Hubble constant, dark
  energy dynamics, and spatial curvature}}.
\newblock \bibinfo{journal}{Mon. Not. Roy. Astron. Soc.} \bibinfo{volume}{504},
  \bibinfo{pages}{300--310}.
\newblock \DOIprefix\doi{10.1093/mnras/stab942},
  \href{http://arxiv.org/abs/2101.08817}{{\tt arXiv:2101.08817}}.
\bibitem[{Capozziello et~al.(2026)Capozziello, Chaudhary, Harko and
  Mustafa}]{Capozziello:2025qmh}
\bibinfo{author}{Capozziello, S.}, \bibinfo{author}{Chaudhary, H.},
  \bibinfo{author}{Harko, T.}, \bibinfo{author}{Mustafa, G.},
  \bibinfo{year}{2026}.
\newblock \bibinfo{title}{{Is dark energy dynamical in the DESI era? A critical
  review}}.
\newblock \bibinfo{journal}{Phys. Dark Univ.} \bibinfo{volume}{51},
  \bibinfo{pages}{102196}.
\newblock \DOIprefix\doi{10.1016/j.dark.2025.102196},
  \href{http://arxiv.org/abs/2512.10585}{{\tt arXiv:2512.10585}}.
\bibitem[{Carloni et~al.(2025)Carloni, Luongo and Muccino}]{Carloni:2024zpl}
\bibinfo{author}{Carloni, Y.}, \bibinfo{author}{Luongo, O.},
  \bibinfo{author}{Muccino, M.}, \bibinfo{year}{2025}.
\newblock \bibinfo{title}{{Does dark energy really revive using DESI 2024
  data?}}
\newblock \bibinfo{journal}{Phys. Rev. D} \bibinfo{volume}{111},
  \bibinfo{pages}{023512}.
\newblock \DOIprefix\doi{10.1103/PhysRevD.111.023512},
  \href{http://arxiv.org/abs/2404.12068}{{\tt arXiv:2404.12068}}.
\bibitem[{Casey et~al.(2024)}]{Casey:2023ghw}
\bibinfo{author}{Casey, C.M.}, et~al., \bibinfo{year}{2024}.
\newblock \bibinfo{title}{{COSMOS-Web: Intrinsically Luminous $z \gtrsim 10$
  Galaxy Candidates Test Early Stellar Mass Assembly}}.
\newblock \bibinfo{journal}{Astrophys. J.} \bibinfo{volume}{965},
  \bibinfo{pages}{98}.
\newblock \DOIprefix\doi{10.3847/1538-4357/ad2075},
  \href{http://arxiv.org/abs/2308.10932}{{\tt arXiv:2308.10932}}.
\bibitem[{Chakraborty et~al.(2025)Chakraborty, Choudhury, Sen and
  Mukherjee}]{Chakraborty:2025yuo}
\bibinfo{author}{Chakraborty, A.}, \bibinfo{author}{Choudhury, T.R.},
  \bibinfo{author}{Sen, A.A.}, \bibinfo{author}{Mukherjee, P.},
  \bibinfo{year}{2025}.
\newblock \bibinfo{title}{{Can an Anti-de Sitter Vacuum in the Dark Energy
  Sector Explain JWST High-Redshift Galaxy and Reionization Observations?}}
  \DOIprefix\doi{10.1093/mnras/stag269},
  \href{http://arxiv.org/abs/2509.02431}{{\tt arXiv:2509.02431}}.
\bibitem[{Chaudhary et~al.(2025)Chaudhary, Capozziello, Sharma and
  Mustafa}]{Chaudhary:2025pcc}
\bibinfo{author}{Chaudhary, H.}, \bibinfo{author}{Capozziello, S.},
  \bibinfo{author}{Sharma, V.K.}, \bibinfo{author}{Mustafa, G.},
  \bibinfo{year}{2025}.
\newblock \bibinfo{title}{{Does DESI DR2 Challenge {\ensuremath{\Lambda}}CDM
  Paradigm?}}
\newblock \bibinfo{journal}{Astrophys. J.} \bibinfo{volume}{992},
  \bibinfo{pages}{194}.
\newblock \DOIprefix\doi{10.3847/1538-4357/ae0458},
  \href{http://arxiv.org/abs/2507.21607}{{\tt arXiv:2507.21607}}.
\bibitem[{Chudaykin et~al.(2021)Chudaykin, Dolgikh and
  Ivanov}]{Chudaykin:2020ghx}
\bibinfo{author}{Chudaykin, A.}, \bibinfo{author}{Dolgikh, K.},
  \bibinfo{author}{Ivanov, M.M.}, \bibinfo{year}{2021}.
\newblock \bibinfo{title}{{Constraints on the curvature of the Universe and
  dynamical dark energy from the Full-shape and BAO data}}.
\newblock \bibinfo{journal}{Phys. Rev. D} \bibinfo{volume}{103},
  \bibinfo{pages}{023507}.
\newblock \DOIprefix\doi{10.1103/PhysRevD.103.023507},
  \href{http://arxiv.org/abs/2009.10106}{{\tt arXiv:2009.10106}}.
\bibitem[{Chudaykin et~al.(2026)Chudaykin, Ivanov and
  Philcox}]{Chudaykin:2025aux}
\bibinfo{author}{Chudaykin, A.}, \bibinfo{author}{Ivanov, M.M.},
  \bibinfo{author}{Philcox, O.H.E.}, \bibinfo{year}{2026}.
\newblock \bibinfo{title}{{Reanalyzing DESI DR1. I. {\ensuremath{\Lambda}}CDM
  constraints from the power spectrum and bispectrum}}.
\newblock \bibinfo{journal}{Phys. Rev. D} \bibinfo{volume}{113},
  \bibinfo{pages}{063502}.
\newblock \DOIprefix\doi{10.1103/qsnt-dppc},
  \href{http://arxiv.org/abs/2507.13433}{{\tt arXiv:2507.13433}}.
\bibitem[{Colg{\'a}in et~al.(2026)Colg{\'a}in, Dainotti, Capozziello,
  Pourojaghi, Sheikh-Jabbari and Stojkovic}]{Colgain:2024xqj}
\bibinfo{author}{Colg{\'a}in, E.{\'O}.}, \bibinfo{author}{Dainotti, M.G.},
  \bibinfo{author}{Capozziello, S.}, \bibinfo{author}{Pourojaghi, S.},
  \bibinfo{author}{Sheikh-Jabbari, M.M.}, \bibinfo{author}{Stojkovic, D.},
  \bibinfo{year}{2026}.
\newblock \bibinfo{title}{{Does DESI 2024 confirm {\ensuremath{\Lambda}}CDM?}}
\newblock \bibinfo{journal}{JHEAp} \bibinfo{volume}{49},
  \bibinfo{pages}{100428}.
\newblock \DOIprefix\doi{10.1016/j.jheap.2025.100428},
  \href{http://arxiv.org/abs/2404.08633}{{\tt arXiv:2404.08633}}.
\bibitem[{Colg{\'a}in et~al.(2025a)Colg{\'a}in, Pourojaghi and
  Sheikh-Jabbari}]{Colgain:2024ksa}
\bibinfo{author}{Colg{\'a}in, E.{\'O}.}, \bibinfo{author}{Pourojaghi, S.},
  \bibinfo{author}{Sheikh-Jabbari, M.M.}, \bibinfo{year}{2025}a.
\newblock \bibinfo{title}{{Implications of DES 5YR SNe Dataset for $\Lambda
  $CDM}}.
\newblock \bibinfo{journal}{Eur. Phys. J. C} \bibinfo{volume}{85},
  \bibinfo{pages}{286}.
\newblock \DOIprefix\doi{10.1140/epjc/s10052-025-13995-4},
  \href{http://arxiv.org/abs/2406.06389}{{\tt arXiv:2406.06389}}.
\bibitem[{Colg{\'a}in and Sheikh-Jabbari(2025)}]{Colgain:2024mtg}
\bibinfo{author}{Colg{\'a}in, E.{\'O}.}, \bibinfo{author}{Sheikh-Jabbari,
  M.M.}, \bibinfo{year}{2025}.
\newblock \bibinfo{title}{{DESI and SNe: Dynamical Dark Energy, $\Omega_m$
  Tension or Systematics?}}
\newblock \bibinfo{journal}{Mon. Not. Roy. Astron. Soc.} \bibinfo{volume}{542},
  \bibinfo{pages}{L24--L30}.
\newblock \DOIprefix\doi{10.1093/mnrasl/slaf042},
  \href{http://arxiv.org/abs/2412.12905}{{\tt arXiv:2412.12905}}.
\bibitem[{Colg{\'a}in et~al.(2025b)Colg{\'a}in, Sheikh-Jabbari and
  Yin}]{Colgain:2024clf}
\bibinfo{author}{Colg{\'a}in, E.{\'O}.}, \bibinfo{author}{Sheikh-Jabbari,
  M.M.}, \bibinfo{author}{Yin, L.}, \bibinfo{year}{2025}b.
\newblock \bibinfo{title}{{Do high redshift QSOs and GRBs corroborate JWST?}}
\newblock \bibinfo{journal}{Phys. Dark Univ.} \bibinfo{volume}{49},
  \bibinfo{pages}{101975}.
\newblock \DOIprefix\doi{10.1016/j.dark.2025.101975},
  \href{http://arxiv.org/abs/2405.19953}{{\tt arXiv:2405.19953}}.
\bibitem[{Conroy and Wechsler(2009)}]{Conroy:2008dx}
\bibinfo{author}{Conroy, C.}, \bibinfo{author}{Wechsler, R.H.},
  \bibinfo{year}{2009}.
\newblock \bibinfo{title}{{Connecting Galaxies, Halos, and Star Formation Rates
  Across Cosmic Time}}.
\newblock \bibinfo{journal}{Astrophys. J.} \bibinfo{volume}{696},
  \bibinfo{pages}{620--635}.
\newblock \DOIprefix\doi{10.1088/0004-637X/696/1/620},
  \href{http://arxiv.org/abs/0805.3346}{{\tt arXiv:0805.3346}}.
\bibitem[{Costa et~al.(2023)Costa, Ren and Yin}]{Costa:2023cmu}
\bibinfo{author}{Costa, A.A.}, \bibinfo{author}{Ren, Z.}, \bibinfo{author}{Yin,
  Z.}, \bibinfo{year}{2023}.
\newblock \bibinfo{title}{{A bias using the ages of the oldest astrophysical
  objects to address the Hubble tension}}.
\newblock \bibinfo{journal}{Eur. Phys. J. C} \bibinfo{volume}{83},
  \bibinfo{pages}{875}.
\newblock \DOIprefix\doi{10.1140/epjc/s10052-023-12038-0},
  \href{http://arxiv.org/abs/2306.01234}{{\tt arXiv:2306.01234}}.
\bibitem[{Cui et~al.(2012)Cui, Borgani, Dolag, Murante and
  Tornatore}]{Cui:2011xc}
\bibinfo{author}{Cui, W.}, \bibinfo{author}{Borgani, S.},
  \bibinfo{author}{Dolag, K.}, \bibinfo{author}{Murante, G.},
  \bibinfo{author}{Tornatore, L.}, \bibinfo{year}{2012}.
\newblock \bibinfo{title}{{The effects of baryons on the halo mass function}}.
\newblock \bibinfo{journal}{Mon. Not. Roy. Astron. Soc.} \bibinfo{volume}{423},
  \bibinfo{pages}{2279}.
\newblock \DOIprefix\doi{10.1111/j.1365-2966.2012.21037.x},
  \href{http://arxiv.org/abs/1111.3066}{{\tt arXiv:1111.3066}}.
\bibitem[{Das et~al.(2025)Das, Mondol, Singh and Laha}]{Das:2025bnr}
\bibinfo{author}{Das, S.}, \bibinfo{author}{Mondol, R.},
  \bibinfo{author}{Singh, A.}, \bibinfo{author}{Laha, R.},
  \bibinfo{year}{2025}.
\newblock \bibinfo{title}{{Dark Secrets of Baryons: Illuminating Dark
  Matter-Baryon Interactions with JWST}}
  \href{http://arxiv.org/abs/2511.02906}{{\tt arXiv:2511.02906}}.
\bibitem[{Dayal and Giri(2024)}]{Dayal:2023nwi}
\bibinfo{author}{Dayal, P.}, \bibinfo{author}{Giri, S.K.},
  \bibinfo{year}{2024}.
\newblock \bibinfo{title}{{Warm dark matter constraints from the JWST}}.
\newblock \bibinfo{journal}{Mon. Not. Roy. Astron. Soc.} \bibinfo{volume}{528},
  \bibinfo{pages}{2784--2789}.
\newblock \DOIprefix\doi{10.1093/mnras/stae176},
  \href{http://arxiv.org/abs/2303.14239}{{\tt arXiv:2303.14239}}.
\bibitem[{Dekel et~al.(2023)Dekel, Sarkar, Birnboim, Mandelker and
  Li}]{Dekel:2023ddd}
\bibinfo{author}{Dekel, A.}, \bibinfo{author}{Sarkar, K.C.},
  \bibinfo{author}{Birnboim, Y.}, \bibinfo{author}{Mandelker, N.},
  \bibinfo{author}{Li, Z.}, \bibinfo{year}{2023}.
\newblock \bibinfo{title}{{Efficient formation of massive galaxies at cosmic
  dawn by feedback-free starbursts}}.
\newblock \bibinfo{journal}{Mon. Not. Roy. Astron. Soc.} \bibinfo{volume}{523},
  \bibinfo{pages}{3201--3218}.
\newblock \DOIprefix\doi{10.1093/mnras/stad1557},
  \href{http://arxiv.org/abs/2303.04827}{{\tt arXiv:2303.04827}}.
\bibitem[{Dhawan et~al.(2021)Dhawan, Alsing and Vagnozzi}]{Dhawan:2021mel}
\bibinfo{author}{Dhawan, S.}, \bibinfo{author}{Alsing, J.},
  \bibinfo{author}{Vagnozzi, S.}, \bibinfo{year}{2021}.
\newblock \bibinfo{title}{{Non-parametric spatial curvature inference using
  late-Universe cosmological probes}}.
\newblock \bibinfo{journal}{Mon. Not. Roy. Astron. Soc.} \bibinfo{volume}{506},
  \bibinfo{pages}{L1--L5}.
\newblock \DOIprefix\doi{10.1093/mnrasl/slab058},
  \href{http://arxiv.org/abs/2104.02485}{{\tt arXiv:2104.02485}}.
\bibitem[{Di~Valentino et~al.(2021a)Di~Valentino, Melchiorri, Mena, Pan and
  Yang}]{DiValentino:2020kpf}
\bibinfo{author}{Di~Valentino, E.}, \bibinfo{author}{Melchiorri, A.},
  \bibinfo{author}{Mena, O.}, \bibinfo{author}{Pan, S.}, \bibinfo{author}{Yang,
  W.}, \bibinfo{year}{2021}a.
\newblock \bibinfo{title}{{Interacting Dark Energy in a closed universe}}.
\newblock \bibinfo{journal}{Mon. Not. Roy. Astron. Soc.} \bibinfo{volume}{502},
  \bibinfo{pages}{L23--L28}.
\newblock \DOIprefix\doi{10.1093/mnrasl/slaa207},
  \href{http://arxiv.org/abs/2011.00283}{{\tt arXiv:2011.00283}}.
\bibitem[{Di~Valentino et~al.(2019)Di~Valentino, Melchiorri and
  Silk}]{DiValentino:2019qzk}
\bibinfo{author}{Di~Valentino, E.}, \bibinfo{author}{Melchiorri, A.},
  \bibinfo{author}{Silk, J.}, \bibinfo{year}{2019}.
\newblock \bibinfo{title}{{Planck evidence for a closed Universe and a possible
  crisis for cosmology}}.
\newblock \bibinfo{journal}{Nature Astron.} \bibinfo{volume}{4},
  \bibinfo{pages}{196--203}.
\newblock \DOIprefix\doi{10.1038/s41550-019-0906-9},
  \href{http://arxiv.org/abs/1911.02087}{{\tt arXiv:1911.02087}}.
\bibitem[{Di~Valentino et~al.(2021b)Di~Valentino, Melchiorri and
  Silk}]{DiValentino:2020hov}
\bibinfo{author}{Di~Valentino, E.}, \bibinfo{author}{Melchiorri, A.},
  \bibinfo{author}{Silk, J.}, \bibinfo{year}{2021}b.
\newblock \bibinfo{title}{{Investigating Cosmic Discordance}}.
\newblock \bibinfo{journal}{Astrophys. J. Lett.} \bibinfo{volume}{908},
  \bibinfo{pages}{L9}.
\newblock \DOIprefix\doi{10.3847/2041-8213/abe1c4},
  \href{http://arxiv.org/abs/2003.04935}{{\tt arXiv:2003.04935}}.
\bibitem[{Di~Valentino et~al.(2025)}]{CosmoVerseNetwork:2025alb}
\bibinfo{author}{Di~Valentino, E.}, et~al. (\bibinfo{collaboration}{CosmoVerse
  Network}), \bibinfo{year}{2025}.
\newblock \bibinfo{title}{{The CosmoVerse White Paper: Addressing observational
  tensions in cosmology with systematics and fundamental physics}}.
\newblock \bibinfo{journal}{Phys. Dark Univ.} \bibinfo{volume}{49},
  \bibinfo{pages}{101965}.
\newblock \DOIprefix\doi{10.1016/j.dark.2025.101965},
  \href{http://arxiv.org/abs/2504.01669}{{\tt arXiv:2504.01669}}.
\bibitem[{Dinda(2022)}]{Dinda:2021ffa}
\bibinfo{author}{Dinda, B.R.}, \bibinfo{year}{2022}.
\newblock \bibinfo{title}{{Cosmic expansion parametrization: Implication for
  curvature and H0 tension}}.
\newblock \bibinfo{journal}{Phys. Rev. D} \bibinfo{volume}{105},
  \bibinfo{pages}{063524}.
\newblock \DOIprefix\doi{10.1103/PhysRevD.105.063524},
  \href{http://arxiv.org/abs/2106.02963}{{\tt arXiv:2106.02963}}.
\bibitem[{Du et~al.(2025)Du, Wu, Li and Zhang}]{Du:2024pai}
\bibinfo{author}{Du, G.H.}, \bibinfo{author}{Wu, P.J.}, \bibinfo{author}{Li,
  T.N.}, \bibinfo{author}{Zhang, X.}, \bibinfo{year}{2025}.
\newblock \bibinfo{title}{{Impacts of dark energy on weighing neutrinos after
  DESI BAO}}.
\newblock \bibinfo{journal}{Eur. Phys. J. C} \bibinfo{volume}{85},
  \bibinfo{pages}{392}.
\newblock \DOIprefix\doi{10.1140/epjc/s10052-025-14094-0},
  \href{http://arxiv.org/abs/2407.15640}{{\tt arXiv:2407.15640}}.
\bibitem[{Efstathiou and Gratton(2020)}]{Efstathiou:2020wem}
\bibinfo{author}{Efstathiou, G.}, \bibinfo{author}{Gratton, S.},
  \bibinfo{year}{2020}.
\newblock \bibinfo{title}{{The evidence for a spatially flat Universe}}.
\newblock \bibinfo{journal}{Mon. Not. Roy. Astron. Soc.} \bibinfo{volume}{496},
  \bibinfo{pages}{L91--L95}.
\newblock \DOIprefix\doi{10.1093/mnrasl/slaa093},
  \href{http://arxiv.org/abs/2002.06892}{{\tt arXiv:2002.06892}}.
\bibitem[{Ellis et~al.(2024)Ellis, Fairbairn, H{\"u}tsi, Urrutia, Vaskonen and
  Veerm{\"a}e}]{Ellis:2024wdh}
\bibinfo{author}{Ellis, J.}, \bibinfo{author}{Fairbairn, M.},
  \bibinfo{author}{H{\"u}tsi, G.}, \bibinfo{author}{Urrutia, J.},
  \bibinfo{author}{Vaskonen, V.}, \bibinfo{author}{Veerm{\"a}e, H.},
  \bibinfo{year}{2024}.
\newblock \bibinfo{title}{{Consistency of JWST black hole observations with
  NANOGrav gravitational wave measurements}}.
\newblock \bibinfo{journal}{Astron. Astrophys.} \bibinfo{volume}{691},
  \bibinfo{pages}{A270}.
\newblock \DOIprefix\doi{10.1051/0004-6361/202450846},
  \href{http://arxiv.org/abs/2403.19650}{{\tt arXiv:2403.19650}}.
\bibitem[{Escamilla et~al.(2024)Escamilla, Giar{\`e}, Di~Valentino, Nunes and
  Vagnozzi}]{Escamilla:2023oce}
\bibinfo{author}{Escamilla, L.A.}, \bibinfo{author}{Giar{\`e}, W.},
  \bibinfo{author}{Di~Valentino, E.}, \bibinfo{author}{Nunes, R.C.},
  \bibinfo{author}{Vagnozzi, S.}, \bibinfo{year}{2024}.
\newblock \bibinfo{title}{{The state of the dark energy equation of state circa
  2023}}.
\newblock \bibinfo{journal}{JCAP} \bibinfo{volume}{05}, \bibinfo{pages}{091}.
\newblock \DOIprefix\doi{10.1088/1475-7516/2024/05/091},
  \href{http://arxiv.org/abs/2307.14802}{{\tt arXiv:2307.14802}}.
\bibitem[{Fakhry et~al.(2025)Fakhry, Salmani and Firouzjaee}]{Fakhry:2025tma}
\bibinfo{author}{Fakhry, S.}, \bibinfo{author}{Salmani, R.V.},
  \bibinfo{author}{Firouzjaee, J.T.}, \bibinfo{year}{2025}.
\newblock \bibinfo{title}{{High-redshift galaxies from JWST observations in
  more realistic dark matter halo models}}.
\newblock \bibinfo{journal}{Phys. Rev. D} \bibinfo{volume}{112},
  \bibinfo{pages}{123503}.
\newblock \DOIprefix\doi{10.1103/9cmb-kf3x},
  \href{http://arxiv.org/abs/2507.23742}{{\tt arXiv:2507.23742}}.
\bibitem[{Fakhry et~al.(2026)Fakhry, Shiravand and Del~Popolo}]{Fakhry:2025yeu}
\bibinfo{author}{Fakhry, S.}, \bibinfo{author}{Shiravand, M.},
  \bibinfo{author}{Del~Popolo, A.}, \bibinfo{year}{2026}.
\newblock \bibinfo{title}{{Matching JWST Ultraviolet Luminosity Functions with
  Refined {\ensuremath{\Lambda}}CDM Halo Models}}.
\newblock \bibinfo{journal}{Astrophys. J.} \bibinfo{volume}{998},
  \bibinfo{pages}{178}.
\newblock \DOIprefix\doi{10.3847/1538-4357/ae371b},
  \href{http://arxiv.org/abs/2510.04709}{{\tt arXiv:2510.04709}}.
\bibitem[{Favale et~al.(2023)Favale, G\'omez-Valent and
  Migliaccio}]{Favale:2023lnp}
\bibinfo{author}{Favale, A.}, \bibinfo{author}{G\'omez-Valent, A.},
  \bibinfo{author}{Migliaccio, M.}, \bibinfo{year}{2023}.
\newblock \bibinfo{title}{{Cosmic chronometers to calibrate the ladders and
  measure the curvature of the Universe. A model-independent study}}.
\newblock \bibinfo{journal}{Mon. Not. Roy. Astron. Soc.} \bibinfo{volume}{523},
  \bibinfo{pages}{3406--3422}.
\newblock \DOIprefix\doi{10.1093/mnras/stad1621},
  \href{http://arxiv.org/abs/2301.09591}{{\tt arXiv:2301.09591}}.
\bibitem[{Feng et~al.(2026)Feng, Li, Du, Zhang and Zhang}]{Feng:2026pzs}
\bibinfo{author}{Feng, L.}, \bibinfo{author}{Li, T.N.}, \bibinfo{author}{Du,
  G.H.}, \bibinfo{author}{Zhang, J.F.}, \bibinfo{author}{Zhang, X.},
  \bibinfo{year}{2026}.
\newblock \bibinfo{title}{{Measuring neutrino mass in light of ACT DR6 and DESI
  DR2}}.
\newblock \bibinfo{journal}{Phys. Dark Univ.} \bibinfo{volume}{52},
  \bibinfo{pages}{102296}.
\newblock \DOIprefix\doi{10.1016/j.dark.2026.102296},
  \href{http://arxiv.org/abs/2603.10787}{{\tt arXiv:2603.10787}}.
\bibitem[{Ferrara et~al.(2023)Ferrara, Pallottini and Dayal}]{Ferrara:2022dqw}
\bibinfo{author}{Ferrara, A.}, \bibinfo{author}{Pallottini, A.},
  \bibinfo{author}{Dayal, P.}, \bibinfo{year}{2023}.
\newblock \bibinfo{title}{{On the stunning abundance of super-early, luminous
  galaxies revealed by JWST}}.
\newblock \bibinfo{journal}{Mon. Not. Roy. Astron. Soc.} \bibinfo{volume}{522},
  \bibinfo{pages}{3986--3991}.
\newblock \DOIprefix\doi{10.1093/mnras/stad1095},
  \href{http://arxiv.org/abs/2208.00720}{{\tt arXiv:2208.00720}}.
\bibitem[{Forconi and DI~Valentino(2025)}]{Forconi:2025zzu}
\bibinfo{author}{Forconi, M.}, \bibinfo{author}{DI~Valentino, E.},
  \bibinfo{year}{2025}.
\newblock \bibinfo{title}{{One extension to explain them all, one
  scale-invariant spectrum to test them all, and in one model bind them}}.
\newblock \bibinfo{journal}{Phys. Dark Univ.} \bibinfo{volume}{48},
  \bibinfo{pages}{101904}.
\newblock \DOIprefix\doi{10.1016/j.dark.2025.101904},
  \href{http://arxiv.org/abs/2503.04705}{{\tt arXiv:2503.04705}}.
\bibitem[{Forconi et~al.(2024)Forconi, Giar{\`e}, Mena, Ruchika, Di~Valentino,
  Melchiorri and Nunes}]{Forconi:2023hsj}
\bibinfo{author}{Forconi, M.}, \bibinfo{author}{Giar{\`e}, W.},
  \bibinfo{author}{Mena, O.}, \bibinfo{author}{Ruchika},
  \bibinfo{author}{Di~Valentino, E.}, \bibinfo{author}{Melchiorri, A.},
  \bibinfo{author}{Nunes, R.C.}, \bibinfo{year}{2024}.
\newblock \bibinfo{title}{{A double take on early and interacting dark energy
  from JWST}}.
\newblock \bibinfo{journal}{JCAP} \bibinfo{volume}{05}, \bibinfo{pages}{097}.
\newblock \DOIprefix\doi{10.1088/1475-7516/2024/05/097},
  \href{http://arxiv.org/abs/2312.11074}{{\tt arXiv:2312.11074}}.
\bibitem[{Forconi et~al.(2023)Forconi, Ruchika, Melchiorri, Mena and
  Menci}]{Forconi:2023izg}
\bibinfo{author}{Forconi, M.}, \bibinfo{author}{Ruchika},
  \bibinfo{author}{Melchiorri, A.}, \bibinfo{author}{Mena, O.},
  \bibinfo{author}{Menci, N.}, \bibinfo{year}{2023}.
\newblock \bibinfo{title}{{Do the early galaxies observed by JWST disagree with
  Planck's CMB polarization measurements?}}
\newblock \bibinfo{journal}{JCAP} \bibinfo{volume}{10}, \bibinfo{pages}{012}.
\newblock \DOIprefix\doi{10.1088/1475-7516/2023/10/012},
  \href{http://arxiv.org/abs/2306.07781}{{\tt arXiv:2306.07781}}.
\bibitem[{Foreman-Mackey et~al.(2013)Foreman-Mackey, Hogg, Lang and
  Goodman}]{Foreman-Mackey:2012any}
\bibinfo{author}{Foreman-Mackey, D.}, \bibinfo{author}{Hogg, D.W.},
  \bibinfo{author}{Lang, D.}, \bibinfo{author}{Goodman, J.},
  \bibinfo{year}{2013}.
\newblock \bibinfo{title}{{emcee: The MCMC Hammer}}.
\newblock \bibinfo{journal}{Publ. Astron. Soc. Pac.} \bibinfo{volume}{125},
  \bibinfo{pages}{306--312}.
\newblock \DOIprefix\doi{10.1086/670067},
  \href{http://arxiv.org/abs/1202.3665}{{\tt arXiv:1202.3665}}.
\bibitem[{Gandolfi et~al.(2022)Gandolfi, Lapi, Ronconi and
  Danese}]{Gandolfi:2022bcm}
\bibinfo{author}{Gandolfi, G.}, \bibinfo{author}{Lapi, A.},
  \bibinfo{author}{Ronconi, T.}, \bibinfo{author}{Danese, L.},
  \bibinfo{year}{2022}.
\newblock \bibinfo{title}{{Astroparticle Constraints from the Cosmic Star
  Formation Rate Density at High Redshift: Current Status and Forecasts for
  JWST}}.
\newblock \bibinfo{journal}{Universe} \bibinfo{volume}{8},
  \bibinfo{pages}{589}.
\newblock \DOIprefix\doi{10.3390/universe8110589},
  \href{http://arxiv.org/abs/2211.02840}{{\tt arXiv:2211.02840}}.
\bibitem[{Gardner et~al.(2006)}]{Gardner:2006ky}
\bibinfo{author}{Gardner, J.P.}, et~al., \bibinfo{year}{2006}.
\newblock \bibinfo{title}{{The James Webb Space Telescope}}.
\newblock \bibinfo{journal}{Space Sci. Rev.} \bibinfo{volume}{123},
  \bibinfo{pages}{485}.
\newblock \DOIprefix\doi{10.1007/s11214-006-8315-7},
  \href{http://arxiv.org/abs/astro-ph/0606175}{{\tt arXiv:astro-ph/0606175}}.
\bibitem[{Giar{\`e}(2025)}]{Giare:2024oil}
\bibinfo{author}{Giar{\`e}, W.}, \bibinfo{year}{2025}.
\newblock \bibinfo{title}{{Dynamical dark energy beyond Planck? Constraints
  from multiple CMB probes, DESI BAO, and type-Ia supernovae}}.
\newblock \bibinfo{journal}{Phys. Rev. D} \bibinfo{volume}{112},
  \bibinfo{pages}{023508}.
\newblock \DOIprefix\doi{10.1103/ss37-cxhn},
  \href{http://arxiv.org/abs/2409.17074}{{\tt arXiv:2409.17074}}.
\bibitem[{Giar\`e et~al.(2024)Giar\`e, Di~Valentino and
  Melchiorri}]{Giare:2023ejv}
\bibinfo{author}{Giar\`e, W.}, \bibinfo{author}{Di~Valentino, E.},
  \bibinfo{author}{Melchiorri, A.}, \bibinfo{year}{2024}.
\newblock \bibinfo{title}{{Measuring the reionization optical depth without
  large-scale CMB polarization}}.
\newblock \bibinfo{journal}{Phys. Rev. D} \bibinfo{volume}{109},
  \bibinfo{pages}{103519}.
\newblock \DOIprefix\doi{10.1103/PhysRevD.109.103519},
  \href{http://arxiv.org/abs/2312.06482}{{\tt arXiv:2312.06482}}.
\bibitem[{Giar{\`e} et~al.(2025)Giar{\`e}, Mahassen, Di~Valentino and
  Pan}]{Giare:2025pzu}
\bibinfo{author}{Giar{\`e}, W.}, \bibinfo{author}{Mahassen, T.},
  \bibinfo{author}{Di~Valentino, E.}, \bibinfo{author}{Pan, S.},
  \bibinfo{year}{2025}.
\newblock \bibinfo{title}{{An overview of what current data can (and cannot
  yet) say about evolving dark energy}}.
\newblock \bibinfo{journal}{Phys. Dark Univ.} \bibinfo{volume}{48},
  \bibinfo{pages}{101906}.
\newblock \DOIprefix\doi{10.1016/j.dark.2025.101906},
  \href{http://arxiv.org/abs/2502.10264}{{\tt arXiv:2502.10264}}.
\bibitem[{Giar{\`e} et~al.(2024a)Giar{\`e}, Najafi, Pan, Di~Valentino and
  Firouzjaee}]{Giare:2024gpk}
\bibinfo{author}{Giar{\`e}, W.}, \bibinfo{author}{Najafi, M.},
  \bibinfo{author}{Pan, S.}, \bibinfo{author}{Di~Valentino, E.},
  \bibinfo{author}{Firouzjaee, J.T.}, \bibinfo{year}{2024}a.
\newblock \bibinfo{title}{{Robust preference for Dynamical Dark Energy in DESI
  BAO and SN measurements}}.
\newblock \bibinfo{journal}{JCAP} \bibinfo{volume}{10}, \bibinfo{pages}{035}.
\newblock \DOIprefix\doi{10.1088/1475-7516/2024/10/035},
  \href{http://arxiv.org/abs/2407.16689}{{\tt arXiv:2407.16689}}.
\bibitem[{Giar{\`e} et~al.(2024b)Giar{\`e}, Sabogal, Nunes and
  Di~Valentino}]{Giare:2024smz}
\bibinfo{author}{Giar{\`e}, W.}, \bibinfo{author}{Sabogal, M.A.},
  \bibinfo{author}{Nunes, R.C.}, \bibinfo{author}{Di~Valentino, E.},
  \bibinfo{year}{2024}b.
\newblock \bibinfo{title}{{Interacting Dark Energy after DESI Baryon Acoustic
  Oscillation Measurements}}.
\newblock \bibinfo{journal}{Phys. Rev. Lett.} \bibinfo{volume}{133},
  \bibinfo{pages}{251003}.
\newblock \DOIprefix\doi{10.1103/PhysRevLett.133.251003},
  \href{http://arxiv.org/abs/2404.15232}{{\tt arXiv:2404.15232}}.
\bibitem[{Glanville et~al.(2022)Glanville, Howlett and
  Davis}]{Glanville:2022xes}
\bibinfo{author}{Glanville, A.}, \bibinfo{author}{Howlett, C.},
  \bibinfo{author}{Davis, T.M.}, \bibinfo{year}{2022}.
\newblock \bibinfo{title}{{Full-shape galaxy power spectra and the curvature
  tension}}.
\newblock \bibinfo{journal}{Mon. Not. Roy. Astron. Soc.} \bibinfo{volume}{517},
  \bibinfo{pages}{3087--3100}.
\newblock \DOIprefix\doi{10.1093/mnras/stac2891},
  \href{http://arxiv.org/abs/2205.05892}{{\tt arXiv:2205.05892}}.
\bibitem[{Gonzalez et~al.(2021)Gonzalez, Benetti, von Marttens and
  Alcaniz}]{Gonzalez:2021ojp}
\bibinfo{author}{Gonzalez, J.E.}, \bibinfo{author}{Benetti, M.},
  \bibinfo{author}{von Marttens, R.}, \bibinfo{author}{Alcaniz, J.},
  \bibinfo{year}{2021}.
\newblock \bibinfo{title}{{Testing the consistency between cosmological data:
  the impact of spatial curvature and the dark energy EoS}}.
\newblock \bibinfo{journal}{JCAP} \bibinfo{volume}{11}, \bibinfo{pages}{060}.
\newblock \DOIprefix\doi{10.1088/1475-7516/2021/11/060},
  \href{http://arxiv.org/abs/2104.13455}{{\tt arXiv:2104.13455}}.
\bibitem[{Green and Meyers(2025)}]{Green:2024xbb}
\bibinfo{author}{Green, D.}, \bibinfo{author}{Meyers, J.},
  \bibinfo{year}{2025}.
\newblock \bibinfo{title}{{Cosmological preference for a negative neutrino
  mass}}.
\newblock \bibinfo{journal}{Phys. Rev. D} \bibinfo{volume}{111},
  \bibinfo{pages}{083507}.
\newblock \DOIprefix\doi{10.1103/PhysRevD.111.083507},
  \href{http://arxiv.org/abs/2407.07878}{{\tt arXiv:2407.07878}}.
\bibitem[{Guo et~al.(2024)Guo, Khlopov, Liu, Wu, Wu and Zhu}]{Guo:2023hyp}
\bibinfo{author}{Guo, S.Y.}, \bibinfo{author}{Khlopov, M.},
  \bibinfo{author}{Liu, X.}, \bibinfo{author}{Wu, L.}, \bibinfo{author}{Wu,
  Y.}, \bibinfo{author}{Zhu, B.}, \bibinfo{year}{2024}.
\newblock \bibinfo{title}{{Footprints of axion-like particle in pulsar timing
  array data and James Webb Space Telescope observations}}.
\newblock \bibinfo{journal}{Sci. China Phys. Mech. Astron.}
  \bibinfo{volume}{67}, \bibinfo{pages}{111011}.
\newblock \DOIprefix\doi{10.1007/s11433-024-2445-1},
  \href{http://arxiv.org/abs/2306.17022}{{\tt arXiv:2306.17022}}.
\bibitem[{Gupta(2023)}]{Gupta:2023mgg}
\bibinfo{author}{Gupta, R.P.}, \bibinfo{year}{2023}.
\newblock \bibinfo{title}{{JWST early Universe observations and
  {\ensuremath{\Lambda}}CDM cosmology}}.
\newblock \bibinfo{journal}{Mon. Not. Roy. Astron. Soc.} \bibinfo{volume}{524},
  \bibinfo{pages}{3385--3395}.
\newblock \DOIprefix\doi{10.1093/mnras/stad2032},
  \href{http://arxiv.org/abs/2309.13100}{{\tt arXiv:2309.13100}}.
\bibitem[{Handley(2021)}]{Handley:2019tkm}
\bibinfo{author}{Handley, W.}, \bibinfo{year}{2021}.
\newblock \bibinfo{title}{{Curvature tension: evidence for a closed universe}}.
\newblock \bibinfo{journal}{Phys. Rev. D} \bibinfo{volume}{103},
  \bibinfo{pages}{L041301}.
\newblock \DOIprefix\doi{10.1103/PhysRevD.103.L041301},
  \href{http://arxiv.org/abs/1908.09139}{{\tt arXiv:1908.09139}}.
\bibitem[{Haslbauer et~al.(2022)Haslbauer, Kroupa, Zonoozi and
  Haghi}]{Haslbauer:2022vnq}
\bibinfo{author}{Haslbauer, M.}, \bibinfo{author}{Kroupa, P.},
  \bibinfo{author}{Zonoozi, A.H.}, \bibinfo{author}{Haghi, H.},
  \bibinfo{year}{2022}.
\newblock \bibinfo{title}{{Has JWST Already Falsified Dark-matter-driven Galaxy
  Formation?}}
\newblock \bibinfo{journal}{Astrophys. J. Lett.} \bibinfo{volume}{939},
  \bibinfo{pages}{L31}.
\newblock \DOIprefix\doi{10.3847/2041-8213/ac9a50},
  \href{http://arxiv.org/abs/2210.14915}{{\tt arXiv:2210.14915}}.
\bibitem[{Hegde et~al.(2024)Hegde, Wyatt and Furlanetto}]{Hegde:2024kph}
\bibinfo{author}{Hegde, S.}, \bibinfo{author}{Wyatt, M.M.},
  \bibinfo{author}{Furlanetto, S.R.}, \bibinfo{year}{2024}.
\newblock \bibinfo{title}{{A hidden population of active galactic nuclei can
  explain the overabundance of luminous z {\ensuremath{>}} 10 objects observed
  by JWST}}.
\newblock \bibinfo{journal}{JCAP} \bibinfo{volume}{08}, \bibinfo{pages}{025}.
\newblock \DOIprefix\doi{10.1088/1475-7516/2024/08/025},
  \href{http://arxiv.org/abs/2405.01629}{{\tt arXiv:2405.01629}}.
\bibitem[{Huang et~al.(2024a)Huang, Cai, Jiang, Zhang and Piao}]{Huang:2023chx}
\bibinfo{author}{Huang, H.L.}, \bibinfo{author}{Cai, Y.},
  \bibinfo{author}{Jiang, J.Q.}, \bibinfo{author}{Zhang, J.},
  \bibinfo{author}{Piao, Y.S.}, \bibinfo{year}{2024}a.
\newblock \bibinfo{title}{{Supermassive Primordial Black Holes for Nano-Hertz
  Gravitational Waves and High-redshift JWST Galaxies}}.
\newblock \bibinfo{journal}{Res. Astron. Astrophys.} \bibinfo{volume}{24},
  \bibinfo{pages}{091001}.
\newblock \DOIprefix\doi{10.1088/1674-4527/ad683d},
  \href{http://arxiv.org/abs/2306.17577}{{\tt arXiv:2306.17577}}.
\bibitem[{Huang et~al.(2024b)Huang, Jiang and Piao}]{Huang:2024aog}
\bibinfo{author}{Huang, H.L.}, \bibinfo{author}{Jiang, J.Q.},
  \bibinfo{author}{Piao, Y.S.}, \bibinfo{year}{2024}b.
\newblock \bibinfo{title}{{High-redshift JWST massive galaxies and the initial
  clustering of supermassive primordial black holes}}.
\newblock \bibinfo{journal}{Phys. Rev. D} \bibinfo{volume}{110},
  \bibinfo{pages}{103540}.
\newblock \DOIprefix\doi{10.1103/PhysRevD.110.103540},
  \href{http://arxiv.org/abs/2407.15781}{{\tt arXiv:2407.15781}}.
\bibitem[{H{\"u}tsi et~al.(2023)H{\"u}tsi, Raidal, Urrutia, Vaskonen and
  Veerm{\"a}e}]{Hutsi:2022fzw}
\bibinfo{author}{H{\"u}tsi, G.}, \bibinfo{author}{Raidal, M.},
  \bibinfo{author}{Urrutia, J.}, \bibinfo{author}{Vaskonen, V.},
  \bibinfo{author}{Veerm{\"a}e, H.}, \bibinfo{year}{2023}.
\newblock \bibinfo{title}{{Did JWST observe imprints of axion miniclusters or
  primordial black holes?}}
\newblock \bibinfo{journal}{Phys. Rev. D} \bibinfo{volume}{107},
  \bibinfo{pages}{043502}.
\newblock \DOIprefix\doi{10.1103/PhysRevD.107.043502},
  \href{http://arxiv.org/abs/2211.02651}{{\tt arXiv:2211.02651}}.
\bibitem[{Hutter et~al.(2025)Hutter, Cueto, Dayal, Gottl{\"o}ber, Trebitsch and
  Yepes}]{Hutter:2024cvr}
\bibinfo{author}{Hutter, A.}, \bibinfo{author}{Cueto, E.R.},
  \bibinfo{author}{Dayal, P.}, \bibinfo{author}{Gottl{\"o}ber, S.},
  \bibinfo{author}{Trebitsch, M.}, \bibinfo{author}{Yepes, G.},
  \bibinfo{year}{2025}.
\newblock \bibinfo{title}{{ASTRAEUS - X. Indications of a top-heavy initial
  mass function in highly star-forming galaxies from JWST observations at z
  {\ensuremath{>}} 10}}.
\newblock \bibinfo{journal}{Astron. Astrophys.} \bibinfo{volume}{694},
  \bibinfo{pages}{A254}.
\newblock \DOIprefix\doi{10.1051/0004-6361/202452460},
  \href{http://arxiv.org/abs/2410.00730}{{\tt arXiv:2410.00730}}.
\bibitem[{Ilie et~al.(2023)Ilie, Paulin and Freese}]{Ilie:2023zfv}
\bibinfo{author}{Ilie, C.}, \bibinfo{author}{Paulin, J.},
  \bibinfo{author}{Freese, K.}, \bibinfo{year}{2023}.
\newblock \bibinfo{title}{{Supermassive Dark Star candidates seen by JWST}}.
\newblock \bibinfo{journal}{Proc. Nat. Acad. Sci.} \bibinfo{volume}{120},
  \bibinfo{pages}{e2305762120}.
\newblock \DOIprefix\doi{10.1073/pnas.2305762120},
  \href{http://arxiv.org/abs/2304.01173}{{\tt arXiv:2304.01173}}.
\bibitem[{Iocco and Visinelli(2024)}]{Iocco:2024rez}
\bibinfo{author}{Iocco, F.}, \bibinfo{author}{Visinelli, L.},
  \bibinfo{year}{2024}.
\newblock \bibinfo{title}{{Compatibility of JWST results with exotic halos}}.
\newblock \bibinfo{journal}{Phys. Dark Univ.} \bibinfo{volume}{44},
  \bibinfo{pages}{101496}.
\newblock \DOIprefix\doi{10.1016/j.dark.2024.101496},
  \href{http://arxiv.org/abs/2403.13068}{{\tt arXiv:2403.13068}}.
\bibitem[{Jiang et~al.(2025a)Jiang, Giar{\`e}, Gariazzo, Dainotti,
  Di~Valentino, Mena, Pedrotti, da~Costa and Vagnozzi}]{Jiang:2024viw}
\bibinfo{author}{Jiang, J.Q.}, \bibinfo{author}{Giar{\`e}, W.},
  \bibinfo{author}{Gariazzo, S.}, \bibinfo{author}{Dainotti, M.G.},
  \bibinfo{author}{Di~Valentino, E.}, \bibinfo{author}{Mena, O.},
  \bibinfo{author}{Pedrotti, D.}, \bibinfo{author}{da~Costa, S.S.},
  \bibinfo{author}{Vagnozzi, S.}, \bibinfo{year}{2025}a.
\newblock \bibinfo{title}{{Neutrino cosmology after DESI: tightest mass upper
  limits, preference for the normal ordering, and tension with terrestrial
  observations}}.
\newblock \bibinfo{journal}{JCAP} \bibinfo{volume}{01}, \bibinfo{pages}{153}.
\newblock \DOIprefix\doi{10.1088/1475-7516/2025/01/153},
  \href{http://arxiv.org/abs/2407.18047}{{\tt arXiv:2407.18047}}.
\bibitem[{Jiang et~al.(2025b)Jiang, Liu, Zhan and Hu}]{Jiang:2024tll}
\bibinfo{author}{Jiang, J.Q.}, \bibinfo{author}{Liu, W.},
  \bibinfo{author}{Zhan, H.}, \bibinfo{author}{Hu, B.}, \bibinfo{year}{2025}b.
\newblock \bibinfo{title}{{Explanation of high redshift luminous galaxies from
  JWST by an early dark energy model}}.
\newblock \bibinfo{journal}{Phys. Rev. D} \bibinfo{volume}{111},
  \bibinfo{pages}{023519}.
\newblock \DOIprefix\doi{10.1103/PhysRevD.111.023519},
  \href{http://arxiv.org/abs/2409.19941}{{\tt arXiv:2409.19941}}.
\bibitem[{Jiang et~al.(2024)Jiang, Pedrotti, da~Costa and
  Vagnozzi}]{Jiang:2024xnu}
\bibinfo{author}{Jiang, J.Q.}, \bibinfo{author}{Pedrotti, D.},
  \bibinfo{author}{da~Costa, S.S.}, \bibinfo{author}{Vagnozzi, S.},
  \bibinfo{year}{2024}.
\newblock \bibinfo{title}{{Nonparametric late-time expansion history
  reconstruction and implications for the Hubble tension in light of recent
  DESI and type Ia supernovae data}}.
\newblock \bibinfo{journal}{Phys. Rev. D} \bibinfo{volume}{110},
  \bibinfo{pages}{123519}.
\newblock \DOIprefix\doi{10.1103/PhysRevD.110.123519},
  \href{http://arxiv.org/abs/2408.02365}{{\tt arXiv:2408.02365}}.
\bibitem[{Jiao et~al.(2023)Jiao, Brandenberger and Refregier}]{Jiao:2023wcn}
\bibinfo{author}{Jiao, H.}, \bibinfo{author}{Brandenberger, R.},
  \bibinfo{author}{Refregier, A.}, \bibinfo{year}{2023}.
\newblock \bibinfo{title}{{Early structure formation from cosmic string loops
  in light of early JWST observations}}.
\newblock \bibinfo{journal}{Phys. Rev. D} \bibinfo{volume}{108},
  \bibinfo{pages}{043510}.
\newblock \DOIprefix\doi{10.1103/PhysRevD.108.043510},
  \href{http://arxiv.org/abs/2304.06429}{{\tt arXiv:2304.06429}}.
\bibitem[{Kennicutt and Evans(2012)}]{Kennicutt:2012ea}
\bibinfo{author}{Kennicutt, Jr., R.C.}, \bibinfo{author}{Evans, II, N.J.},
  \bibinfo{year}{2012}.
\newblock \bibinfo{title}{{Star Formation in the Milky Way and Nearby
  Galaxies}}.
\newblock \bibinfo{journal}{Ann. Rev. Astron. Astrophys.} \bibinfo{volume}{50},
  \bibinfo{pages}{531--608}.
\newblock \DOIprefix\doi{10.1146/annurev-astro-081811-125610},
  \href{http://arxiv.org/abs/1204.3552}{{\tt arXiv:1204.3552}}.
\bibitem[{Labb\'{e} et~al.(2023)}]{Labbe:2022ahb}
\bibinfo{author}{Labb\'{e}, I.}, et~al., \bibinfo{year}{2023}.
\newblock \bibinfo{title}{{A population of red candidate massive galaxies
  {\textasciitilde}600 Myr after the Big Bang}}.
\newblock \bibinfo{journal}{Nature} \bibinfo{volume}{616},
  \bibinfo{pages}{266--269}.
\newblock \DOIprefix\doi{10.1038/s41586-023-05786-2},
  \href{http://arxiv.org/abs/2207.12446}{{\tt arXiv:2207.12446}}.
\bibitem[{Lee(2025)}]{Lee:2025kbn}
\bibinfo{author}{Lee, S.}, \bibinfo{year}{2025}.
\newblock \bibinfo{title}{{The impact of {\ensuremath{\Omega}}m0 prior bias on
  cosmological parameter estimation: reconciling DESI DR2 BAO and
  Pantheon~+~SNe Data}}.
\newblock \bibinfo{journal}{Mon. Not. Roy. Astron. Soc.} \bibinfo{volume}{544},
  \bibinfo{pages}{3388--3393}.
\newblock \DOIprefix\doi{10.1093/mnras/staf1890},
  \href{http://arxiv.org/abs/2506.16022}{{\tt arXiv:2506.16022}}.
\bibitem[{Lee(2026)}]{Lee:2025hjw}
\bibinfo{author}{Lee, S.}, \bibinfo{year}{2026}.
\newblock \bibinfo{title}{{Probing time-varying dark energy with DESI: the
  crucial role of precision matter density ($\Omega _\mathrm{{m} 0}$)
  measurements}}.
\newblock \bibinfo{journal}{Eur. Phys. J. C} \bibinfo{volume}{86},
  \bibinfo{pages}{297}.
\newblock \DOIprefix\doi{10.1140/epjc/s10052-026-15541-2},
  \href{http://arxiv.org/abs/2505.19052}{{\tt arXiv:2505.19052}}.
\bibitem[{Lei et~al.(2025)Lei, Wang, Yuan, Wang, Groenewegen and
  Fan}]{Lei:2025ooq}
\bibinfo{author}{Lei, L.}, \bibinfo{author}{Wang, Y.Y.}, \bibinfo{author}{Yuan,
  G.W.}, \bibinfo{author}{Wang, T.L.}, \bibinfo{author}{Groenewegen, M.A.T.},
  \bibinfo{author}{Fan, Y.Z.}, \bibinfo{year}{2025}.
\newblock \bibinfo{title}{{Can Dark Stars Account for the Star Formation
  Efficiency Excess at Very High Redshifts?}}
\newblock \bibinfo{journal}{Astrophys. J.} \bibinfo{volume}{980},
  \bibinfo{pages}{249}.
\newblock \DOIprefix\doi{10.3847/1538-4357/ada93b},
  \href{http://arxiv.org/abs/2501.07119}{{\tt arXiv:2501.07119}}.
\bibitem[{Lei et~al.(2026)Lei, Wang, Wang, Wang, Yuan, Lin and
  Fan}]{Lei:2025zjx}
\bibinfo{author}{Lei, L.}, \bibinfo{author}{Wang, Z.F.}, \bibinfo{author}{Wang,
  T.L.}, \bibinfo{author}{Wang, Y.Y.}, \bibinfo{author}{Yuan, G.W.},
  \bibinfo{author}{Lin, W.L.}, \bibinfo{author}{Fan, Y.Z.},
  \bibinfo{year}{2026}.
\newblock \bibinfo{title}{{Stringent constraint on the CCC+TL cosmology with
  $H(z)$ Measurements}}.
\newblock \bibinfo{journal}{Mon. Not. Roy. Astron. Soc.} \bibinfo{volume}{547},
  \bibinfo{pages}{1--7}.
\newblock \DOIprefix\doi{10.1093/mnras/stag430},
  \href{http://arxiv.org/abs/2508.04277}{{\tt arXiv:2508.04277}}.
\bibitem[{Lei et~al.(2024)}]{Lei:2023mke}
\bibinfo{author}{Lei, L.}, et~al., \bibinfo{year}{2024}.
\newblock \bibinfo{title}{{Black holes as the source of dark energy: A
  stringent test with high-redshift JWST AGNs}}.
\newblock \bibinfo{journal}{Sci. China Phys. Mech. Astron.}
  \bibinfo{volume}{67}, \bibinfo{pages}{229811}.
\newblock \DOIprefix\doi{10.1007/s11433-023-2233-2},
  \href{http://arxiv.org/abs/2305.03408}{{\tt arXiv:2305.03408}}.
\bibitem[{Leroy et~al.(2008)Leroy, Walter, Brinks, Bigiel, de~Blok, Madore and
  Thornley}]{Leroy:2008kh}
\bibinfo{author}{Leroy, A.K.}, \bibinfo{author}{Walter, F.},
  \bibinfo{author}{Brinks, E.}, \bibinfo{author}{Bigiel, F.},
  \bibinfo{author}{de~Blok, W.J.G.}, \bibinfo{author}{Madore, B.},
  \bibinfo{author}{Thornley, M.D.}, \bibinfo{year}{2008}.
\newblock \bibinfo{title}{{The Star Formation Efficiency in Nearby Galaxies:
  Measuring Where Gas Forms Stars Effectively}}.
\newblock \bibinfo{journal}{Astron. J.} \bibinfo{volume}{136},
  \bibinfo{pages}{2782--2845}.
\newblock \DOIprefix\doi{10.1088/0004-6256/136/6/2782},
  \href{http://arxiv.org/abs/0810.2556}{{\tt arXiv:0810.2556}}.
\bibitem[{Lewis(2025)}]{Lewis:2019xzd}
\bibinfo{author}{Lewis, A.}, \bibinfo{year}{2025}.
\newblock \bibinfo{title}{{GetDist: a Python package for analysing Monte Carlo
  samples}}.
\newblock \bibinfo{journal}{JCAP} \bibinfo{volume}{08}, \bibinfo{pages}{025}.
\newblock \DOIprefix\doi{10.1088/1475-7516/2025/08/025},
  \href{http://arxiv.org/abs/1910.13970}{{\tt arXiv:1910.13970}}.
\bibitem[{Li et~al.(2026)Li, Du, Zhou, Li, Zhang and Zhang}]{Li:2025vuh}
\bibinfo{author}{Li, T.N.}, \bibinfo{author}{Du, G.H.}, \bibinfo{author}{Zhou,
  S.H.}, \bibinfo{author}{Li, Y.H.}, \bibinfo{author}{Zhang, J.F.},
  \bibinfo{author}{Zhang, X.}, \bibinfo{year}{2026}.
\newblock \bibinfo{title}{{Robust evidence for dynamical dark energy in light
  of DESI DR2 and joint ACT, SPT, and Planck data}}.
\newblock \bibinfo{journal}{Phys. Dark Univ.} \bibinfo{volume}{52},
  \bibinfo{pages}{102254}.
\newblock \DOIprefix\doi{10.1016/j.dark.2026.102254},
  \href{http://arxiv.org/abs/2511.22512}{{\tt arXiv:2511.22512}}.
\bibitem[{Li et~al.(2024)Li, Wu, Du, Jin, Li, Zhang and Zhang}]{Li:2024qso}
\bibinfo{author}{Li, T.N.}, \bibinfo{author}{Wu, P.J.}, \bibinfo{author}{Du,
  G.H.}, \bibinfo{author}{Jin, S.J.}, \bibinfo{author}{Li, H.L.},
  \bibinfo{author}{Zhang, J.F.}, \bibinfo{author}{Zhang, X.},
  \bibinfo{year}{2024}.
\newblock \bibinfo{title}{{Constraints on Interacting Dark Energy Models from
  the DESI Baryon Acoustic Oscillation and DES Supernovae Data}}.
\newblock \bibinfo{journal}{Astrophys. J.} \bibinfo{volume}{976},
  \bibinfo{pages}{1}.
\newblock \DOIprefix\doi{10.3847/1538-4357/ad87f0},
  \href{http://arxiv.org/abs/2407.14934}{{\tt arXiv:2407.14934}}.
\bibitem[{Lin et~al.(2024)Lin, Gong, Yue and Chen}]{Lin:2023ewc}
\bibinfo{author}{Lin, H.}, \bibinfo{author}{Gong, Y.}, \bibinfo{author}{Yue,
  B.}, \bibinfo{author}{Chen, X.}, \bibinfo{year}{2024}.
\newblock \bibinfo{title}{{Implications of the Stellar Mass Density of High-z
  Massive Galaxies from JWST on Warm Dark Matter}}.
\newblock \bibinfo{journal}{Res. Astron. Astrophys.} \bibinfo{volume}{24},
  \bibinfo{pages}{015009}.
\newblock \DOIprefix\doi{10.1088/1674-4527/ad0864},
  \href{http://arxiv.org/abs/2306.05648}{{\tt arXiv:2306.05648}}.
\bibitem[{Lin et~al.(2021)Lin, Chen and Mack}]{Lin:2021sfs}
\bibinfo{author}{Lin, W.}, \bibinfo{author}{Chen, X.}, \bibinfo{author}{Mack,
  K.J.}, \bibinfo{year}{2021}.
\newblock \bibinfo{title}{{Early Universe Physics Insensitive and Uncalibrated
  Cosmic Standards: Constraints on {\ensuremath{\Omega}}m and Implications for
  the Hubble Tension}}.
\newblock \bibinfo{journal}{Astrophys. J.} \bibinfo{volume}{920},
  \bibinfo{pages}{159}.
\newblock \DOIprefix\doi{10.3847/1538-4357/ac12cf},
  \href{http://arxiv.org/abs/2102.05701}{{\tt arXiv:2102.05701}}.
\bibitem[{Liu and Bromm(2022)}]{Liu:2022bvr}
\bibinfo{author}{Liu, B.}, \bibinfo{author}{Bromm, V.}, \bibinfo{year}{2022}.
\newblock \bibinfo{title}{{Accelerating Early Massive Galaxy Formation with
  Primordial Black Holes}}.
\newblock \bibinfo{journal}{Astrophys. J. Lett.} \bibinfo{volume}{937},
  \bibinfo{pages}{L30}.
\newblock \DOIprefix\doi{10.3847/2041-8213/ac927f},
  \href{http://arxiv.org/abs/2208.13178}{{\tt arXiv:2208.13178}}.
\bibitem[{Liu et~al.(2025)Liu, Wang, Wu, Cao and Wang}]{Liu:2024yib}
\bibinfo{author}{Liu, T.}, \bibinfo{author}{Wang, S.}, \bibinfo{author}{Wu,
  H.}, \bibinfo{author}{Cao, S.}, \bibinfo{author}{Wang, J.},
  \bibinfo{year}{2025}.
\newblock \bibinfo{title}{{Newest Measurements of Cosmic Curvature with
  BOSS/eBOSS and DESI DR1 Baryon Acoustic Oscillation Observations}}.
\newblock \bibinfo{journal}{Astrophys. J. Lett.} \bibinfo{volume}{981},
  \bibinfo{pages}{L24}.
\newblock \DOIprefix\doi{10.3847/2041-8213/adb7de},
  \href{http://arxiv.org/abs/2411.14154}{{\tt arXiv:2411.14154}}.
\bibitem[{Loverde and Weiner(2024)}]{Loverde:2024nfi}
\bibinfo{author}{Loverde, M.}, \bibinfo{author}{Weiner, Z.J.},
  \bibinfo{year}{2024}.
\newblock \bibinfo{title}{{Massive neutrinos and cosmic composition}}.
\newblock \bibinfo{journal}{JCAP} \bibinfo{volume}{12}, \bibinfo{pages}{048}.
\newblock \DOIprefix\doi{10.1088/1475-7516/2024/12/048},
  \href{http://arxiv.org/abs/2410.00090}{{\tt arXiv:2410.00090}}.
\bibitem[{Lu et~al.(2024)Lu, Frenk, Bose, Lacey, Cole, Baugh and
  Helly}]{Lu:2024oli}
\bibinfo{author}{Lu, S.}, \bibinfo{author}{Frenk, C.S.}, \bibinfo{author}{Bose,
  S.}, \bibinfo{author}{Lacey, C.G.}, \bibinfo{author}{Cole, S.},
  \bibinfo{author}{Baugh, C.M.}, \bibinfo{author}{Helly, J.C.},
  \bibinfo{year}{2024}.
\newblock \bibinfo{title}{{A comparison of pre-existing
  {\ensuremath{\Lambda}}CDM predictions with the abundance of JWST galaxies at
  high redshift}}.
\newblock \bibinfo{journal}{Mon. Not. Roy. Astron. Soc.} \bibinfo{volume}{536},
  \bibinfo{pages}{1018--1034}.
\newblock \DOIprefix\doi{10.1093/mnras/stae2646},
  \href{http://arxiv.org/abs/2406.02672}{{\tt arXiv:2406.02672}}.
\bibitem[{Luberto et~al.(2025)Luberto, Furlanetto and
  Mirocha}]{Luberto:2025bct}
\bibinfo{author}{Luberto, J.}, \bibinfo{author}{Furlanetto, S.},
  \bibinfo{author}{Mirocha, J.}, \bibinfo{year}{2025}.
\newblock \bibinfo{title}{{A physically-motivated template set for high-z
  galaxy SED fitting}}.
\newblock \bibinfo{journal}{JCAP} \bibinfo{volume}{10}, \bibinfo{pages}{084}.
\newblock \DOIprefix\doi{10.1088/1475-7516/2025/10/084},
  \href{http://arxiv.org/abs/2409.20519}{{\tt arXiv:2409.20519}}.
\bibitem[{Luongo and Muccino(2024)}]{Luongo:2024fww}
\bibinfo{author}{Luongo, O.}, \bibinfo{author}{Muccino, M.},
  \bibinfo{year}{2024}.
\newblock \bibinfo{title}{{Model-independent cosmographic constraints from DESI
  2024}}.
\newblock \bibinfo{journal}{Astron. Astrophys.} \bibinfo{volume}{690},
  \bibinfo{pages}{A40}.
\newblock \DOIprefix\doi{10.1051/0004-6361/202450512},
  \href{http://arxiv.org/abs/2404.07070}{{\tt arXiv:2404.07070}}.
\bibitem[{Lynch and Knox(2025)}]{Lynch:2025ine}
\bibinfo{author}{Lynch, G.P.}, \bibinfo{author}{Knox, L.},
  \bibinfo{year}{2025}.
\newblock \bibinfo{title}{{What{\textquoteright}s the matter with
  {\ensuremath{\Sigma}}m{\ensuremath{\nu}}?}}
\newblock \bibinfo{journal}{Phys. Rev. D} \bibinfo{volume}{112},
  \bibinfo{pages}{083543}.
\newblock \DOIprefix\doi{10.1103/613p-pph2},
  \href{http://arxiv.org/abs/2503.14470}{{\tt arXiv:2503.14470}}.
\bibitem[{Maio and Viel(2023)}]{Maio:2022lzg}
\bibinfo{author}{Maio, U.}, \bibinfo{author}{Viel, M.}, \bibinfo{year}{2023}.
\newblock \bibinfo{title}{{JWST high-redshift galaxy constraints on warm and
  cold dark matter models}}.
\newblock \bibinfo{journal}{Astron. Astrophys.} \bibinfo{volume}{672},
  \bibinfo{pages}{A71}.
\newblock \DOIprefix\doi{10.1051/0004-6361/202345851},
  \href{http://arxiv.org/abs/2211.03620}{{\tt arXiv:2211.03620}}.
\bibitem[{McCarthy et~al.(2017)McCarthy, Schaye, Bird and
  Le~Brun}]{McCarthy:2016mry}
\bibinfo{author}{McCarthy, I.G.}, \bibinfo{author}{Schaye, J.},
  \bibinfo{author}{Bird, S.}, \bibinfo{author}{Le~Brun, A.M.C.},
  \bibinfo{year}{2017}.
\newblock \bibinfo{title}{{The BAHAMAS project: Calibrated hydrodynamical
  simulations for large-scale structure cosmology}}.
\newblock \bibinfo{journal}{Mon. Not. Roy. Astron. Soc.} \bibinfo{volume}{465},
  \bibinfo{pages}{2936--2965}.
\newblock \DOIprefix\doi{10.1093/mnras/stw2792},
  \href{http://arxiv.org/abs/1603.02702}{{\tt arXiv:1603.02702}}.
\bibitem[{Mehta and Mukherjee(2026)}]{Mehta:2025xwf}
\bibinfo{author}{Mehta, H.}, \bibinfo{author}{Mukherjee, S.},
  \bibinfo{year}{2026}.
\newblock \bibinfo{title}{{Modified Cosmology or Modified Galaxy Astrophysics
  is Driving the z~{\ensuremath{>}}~6 JWST Results? Cosmic Microwave Background
  Experiments Can Discover the Origin in the Near Future}}.
\newblock \bibinfo{journal}{Astrophys. J.} \bibinfo{volume}{998},
  \bibinfo{pages}{143}.
\newblock \DOIprefix\doi{10.3847/1538-4357/ae3170},
  \href{http://arxiv.org/abs/2509.21952}{{\tt arXiv:2509.21952}}.
\bibitem[{Menci et~al.(2024a)Menci, Adil, Mukhopadhyay, Sen and
  Vagnozzi}]{Menci:2024rbq}
\bibinfo{author}{Menci, N.}, \bibinfo{author}{Adil, S.A.},
  \bibinfo{author}{Mukhopadhyay, U.}, \bibinfo{author}{Sen, A.A.},
  \bibinfo{author}{Vagnozzi, S.}, \bibinfo{year}{2024}a.
\newblock \bibinfo{title}{{Negative cosmological constant in the dark energy
  sector: tests from JWST photometric and spectroscopic observations of
  high-redshift galaxies}}.
\newblock \bibinfo{journal}{JCAP} \bibinfo{volume}{07}, \bibinfo{pages}{072}.
\newblock \DOIprefix\doi{10.1088/1475-7516/2024/07/072},
  \href{http://arxiv.org/abs/2401.12659}{{\tt arXiv:2401.12659}}.
\bibitem[{Menci et~al.(2026)Menci, Castellano, Mukherjee, Roberts, Santini, Sen
  and Shankar}]{Menci:2026ajy}
\bibinfo{author}{Menci, N.}, \bibinfo{author}{Castellano, M.},
  \bibinfo{author}{Mukherjee, P.}, \bibinfo{author}{Roberts, D.},
  \bibinfo{author}{Santini, P.}, \bibinfo{author}{Sen, A.A.},
  \bibinfo{author}{Shankar, F.}, \bibinfo{year}{2026}.
\newblock \bibinfo{title}{{Early growth of massive black holes in dynamical
  dark energy models with negative cosmological constant}}.
\newblock \bibinfo{journal}{Astron. Astrophys.} \bibinfo{volume}{707},
  \bibinfo{pages}{A300}.
\newblock \DOIprefix\doi{10.1051/0004-6361/202558610},
  \href{http://arxiv.org/abs/2602.05921}{{\tt arXiv:2602.05921}}.
\bibitem[{Menci et~al.(2022)Menci, Castellano, Santini, Merlin, Fontana and
  Shankar}]{Menci:2022wia}
\bibinfo{author}{Menci, N.}, \bibinfo{author}{Castellano, M.},
  \bibinfo{author}{Santini, P.}, \bibinfo{author}{Merlin, E.},
  \bibinfo{author}{Fontana, A.}, \bibinfo{author}{Shankar, F.},
  \bibinfo{year}{2022}.
\newblock \bibinfo{title}{{High-redshift Galaxies from Early JWST Observations:
  Constraints on Dark Energy Models}}.
\newblock \bibinfo{journal}{Astrophys. J. Lett.} \bibinfo{volume}{938},
  \bibinfo{pages}{L5}.
\newblock \DOIprefix\doi{10.3847/2041-8213/ac96e9},
  \href{http://arxiv.org/abs/2208.11471}{{\tt arXiv:2208.11471}}.
\bibitem[{Menci et~al.(2024b)Menci, Sen and Castellano}]{Menci:2024hop}
\bibinfo{author}{Menci, N.}, \bibinfo{author}{Sen, A.A.},
  \bibinfo{author}{Castellano, M.}, \bibinfo{year}{2024}b.
\newblock \bibinfo{title}{{The Excess of JWST Bright Galaxies: A Possible
  Origin in the Ground State of Dynamical Dark Energy in the Light of DESI 2024
  Data}}.
\newblock \bibinfo{journal}{Astrophys. J.} \bibinfo{volume}{976},
  \bibinfo{pages}{227}.
\newblock \DOIprefix\doi{10.3847/1538-4357/ad8d5b},
  \href{http://arxiv.org/abs/2410.22940}{{\tt arXiv:2410.22940}}.
\bibitem[{Moster et~al.(2013)Moster, Naab and White}]{Moster:2012fv}
\bibinfo{author}{Moster, B.P.}, \bibinfo{author}{Naab, T.},
  \bibinfo{author}{White, S.D.M.}, \bibinfo{year}{2013}.
\newblock \bibinfo{title}{{Galactic star formation and accretion histories from
  matching galaxies to dark matter haloes}}.
\newblock \bibinfo{journal}{Mon. Not. Roy. Astron. Soc.} \bibinfo{volume}{428},
  \bibinfo{pages}{3121}.
\newblock \DOIprefix\doi{10.1093/mnras/sts261},
  \href{http://arxiv.org/abs/1205.5807}{{\tt arXiv:1205.5807}}.
\bibitem[{Murray et~al.(2013)Murray, Power and Robotham}]{Murray:2013qza}
\bibinfo{author}{Murray, S.}, \bibinfo{author}{Power, C.},
  \bibinfo{author}{Robotham, A.S.G.}, \bibinfo{year}{2013}.
\newblock \bibinfo{title}{{HMFcalc: An online tool for calculating dark matter
  halo mass functions}}.
\newblock \bibinfo{journal}{Astron. Comput.} \bibinfo{volume}{3-4},
  \bibinfo{pages}{23--34}.
\newblock \DOIprefix\doi{10.1016/j.ascom.2013.11.001},
  \href{http://arxiv.org/abs/1306.6721}{{\tt arXiv:1306.6721}}.
\bibitem[{Naredo-Tuero et~al.(2024)Naredo-Tuero, Escudero,
  Fern{\'a}ndez-Mart{\'\i}nez, Marcano and Poulin}]{Naredo-Tuero:2024sgf}
\bibinfo{author}{Naredo-Tuero, D.}, \bibinfo{author}{Escudero, M.},
  \bibinfo{author}{Fern{\'a}ndez-Mart{\'\i}nez, E.}, \bibinfo{author}{Marcano,
  X.}, \bibinfo{author}{Poulin, V.}, \bibinfo{year}{2024}.
\newblock \bibinfo{title}{{Critical look at the cosmological neutrino mass
  bound}}.
\newblock \bibinfo{journal}{Phys. Rev. D} \bibinfo{volume}{110},
  \bibinfo{pages}{123537}.
\newblock \DOIprefix\doi{10.1103/PhysRevD.110.123537},
  \href{http://arxiv.org/abs/2407.13831}{{\tt arXiv:2407.13831}}.
\bibitem[{Pacucci et~al.(2023)Pacucci, Nguyen, Carniani, Maiolino and
  Fan}]{Pacucci:2023oci}
\bibinfo{author}{Pacucci, F.}, \bibinfo{author}{Nguyen, B.},
  \bibinfo{author}{Carniani, S.}, \bibinfo{author}{Maiolino, R.},
  \bibinfo{author}{Fan, X.}, \bibinfo{year}{2023}.
\newblock \bibinfo{title}{{JWST CEERS and JADES Active Galaxies at z =
  4{\textendash}7 Violate the Local M $_{\odot}${\textendash}M $_{\star}$
  Relation at {\ensuremath{>}}3{\ensuremath{\sigma}}: Implications for Low-mass
  Black Holes and Seeding Models}}.
\newblock \bibinfo{journal}{Astrophys. J. Lett.} \bibinfo{volume}{957},
  \bibinfo{pages}{L3}.
\newblock \DOIprefix\doi{10.3847/2041-8213/ad0158},
  \href{http://arxiv.org/abs/2308.12331}{{\tt arXiv:2308.12331}}.
\bibitem[{Padmanabhan and Loeb(2023)}]{Padmanabhan:2023esp}
\bibinfo{author}{Padmanabhan, H.}, \bibinfo{author}{Loeb, A.},
  \bibinfo{year}{2023}.
\newblock \bibinfo{title}{{Alleviating the Need for Exponential Evolution of
  JWST Galaxies in 10$^{10}$ M $_{\odot}$ Haloes at z {\ensuremath{>}} 10 by a
  Modified {\ensuremath{\Lambda}}CDM Power Spectrum}}.
\newblock \bibinfo{journal}{Astrophys. J. Lett.} \bibinfo{volume}{953},
  \bibinfo{pages}{L4}.
\newblock \DOIprefix\doi{10.3847/2041-8213/acea7a},
  \href{http://arxiv.org/abs/2306.04684}{{\tt arXiv:2306.04684}}.
\bibitem[{Pallottini and Ferrara(2023)}]{Pallottini:2023yqg}
\bibinfo{author}{Pallottini, A.}, \bibinfo{author}{Ferrara, A.},
  \bibinfo{year}{2023}.
\newblock \bibinfo{title}{{Stochastic star formation in early galaxies:
  Implications for the James Webb Space Telescope}}.
\newblock \bibinfo{journal}{Astron. Astrophys.} \bibinfo{volume}{677},
  \bibinfo{pages}{L4}.
\newblock \DOIprefix\doi{10.1051/0004-6361/202347384},
  \href{http://arxiv.org/abs/2307.03219}{{\tt arXiv:2307.03219}}.
\bibitem[{Parashari and Laha(2023)}]{Parashari:2023cui}
\bibinfo{author}{Parashari, P.}, \bibinfo{author}{Laha, R.},
  \bibinfo{year}{2023}.
\newblock \bibinfo{title}{{Primordial power spectrum in light of JWST
  observations of high redshift galaxies}}.
\newblock \bibinfo{journal}{Mon. Not. Roy. Astron. Soc.} \bibinfo{volume}{526},
  \bibinfo{pages}{L63--L69}.
\newblock \DOIprefix\doi{10.1093/mnrasl/slad107},
  \href{http://arxiv.org/abs/2305.00999}{{\tt arXiv:2305.00999}}.
\bibitem[{Pedrotti et~al.(2026)Pedrotti, Escamilla, Marra, Perivolaropoulos and
  Vagnozzi}]{Pedrotti:2025ccw}
\bibinfo{author}{Pedrotti, D.}, \bibinfo{author}{Escamilla, L.A.},
  \bibinfo{author}{Marra, V.}, \bibinfo{author}{Perivolaropoulos, L.},
  \bibinfo{author}{Vagnozzi, S.}, \bibinfo{year}{2026}.
\newblock \bibinfo{title}{{BAO miscalibration cannot rescue late-time solutions
  to the Hubble tension}}.
\newblock \bibinfo{journal}{Phys. Rev. D} \bibinfo{volume}{113},
  \bibinfo{pages}{043507}.
\newblock \DOIprefix\doi{10.1103/pn9j-8whx},
  \href{http://arxiv.org/abs/2510.01974}{{\tt arXiv:2510.01974}}.
\bibitem[{Pedrotti et~al.(2025)Pedrotti, Jiang, Escamilla, da~Costa and
  Vagnozzi}]{Pedrotti:2024kpn}
\bibinfo{author}{Pedrotti, D.}, \bibinfo{author}{Jiang, J.Q.},
  \bibinfo{author}{Escamilla, L.A.}, \bibinfo{author}{da~Costa, S.S.},
  \bibinfo{author}{Vagnozzi, S.}, \bibinfo{year}{2025}.
\newblock \bibinfo{title}{{Multidimensionality of the Hubble tension: The roles
  of {\ensuremath{\Omega}}m and {\ensuremath{\omega}}c}}.
\newblock \bibinfo{journal}{Phys. Rev. D} \bibinfo{volume}{111},
  \bibinfo{pages}{023506}.
\newblock \DOIprefix\doi{10.1103/PhysRevD.111.023506},
  \href{http://arxiv.org/abs/2408.04530}{{\tt arXiv:2408.04530}}.
\bibitem[{van Putten(2024)}]{vanPutten:2023ths}
\bibinfo{author}{van Putten, M.H.P.M.}, \bibinfo{year}{2024}.
\newblock \bibinfo{title}{{The Fast and Furious in JWST high-z galaxies}}.
\newblock \bibinfo{journal}{Phys. Dark Univ.} \bibinfo{volume}{43},
  \bibinfo{pages}{101417}.
\newblock \DOIprefix\doi{10.1016/j.dark.2023.101417},
  \href{http://arxiv.org/abs/2312.16692}{{\tt arXiv:2312.16692}}.
\bibitem[{Qi et~al.(2023)Qi, Meng, Zhang and Zhang}]{Qi:2023oxv}
\bibinfo{author}{Qi, J.Z.}, \bibinfo{author}{Meng, P.}, \bibinfo{author}{Zhang,
  J.F.}, \bibinfo{author}{Zhang, X.}, \bibinfo{year}{2023}.
\newblock \bibinfo{title}{{Model-independent measurement of cosmic curvature
  with the latest H(z) and SNe Ia data: A comprehensive investigation}}.
\newblock \bibinfo{journal}{Phys. Rev. D} \bibinfo{volume}{108},
  \bibinfo{pages}{063522}.
\newblock \DOIprefix\doi{10.1103/PhysRevD.108.063522},
  \href{http://arxiv.org/abs/2302.08889}{{\tt arXiv:2302.08889}}.
\bibitem[{Qin et~al.(2023)Qin, Balu and Wyithe}]{Qin:2023rtf}
\bibinfo{author}{Qin, Y.}, \bibinfo{author}{Balu, S.}, \bibinfo{author}{Wyithe,
  J.S.B.}, \bibinfo{year}{2023}.
\newblock \bibinfo{title}{{Implications of z {\ensuremath{\gtrsim}} 12 JWST
  galaxies for galaxy formation at high redshift}}.
\newblock \bibinfo{journal}{Mon. Not. Roy. Astron. Soc.} \bibinfo{volume}{526},
  \bibinfo{pages}{1324--1342}.
\newblock \DOIprefix\doi{10.1093/mnras/stad2448},
  \href{http://arxiv.org/abs/2305.17959}{{\tt arXiv:2305.17959}}.
\bibitem[{Reed et~al.(2007)Reed, Bower, Frenk, Jenkins and
  Theuns}]{Reed:2006rw}
\bibinfo{author}{Reed, D.}, \bibinfo{author}{Bower, R.},
  \bibinfo{author}{Frenk, C.}, \bibinfo{author}{Jenkins, A.},
  \bibinfo{author}{Theuns, T.}, \bibinfo{year}{2007}.
\newblock \bibinfo{title}{{The halo mass function from the dark ages through
  the present day}}.
\newblock \bibinfo{journal}{Mon. Not. Roy. Astron. Soc.} \bibinfo{volume}{374},
  \bibinfo{pages}{2--15}.
\newblock \DOIprefix\doi{10.1111/j.1365-2966.2006.11204.x},
  \href{http://arxiv.org/abs/astro-ph/0607150}{{\tt arXiv:astro-ph/0607150}}.
\bibitem[{Robitaille et~al.(2013)}]{Astropy:2013muo}
\bibinfo{author}{Robitaille, T.P.}, et~al. (\bibinfo{collaboration}{Astropy}),
  \bibinfo{year}{2013}.
\newblock \bibinfo{title}{{Astropy: A Community Python Package for Astronomy}}.
\newblock \bibinfo{journal}{Astron. Astrophys.} \bibinfo{volume}{558},
  \bibinfo{pages}{A33}.
\newblock \DOIprefix\doi{10.1051/0004-6361/201322068},
  \href{http://arxiv.org/abs/1307.6212}{{\tt arXiv:1307.6212}}.
\bibitem[{{Rodighiero} et~al.(2026){Rodighiero}, {Ferrara}, {Catone},
  {Napolitano}, {Cassata}, {Gandolfi}, {Merlin}, {Grazian}, {Renzini},
  {Bisigello}, {Castellano}, {P{\'e}rez-Gonz{\'a}lez}, {P{\'e}rez-D{\'\i}az},
  {Iani}, {Gruppioni}, {Finkelstein}, {Koekemoer}, {Bianchetti} and
  {Sinigaglia}}]{Rodighiero:2026ghw}
\bibinfo{author}{{Rodighiero}, G.}, \bibinfo{author}{{Ferrara}, A.},
  \bibinfo{author}{{Catone}, M.}, \bibinfo{author}{{Napolitano}, L.},
  \bibinfo{author}{{Cassata}, P.}, \bibinfo{author}{{Gandolfi}, G.},
  \bibinfo{author}{{Merlin}, E.}, \bibinfo{author}{{Grazian}, A.},
  \bibinfo{author}{{Renzini}, A.}, \bibinfo{author}{{Bisigello}, L.},
  \bibinfo{author}{{Castellano}, M.},
  \bibinfo{author}{{P{\'e}rez-Gonz{\'a}lez}, P.G.},
  \bibinfo{author}{{P{\'e}rez-D{\'\i}az}, B.}, \bibinfo{author}{{Iani}, E.},
  \bibinfo{author}{{Gruppioni}, C.}, \bibinfo{author}{{Finkelstein}, S.L.},
  \bibinfo{author}{{Koekemoer}, A.M.}, \bibinfo{author}{{Bianchetti}, A.},
  \bibinfo{author}{{Sinigaglia}, F.}, \bibinfo{year}{2026}.
\newblock \bibinfo{title}{{EGS-z11-R0: a red, dust-rich galaxy at Cosmic
  Dawn}}.
\newblock \bibinfo{journal}{arXiv e-prints} ,
  \bibinfo{pages}{arXiv:2603.15841}\DOIprefix\doi{10.48550/arXiv.2603.15841},
  \href{http://arxiv.org/abs/2603.15841}{{\tt arXiv:2603.15841}}.
\bibitem[{Roy~Choudhury(2025)}]{RoyChoudhury:2025dhe}
\bibinfo{author}{Roy~Choudhury, S.}, \bibinfo{year}{2025}.
\newblock \bibinfo{title}{{Cosmology in Extended Parameter Space with DESI Data
  Release 2 Baryon Acoustic Oscillations: A 2{\ensuremath{\sigma}}+ Detection
  of Nonzero Neutrino Masses with an Update on Dynamical Dark Energy and
  Lensing Anomaly}}.
\newblock \bibinfo{journal}{Astrophys. J. Lett.} \bibinfo{volume}{986},
  \bibinfo{pages}{L31}.
\newblock \DOIprefix\doi{10.3847/2041-8213/ade1cc},
  \href{http://arxiv.org/abs/2504.15340}{{\tt arXiv:2504.15340}}.
\bibitem[{Roy~Choudhury and Hannestad(2020)}]{RoyChoudhury:2019hls}
\bibinfo{author}{Roy~Choudhury, S.}, \bibinfo{author}{Hannestad, S.},
  \bibinfo{year}{2020}.
\newblock \bibinfo{title}{{Updated results on neutrino mass and mass hierarchy
  from cosmology with Planck 2018 likelihoods}}.
\newblock \bibinfo{journal}{JCAP} \bibinfo{volume}{07}, \bibinfo{pages}{037}.
\newblock \DOIprefix\doi{10.1088/1475-7516/2020/07/037},
  \href{http://arxiv.org/abs/1907.12598}{{\tt arXiv:1907.12598}}.
\bibitem[{Roy~Choudhury and Naskar(2019)}]{RoyChoudhury:2018vnm}
\bibinfo{author}{Roy~Choudhury, S.}, \bibinfo{author}{Naskar, A.},
  \bibinfo{year}{2019}.
\newblock \bibinfo{title}{{Strong Bounds on Sum of Neutrino Masses in a 12
  Parameter Extended Scenario with Non-Phantom Dynamical Dark Energy ($w(z)\geq
  -1$) with CPL parameterization}}.
\newblock \bibinfo{journal}{Eur. Phys. J. C} \bibinfo{volume}{79},
  \bibinfo{pages}{262}.
\newblock \DOIprefix\doi{10.1140/epjc/s10052-019-6762-z},
  \href{http://arxiv.org/abs/1807.02860}{{\tt arXiv:1807.02860}}.
\bibitem[{Roy~Choudhury and Okumura(2024)}]{RoyChoudhury:2024wri}
\bibinfo{author}{Roy~Choudhury, S.}, \bibinfo{author}{Okumura, T.},
  \bibinfo{year}{2024}.
\newblock \bibinfo{title}{{Updated Cosmological Constraints in Extended
  Parameter Space with Planck PR4, DESI Baryon Acoustic Oscillations, and
  Supernovae: Dynamical Dark Energy, Neutrino Masses, Lensing Anomaly, and the
  Hubble Tension}}.
\newblock \bibinfo{journal}{Astrophys. J. Lett.} \bibinfo{volume}{976},
  \bibinfo{pages}{L11}.
\newblock \DOIprefix\doi{10.3847/2041-8213/ad8c26},
  \href{http://arxiv.org/abs/2409.13022}{{\tt arXiv:2409.13022}}.
\bibitem[{Roy~Choudhury et~al.(2025)Roy~Choudhury, Okumura and
  Umetsu}]{RoyChoudhury:2025iis}
\bibinfo{author}{Roy~Choudhury, S.}, \bibinfo{author}{Okumura, T.},
  \bibinfo{author}{Umetsu, K.}, \bibinfo{year}{2025}.
\newblock \bibinfo{title}{{Cosmological Constraints on Nonphantom Dynamical
  Dark Energy with DESI Data Release 2 Baryon Acoustic Oscillations: A
  3{\ensuremath{\sigma}}+ Lensing Anomaly}}.
\newblock \bibinfo{journal}{Astrophys. J. Lett.} \bibinfo{volume}{994},
  \bibinfo{pages}{L26}.
\newblock \DOIprefix\doi{10.3847/2041-8213/ae1a64},
  \href{http://arxiv.org/abs/2509.26144}{{\tt arXiv:2509.26144}}.
\bibitem[{Rubin et~al.(2025)}]{Rubin:2023jdq}
\bibinfo{author}{Rubin, D.}, et~al., \bibinfo{year}{2025}.
\newblock \bibinfo{title}{{Union Through UNITY: Cosmology with 2,000 SNe Using
  a Unified Bayesian Framework}}.
\newblock \bibinfo{journal}{Astrophys. J.} \bibinfo{volume}{986},
  \bibinfo{pages}{231}.
\newblock \DOIprefix\doi{10.3847/1538-4357/adc0a5},
  \href{http://arxiv.org/abs/2311.12098}{{\tt arXiv:2311.12098}}.
\bibitem[{Sakr(2023)}]{Sakr:2023hrl}
\bibinfo{author}{Sakr, Z.}, \bibinfo{year}{2023}.
\newblock \bibinfo{title}{{Testing the hypothesis of a matter density
  discrepancy within LCDM model using multiple probes}}.
\newblock \bibinfo{journal}{Phys. Rev. D} \bibinfo{volume}{108},
  \bibinfo{pages}{083519}.
\newblock \DOIprefix\doi{10.1103/PhysRevD.108.083519},
  \href{http://arxiv.org/abs/2305.02846}{{\tt arXiv:2305.02846}}.
\bibitem[{Sammut(2025)}]{Sammut:2025eik}
\bibinfo{author}{Sammut, K.}, \bibinfo{year}{2025}.
\newblock \bibinfo{title}{{Testing Planck 2020 and DESI data on wCDM Models}}
  \href{http://arxiv.org/abs/2507.11237}{{\tt arXiv:2507.11237}}.
\bibitem[{Scherer et~al.(2025)Scherer, Sabogal, Nunes and
  De~Felice}]{Scherer:2025esj}
\bibinfo{author}{Scherer, M.}, \bibinfo{author}{Sabogal, M.A.},
  \bibinfo{author}{Nunes, R.C.}, \bibinfo{author}{De~Felice, A.},
  \bibinfo{year}{2025}.
\newblock \bibinfo{title}{{Challenging the {\ensuremath{\Lambda}}CDM model:
  5{\ensuremath{\sigma}} evidence for a dynamical dark energy late-time
  transition}}.
\newblock \bibinfo{journal}{Phys. Rev. D} \bibinfo{volume}{112},
  \bibinfo{pages}{043513}.
\newblock \DOIprefix\doi{10.1103/n86r-sjgm},
  \href{http://arxiv.org/abs/2504.20664}{{\tt arXiv:2504.20664}}.
\bibitem[{Shen et~al.(2023)Shen, Vogelsberger, Boylan-Kolchin, Tacchella and
  Kannan}]{Shen:2023cva}
\bibinfo{author}{Shen, X.}, \bibinfo{author}{Vogelsberger, M.},
  \bibinfo{author}{Boylan-Kolchin, M.}, \bibinfo{author}{Tacchella, S.},
  \bibinfo{author}{Kannan, R.}, \bibinfo{year}{2023}.
\newblock \bibinfo{title}{{The impact of UV variability on the abundance of
  bright galaxies at $z \geq 9$}} \href{http://arxiv.org/abs/2305.05679}{{\tt
  arXiv:2305.05679}}.
\bibitem[{Shen et~al.(2026)}]{Shen:2025isu}
\bibinfo{author}{Shen, X.}, et~al., \bibinfo{year}{2026}.
\newblock \bibinfo{title}{{The thesan-zoom project: star formation efficiencies
  in high-redshift galaxies}}.
\newblock \bibinfo{journal}{Mon. Not. Roy. Astron. Soc.} \bibinfo{volume}{545},
  \bibinfo{pages}{staf2119}.
\newblock \DOIprefix\doi{10.1093/mnras/staf2119},
  \href{http://arxiv.org/abs/2503.01949}{{\tt arXiv:2503.01949}}.
\bibitem[{Sheth et~al.(2001)Sheth, Mo and Tormen}]{Sheth:1999su}
\bibinfo{author}{Sheth, R.K.}, \bibinfo{author}{Mo, H.J.},
  \bibinfo{author}{Tormen, G.}, \bibinfo{year}{2001}.
\newblock \bibinfo{title}{{Ellipsoidal collapse and an improved model for the
  number and spatial distribution of dark matter haloes}}.
\newblock \bibinfo{journal}{Mon. Not. Roy. Astron. Soc.} \bibinfo{volume}{323},
  \bibinfo{pages}{1}.
\newblock \DOIprefix\doi{10.1046/j.1365-8711.2001.04006.x},
  \href{http://arxiv.org/abs/astro-ph/9907024}{{\tt arXiv:astro-ph/9907024}}.
\bibitem[{Shlivko and Poulin(2026)}]{Shlivko:2026jxa}
\bibinfo{author}{Shlivko, D.}, \bibinfo{author}{Poulin, V.},
  \bibinfo{year}{2026}.
\newblock \bibinfo{title}{{Phantom-Crossing Dark Energy and the $\Omega_m$
  Tug-of-War}} \href{http://arxiv.org/abs/2603.22406}{{\tt arXiv:2603.22406}}.
\bibitem[{Shuntov et~al.(2022)}]{Shuntov:2022qwu}
\bibinfo{author}{Shuntov, M.}, et~al., \bibinfo{year}{2022}.
\newblock \bibinfo{title}{{COSMOS2020: Cosmic evolution of the stellar-to-halo
  mass relation for central and satellite galaxies up to z {\ensuremath{\sim}}
  5}}.
\newblock \bibinfo{journal}{Astron. Astrophys.} \bibinfo{volume}{664},
  \bibinfo{pages}{A61}.
\newblock \DOIprefix\doi{10.1051/0004-6361/202243136},
  \href{http://arxiv.org/abs/2203.10895}{{\tt arXiv:2203.10895}}.
\bibitem[{Specogna et~al.(2025)Specogna, Vardanyan, Giar{\`e} and
  Di~Valentino}]{Specogna:2025ufe}
\bibinfo{author}{Specogna, E.}, \bibinfo{author}{Vardanyan, T.},
  \bibinfo{author}{Giar{\`e}, W.}, \bibinfo{author}{Di~Valentino, E.},
  \bibinfo{year}{2025}.
\newblock \bibinfo{title}{{Slow-rolling down the curvature: a reassessment of
  the Planck constraints on $\phi^2$ inflation in a closed universe}}
  \href{http://arxiv.org/abs/2509.26263}{{\tt arXiv:2509.26263}}.
\bibitem[{Stevens et~al.(2023)Stevens, Khoraminezhad and
  Saito}]{Stevens:2022evv}
\bibinfo{author}{Stevens, J.}, \bibinfo{author}{Khoraminezhad, H.},
  \bibinfo{author}{Saito, S.}, \bibinfo{year}{2023}.
\newblock \bibinfo{title}{{Constraining the spatial curvature with cosmic
  expansion history in a cosmological model with a non-standard sound
  horizon}}.
\newblock \bibinfo{journal}{JCAP} \bibinfo{volume}{07}, \bibinfo{pages}{046}.
\newblock \DOIprefix\doi{10.1088/1475-7516/2023/07/046},
  \href{http://arxiv.org/abs/2212.09804}{{\tt arXiv:2212.09804}}.
\bibitem[{Su et~al.(2023)Su, Li and Feng}]{Su:2023jno}
\bibinfo{author}{Su, B.Y.}, \bibinfo{author}{Li, N.}, \bibinfo{author}{Feng,
  L.}, \bibinfo{year}{2023}.
\newblock \bibinfo{title}{{An inflation model for massive primordial black
  holes to interpret the JWST observations}}
  \href{http://arxiv.org/abs/2306.05364}{{\tt arXiv:2306.05364}}.
\bibitem[{Sun et~al.(2023)Sun, Faucher-Gigu{\`e}re, Hayward, Shen, Wetzel and
  Cochrane}]{Sun:2023ocn}
\bibinfo{author}{Sun, G.}, \bibinfo{author}{Faucher-Gigu{\`e}re, C.A.},
  \bibinfo{author}{Hayward, C.C.}, \bibinfo{author}{Shen, X.},
  \bibinfo{author}{Wetzel, A.}, \bibinfo{author}{Cochrane, R.K.},
  \bibinfo{year}{2023}.
\newblock \bibinfo{title}{{Bursty Star Formation Naturally Explains the
  Abundance of Bright Galaxies at Cosmic Dawn}}.
\newblock \bibinfo{journal}{Astrophys. J. Lett.} \bibinfo{volume}{955},
  \bibinfo{pages}{L35}.
\newblock \DOIprefix\doi{10.3847/2041-8213/acf85a},
  \href{http://arxiv.org/abs/2307.15305}{{\tt arXiv:2307.15305}}.
\bibitem[{Tacchella et~al.(2018)Tacchella, Bose, Conroy, Eisenstein and
  Johnson}]{Tacchella:2018qny}
\bibinfo{author}{Tacchella, S.}, \bibinfo{author}{Bose, S.},
  \bibinfo{author}{Conroy, C.}, \bibinfo{author}{Eisenstein, D.J.},
  \bibinfo{author}{Johnson, B.D.}, \bibinfo{year}{2018}.
\newblock \bibinfo{title}{{A Redshift-independent Efficiency Model: Star
  Formation and Stellar Masses in Dark Matter Halos at z {\ensuremath{\gtrsim}}
  4}}.
\newblock \bibinfo{journal}{Astrophys. J.} \bibinfo{volume}{868},
  \bibinfo{pages}{92}.
\newblock \DOIprefix\doi{10.3847/1538-4357/aae8e0},
  \href{http://arxiv.org/abs/1806.03299}{{\tt arXiv:1806.03299}}.
\bibitem[{Tanseri et~al.(2022)Tanseri, Hagstotz, Vagnozzi, Giusarma and
  Freese}]{Tanseri:2022zfe}
\bibinfo{author}{Tanseri, I.}, \bibinfo{author}{Hagstotz, S.},
  \bibinfo{author}{Vagnozzi, S.}, \bibinfo{author}{Giusarma, E.},
  \bibinfo{author}{Freese, K.}, \bibinfo{year}{2022}.
\newblock \bibinfo{title}{{Updated neutrino mass constraints from galaxy
  clustering and CMB lensing-galaxy cross-correlation measurements}}.
\newblock \bibinfo{journal}{JHEAp} \bibinfo{volume}{36},
  \bibinfo{pages}{1--26}.
\newblock \DOIprefix\doi{10.1016/j.jheap.2022.07.002},
  \href{http://arxiv.org/abs/2207.01913}{{\tt arXiv:2207.01913}}.
\bibitem[{Tinker et~al.(2008)Tinker, Kravtsov, Klypin, Abazajian, Warren,
  Yepes, Gottlober and Holz}]{Tinker:2008ff}
\bibinfo{author}{Tinker, J.L.}, \bibinfo{author}{Kravtsov, A.V.},
  \bibinfo{author}{Klypin, A.}, \bibinfo{author}{Abazajian, K.},
  \bibinfo{author}{Warren, M.S.}, \bibinfo{author}{Yepes, G.},
  \bibinfo{author}{Gottlober, S.}, \bibinfo{author}{Holz, D.E.},
  \bibinfo{year}{2008}.
\newblock \bibinfo{title}{{Toward a halo mass function for precision cosmology:
  The Limits of universality}}.
\newblock \bibinfo{journal}{Astrophys. J.} \bibinfo{volume}{688},
  \bibinfo{pages}{709--728}.
\newblock \DOIprefix\doi{10.1086/591439},
  \href{http://arxiv.org/abs/0803.2706}{{\tt arXiv:0803.2706}}.
\bibitem[{Urrutia et~al.(2025)Urrutia, Ellis, Fairbairn and
  Vaskonen}]{Urrutia:2024hwc}
\bibinfo{author}{Urrutia, J.}, \bibinfo{author}{Ellis, J.},
  \bibinfo{author}{Fairbairn, M.}, \bibinfo{author}{Vaskonen, V.},
  \bibinfo{year}{2025}.
\newblock \bibinfo{title}{{The origin of the JWST supermassive black holes}}.
\newblock \bibinfo{journal}{Astron. Astrophys.} \bibinfo{volume}{703},
  \bibinfo{pages}{A138}.
\newblock \DOIprefix\doi{10.1051/0004-6361/202555389},
  \href{http://arxiv.org/abs/2410.24224}{{\tt arXiv:2410.24224}}.
\bibitem[{Vagnozzi(2020)}]{Vagnozzi:2019ezj}
\bibinfo{author}{Vagnozzi, S.}, \bibinfo{year}{2020}.
\newblock \bibinfo{title}{{New physics in light of the $H_0$ tension: An
  alternative view}}.
\newblock \bibinfo{journal}{Phys. Rev. D} \bibinfo{volume}{102},
  \bibinfo{pages}{023518}.
\newblock \DOIprefix\doi{10.1103/PhysRevD.102.023518},
  \href{http://arxiv.org/abs/1907.07569}{{\tt arXiv:1907.07569}}.
\bibitem[{Vagnozzi et~al.(2018)Vagnozzi, Dhawan, Gerbino, Freese, Goobar and
  Mena}]{Vagnozzi:2018jhn}
\bibinfo{author}{Vagnozzi, S.}, \bibinfo{author}{Dhawan, S.},
  \bibinfo{author}{Gerbino, M.}, \bibinfo{author}{Freese, K.},
  \bibinfo{author}{Goobar, A.}, \bibinfo{author}{Mena, O.},
  \bibinfo{year}{2018}.
\newblock \bibinfo{title}{{Constraints on the sum of the neutrino masses in
  dynamical dark energy models with $w(z) \geq -1$ are tighter than those
  obtained in $\Lambda$CDM}}.
\newblock \bibinfo{journal}{Phys. Rev. D} \bibinfo{volume}{98},
  \bibinfo{pages}{083501}.
\newblock \DOIprefix\doi{10.1103/PhysRevD.98.083501},
  \href{http://arxiv.org/abs/1801.08553}{{\tt arXiv:1801.08553}}.
\bibitem[{Vagnozzi et~al.(2021a)Vagnozzi, Di~Valentino, Gariazzo, Melchiorri,
  Mena and Silk}]{Vagnozzi:2020rcz}
\bibinfo{author}{Vagnozzi, S.}, \bibinfo{author}{Di~Valentino, E.},
  \bibinfo{author}{Gariazzo, S.}, \bibinfo{author}{Melchiorri, A.},
  \bibinfo{author}{Mena, O.}, \bibinfo{author}{Silk, J.},
  \bibinfo{year}{2021}a.
\newblock \bibinfo{title}{{The galaxy power spectrum take on spatial curvature
  and cosmic concordance}}.
\newblock \bibinfo{journal}{Phys. Dark Univ.} \bibinfo{volume}{33},
  \bibinfo{pages}{100851}.
\newblock \DOIprefix\doi{10.1016/j.dark.2021.100851},
  \href{http://arxiv.org/abs/2010.02230}{{\tt arXiv:2010.02230}}.
\bibitem[{Vagnozzi et~al.(2017)Vagnozzi, Giusarma, Mena, Freese, Gerbino, Ho
  and Lattanzi}]{Vagnozzi:2017ovm}
\bibinfo{author}{Vagnozzi, S.}, \bibinfo{author}{Giusarma, E.},
  \bibinfo{author}{Mena, O.}, \bibinfo{author}{Freese, K.},
  \bibinfo{author}{Gerbino, M.}, \bibinfo{author}{Ho, S.},
  \bibinfo{author}{Lattanzi, M.}, \bibinfo{year}{2017}.
\newblock \bibinfo{title}{{Unveiling $\nu$ secrets with cosmological data:
  neutrino masses and mass hierarchy}}.
\newblock \bibinfo{journal}{Phys. Rev. D} \bibinfo{volume}{96},
  \bibinfo{pages}{123503}.
\newblock \DOIprefix\doi{10.1103/PhysRevD.96.123503},
  \href{http://arxiv.org/abs/1701.08172}{{\tt arXiv:1701.08172}}.
\bibitem[{Vagnozzi et~al.(2021b)Vagnozzi, Loeb and Moresco}]{Vagnozzi:2020dfn}
\bibinfo{author}{Vagnozzi, S.}, \bibinfo{author}{Loeb, A.},
  \bibinfo{author}{Moresco, M.}, \bibinfo{year}{2021}b.
\newblock \bibinfo{title}{{Eppur \`e piatto? The Cosmic Chronometers Take on
  Spatial Curvature and Cosmic Concordance}}.
\newblock \bibinfo{journal}{Astrophys. J.} \bibinfo{volume}{908},
  \bibinfo{pages}{84}.
\newblock \DOIprefix\doi{10.3847/1538-4357/abd4df},
  \href{http://arxiv.org/abs/2011.11645}{{\tt arXiv:2011.11645}}.
\bibitem[{Vagnozzi et~al.(2022)Vagnozzi, Pacucci and Loeb}]{Vagnozzi:2021tjv}
\bibinfo{author}{Vagnozzi, S.}, \bibinfo{author}{Pacucci, F.},
  \bibinfo{author}{Loeb, A.}, \bibinfo{year}{2022}.
\newblock \bibinfo{title}{{Implications for the Hubble tension from the ages of
  the oldest astrophysical objects}}.
\newblock \bibinfo{journal}{JHEAp} \bibinfo{volume}{36},
  \bibinfo{pages}{27--35}.
\newblock \DOIprefix\doi{10.1016/j.jheap.2022.07.004},
  \href{http://arxiv.org/abs/2105.10421}{{\tt arXiv:2105.10421}}.
\bibitem[{Wang et~al.(2020)Wang, Qi, Zhang and Zhang}]{Wang:2019yob}
\bibinfo{author}{Wang, B.}, \bibinfo{author}{Qi, J.Z.}, \bibinfo{author}{Zhang,
  J.F.}, \bibinfo{author}{Zhang, X.}, \bibinfo{year}{2020}.
\newblock \bibinfo{title}{{Cosmological Model-independent Constraints on
  Spatial Curvature from Strong Gravitational Lensing and SN Ia Observations}}.
\newblock \bibinfo{journal}{Astrophys. J.} \bibinfo{volume}{898},
  \bibinfo{pages}{100}.
\newblock \DOIprefix\doi{10.3847/1538-4357/ab9b22},
  \href{http://arxiv.org/abs/1910.12173}{{\tt arXiv:1910.12173}}.
\bibitem[{Wang and Liu(2022)}]{Wang:2022jvx}
\bibinfo{author}{Wang, D.}, \bibinfo{author}{Liu, Y.}, \bibinfo{year}{2022}.
\newblock \bibinfo{title}{{JWST high redshift galaxy observations have a strong
  tension with Planck CMB measurements}}
  \href{http://arxiv.org/abs/2301.00347}{{\tt arXiv:2301.00347}}.
\bibitem[{Wang et~al.(2024a)Wang, Mena, Di~Valentino and
  Gariazzo}]{Wang:2024hen}
\bibinfo{author}{Wang, D.}, \bibinfo{author}{Mena, O.},
  \bibinfo{author}{Di~Valentino, E.}, \bibinfo{author}{Gariazzo, S.},
  \bibinfo{year}{2024}a.
\newblock \bibinfo{title}{{Updating neutrino mass constraints with background
  measurements}}.
\newblock \bibinfo{journal}{Phys. Rev. D} \bibinfo{volume}{110},
  \bibinfo{pages}{103536}.
\newblock \DOIprefix\doi{10.1103/PhysRevD.110.103536},
  \href{http://arxiv.org/abs/2405.03368}{{\tt arXiv:2405.03368}}.
\bibitem[{Wang and Piao(2026)}]{Wang:2024dka}
\bibinfo{author}{Wang, H.}, \bibinfo{author}{Piao, Y.S.}, \bibinfo{year}{2026}.
\newblock \bibinfo{title}{{Dark energy in light of DESI DR1 and Hubble
  tension}}.
\newblock \bibinfo{journal}{Phys. Lett. B} \bibinfo{volume}{873},
  \bibinfo{pages}{140180}.
\newblock \DOIprefix\doi{10.1016/j.physletb.2026.140180},
  \href{http://arxiv.org/abs/2404.18579}{{\tt arXiv:2404.18579}}.
\bibitem[{Wang et~al.(2024b)Wang, Huang, Huang and Liu}]{Wang:2023gla}
\bibinfo{author}{Wang, J.}, \bibinfo{author}{Huang, Z.},
  \bibinfo{author}{Huang, L.}, \bibinfo{author}{Liu, J.},
  \bibinfo{year}{2024}b.
\newblock \bibinfo{title}{{Quantifying the Tension between Cosmological Models
  and JWST Red Candidate Massive Galaxies}}.
\newblock \bibinfo{journal}{Res. Astron. Astrophys.} \bibinfo{volume}{24},
  \bibinfo{pages}{045001}.
\newblock \DOIprefix\doi{10.1088/1674-4527/ad2cd3},
  \href{http://arxiv.org/abs/2311.02866}{{\tt arXiv:2311.02866}}.
\bibitem[{Wang et~al.(2025)Wang, Cai, Guo and Wang}]{Wang:2025znm}
\bibinfo{author}{Wang, J.Q.}, \bibinfo{author}{Cai, R.G.},
  \bibinfo{author}{Guo, Z.K.}, \bibinfo{author}{Wang, S.J.},
  \bibinfo{year}{2025}.
\newblock \bibinfo{title}{{Resolving the Planck-DESI tension by non-minimally
  coupled quintessence}} \href{http://arxiv.org/abs/2508.01759}{{\tt
  arXiv:2508.01759}}.
\bibitem[{Wang et~al.(2024c)Wang, Su, Zu, Yang and Feng}]{Wang:2023ros}
\bibinfo{author}{Wang, P.}, \bibinfo{author}{Su, B.Y.}, \bibinfo{author}{Zu,
  L.}, \bibinfo{author}{Yang, Y.}, \bibinfo{author}{Feng, L.},
  \bibinfo{year}{2024}c.
\newblock \bibinfo{title}{{Exploring the dark energy equation of state with
  JWST}}.
\newblock \bibinfo{journal}{Eur. Phys. J. Plus} \bibinfo{volume}{139},
  \bibinfo{pages}{711}.
\newblock \DOIprefix\doi{10.1140/epjp/s13360-024-05276-y},
  \href{http://arxiv.org/abs/2307.11374}{{\tt arXiv:2307.11374}}.
\bibitem[{Wang and Lin(2025)}]{Wang:2025mqz}
\bibinfo{author}{Wang, Y.}, \bibinfo{author}{Lin, W.}, \bibinfo{year}{2025}.
\newblock \bibinfo{title}{{Uncalibrated Cosmic Standards as a Robust Test on
  Late-time Cosmological Models}}.
\newblock \bibinfo{journal}{Astrophys. J.} \bibinfo{volume}{989},
  \bibinfo{pages}{120}.
\newblock \DOIprefix\doi{10.3847/1538-4357/adf336},
  \href{http://arxiv.org/abs/2506.04333}{{\tt arXiv:2506.04333}}.
\bibitem[{Wang et~al.(2023)Wang, Lei, Yuan and Fan}]{Wang:2023xmm}
\bibinfo{author}{Wang, Y.Y.}, \bibinfo{author}{Lei, L.}, \bibinfo{author}{Yuan,
  G.W.}, \bibinfo{author}{Fan, Y.Z.}, \bibinfo{year}{2023}.
\newblock \bibinfo{title}{{Modeling the JWST High-redshift Galaxies with a
  General Formation Scenario and the Consistency with the
  {\ensuremath{\Lambda}}CDM Model}}.
\newblock \bibinfo{journal}{Astrophys. J. Lett.} \bibinfo{volume}{954},
  \bibinfo{pages}{L48}.
\newblock \DOIprefix\doi{10.3847/2041-8213/acf46c},
  \href{http://arxiv.org/abs/2307.12487}{{\tt arXiv:2307.12487}}.
\bibitem[{Wang et~al.(2024d)Wang, Lei, Feng and Fan}]{Wang:2024hce}
\bibinfo{author}{Wang, Z.F.}, \bibinfo{author}{Lei, L.}, \bibinfo{author}{Feng,
  L.}, \bibinfo{author}{Fan, Y.Z.}, \bibinfo{year}{2024}d.
\newblock \bibinfo{title}{{JWST Observations Constrain the Time Evolution of
  Fine Structure Constants and Dark Energy-electromagnetic Coupling}}.
\newblock \bibinfo{journal}{Res. Astron. Astrophys.} \bibinfo{volume}{24},
  \bibinfo{pages}{125012}.
\newblock \DOIprefix\doi{10.1088/1674-4527/ad9198},
  \href{http://arxiv.org/abs/2411.08774}{{\tt arXiv:2411.08774}}.
\bibitem[{Watson et~al.(2013)Watson, Iliev, D'Aloisio, Knebe, Shapiro and
  Yepes}]{Watson:2012mt}
\bibinfo{author}{Watson, W.A.}, \bibinfo{author}{Iliev, I.T.},
  \bibinfo{author}{D'Aloisio, A.}, \bibinfo{author}{Knebe, A.},
  \bibinfo{author}{Shapiro, P.R.}, \bibinfo{author}{Yepes, G.},
  \bibinfo{year}{2013}.
\newblock \bibinfo{title}{{The halo mass function through the cosmic ages}}.
\newblock \bibinfo{journal}{Mon. Not. Roy. Astron. Soc.} \bibinfo{volume}{433},
  \bibinfo{pages}{1230}.
\newblock \DOIprefix\doi{10.1093/mnras/stt791},
  \href{http://arxiv.org/abs/1212.0095}{{\tt arXiv:1212.0095}}.
\bibitem[{Wechsler and Tinker(2018)}]{Wechsler:2018pic}
\bibinfo{author}{Wechsler, R.H.}, \bibinfo{author}{Tinker, J.L.},
  \bibinfo{year}{2018}.
\newblock \bibinfo{title}{{The Connection between Galaxies and their Dark
  Matter Halos}}.
\newblock \bibinfo{journal}{Ann. Rev. Astron. Astrophys.} \bibinfo{volume}{56},
  \bibinfo{pages}{435--487}.
\newblock \DOIprefix\doi{10.1146/annurev-astro-081817-051756},
  \href{http://arxiv.org/abs/1804.03097}{{\tt arXiv:1804.03097}}.
\bibitem[{Wei and Melia(2022)}]{Wei:2022plg}
\bibinfo{author}{Wei, J.J.}, \bibinfo{author}{Melia, F.}, \bibinfo{year}{2022}.
\newblock \bibinfo{title}{{Exploring the Hubble Tension and Spatial Curvature
  from the Ages of Old Astrophysical Objects}}.
\newblock \bibinfo{journal}{Astrophys. J.} \bibinfo{volume}{928},
  \bibinfo{pages}{165}.
\newblock \DOIprefix\doi{10.3847/1538-4357/ac562c},
  \href{http://arxiv.org/abs/2202.07865}{{\tt arXiv:2202.07865}}.
\bibitem[{Weiner(2026)}]{Weiner:2026sfm}
\bibinfo{author}{Weiner, Z.J.}, \bibinfo{year}{2026}.
\newblock \bibinfo{title}{{High-redshift physics from the acoustic scale}}
  \href{http://arxiv.org/abs/2603.18131}{{\tt arXiv:2603.18131}}.
\bibitem[{Wu et~al.(2023)Wu, Qi and Zhang}]{Wu:2022fmr}
\bibinfo{author}{Wu, P.J.}, \bibinfo{author}{Qi, J.Z.}, \bibinfo{author}{Zhang,
  X.}, \bibinfo{year}{2023}.
\newblock \bibinfo{title}{{Null test for cosmic curvature using Gaussian
  process*}}.
\newblock \bibinfo{journal}{Chin. Phys. C} \bibinfo{volume}{47},
  \bibinfo{pages}{055106}.
\newblock \DOIprefix\doi{10.1088/1674-1137/acc647},
  \href{http://arxiv.org/abs/2209.08502}{{\tt arXiv:2209.08502}}.
\bibitem[{Wu and Zhang(2025)}]{Wu:2024faw}
\bibinfo{author}{Wu, P.J.}, \bibinfo{author}{Zhang, X.}, \bibinfo{year}{2025}.
\newblock \bibinfo{title}{{Measuring cosmic curvature with non-CMB
  observations}}.
\newblock \bibinfo{journal}{Phys. Rev. D} \bibinfo{volume}{112},
  \bibinfo{pages}{063514}.
\newblock \DOIprefix\doi{10.1103/sn3q-q589},
  \href{http://arxiv.org/abs/2411.06356}{{\tt arXiv:2411.06356}}.
\bibitem[{Xiao et~al.(2024)}]{Xiao:2023ghw}
\bibinfo{author}{Xiao, M.}, et~al., \bibinfo{year}{2024}.
\newblock \bibinfo{title}{{Accelerated formation of ultra-massive galaxies in
  the first billion years}}.
\newblock \bibinfo{journal}{Nature} \bibinfo{volume}{635},
  \bibinfo{pages}{311--315}.
\newblock \DOIprefix\doi{10.1038/s41586-024-08094-5},
  \href{http://arxiv.org/abs/2309.02492}{{\tt arXiv:2309.02492}}.
\bibitem[{Yadav et~al.(2025)Yadav, Dixit, Barak and Pradhan}]{Yadav:2025vgo}
\bibinfo{author}{Yadav, M.}, \bibinfo{author}{Dixit, A.},
  \bibinfo{author}{Barak, M.S.}, \bibinfo{author}{Pradhan, A.},
  \bibinfo{year}{2025}.
\newblock \bibinfo{title}{{Investigating the wCDM model with latest DESI BAO
  observations}}.
\newblock \bibinfo{journal}{Eur. Phys. J. C} \bibinfo{volume}{85},
  \bibinfo{pages}{1013}.
\newblock \DOIprefix\doi{10.1140/epjc/s10052-025-14720-x},
  \href{http://arxiv.org/abs/2510.09074}{{\tt arXiv:2510.09074}}.
\bibitem[{Yang et~al.(2023)Yang, Giar\`e, Pan, Di~Valentino, Melchiorri and
  Silk}]{Yang:2022kho}
\bibinfo{author}{Yang, W.}, \bibinfo{author}{Giar\`e, W.},
  \bibinfo{author}{Pan, S.}, \bibinfo{author}{Di~Valentino, E.},
  \bibinfo{author}{Melchiorri, A.}, \bibinfo{author}{Silk, J.},
  \bibinfo{year}{2023}.
\newblock \bibinfo{title}{{Revealing the effects of curvature on the
  cosmological models}}.
\newblock \bibinfo{journal}{Phys. Rev. D} \bibinfo{volume}{107},
  \bibinfo{pages}{063509}.
\newblock \DOIprefix\doi{10.1103/PhysRevD.107.063509},
  \href{http://arxiv.org/abs/2210.09865}{{\tt arXiv:2210.09865}}.
\bibitem[{Yang et~al.(2021)Yang, Pan, Di~Valentino, Mena and
  Melchiorri}]{Yang:2021hxg}
\bibinfo{author}{Yang, W.}, \bibinfo{author}{Pan, S.},
  \bibinfo{author}{Di~Valentino, E.}, \bibinfo{author}{Mena, O.},
  \bibinfo{author}{Melchiorri, A.}, \bibinfo{year}{2021}.
\newblock \bibinfo{title}{{2021-H0 odyssey: closed, phantom and interacting
  dark energy cosmologies}}.
\newblock \bibinfo{journal}{JCAP} \bibinfo{volume}{10}, \bibinfo{pages}{008}.
\newblock \DOIprefix\doi{10.1088/1475-7516/2021/10/008},
  \href{http://arxiv.org/abs/2101.03129}{{\tt arXiv:2101.03129}}.
\bibitem[{Yang et~al.(2024)Yang, Ren, Wang, Lu, Zhang, Cai and
  Saridakis}]{Yang:2024kdo}
\bibinfo{author}{Yang, Y.}, \bibinfo{author}{Ren, X.}, \bibinfo{author}{Wang,
  Q.}, \bibinfo{author}{Lu, Z.}, \bibinfo{author}{Zhang, D.},
  \bibinfo{author}{Cai, Y.F.}, \bibinfo{author}{Saridakis, E.N.},
  \bibinfo{year}{2024}.
\newblock \bibinfo{title}{{Quintom cosmology and modified gravity after DESI
  2024}}.
\newblock \bibinfo{journal}{Sci. Bull.} \bibinfo{volume}{69},
  \bibinfo{pages}{2698--2704}.
\newblock \DOIprefix\doi{10.1016/j.scib.2024.07.029},
  \href{http://arxiv.org/abs/2404.19437}{{\tt arXiv:2404.19437}}.
\bibitem[{Yuan et~al.(2024)Yuan, Lei, Wang, Wang, Wang, Chen, Shen, Cai and
  Fan}]{Yuan:2023bvh}
\bibinfo{author}{Yuan, G.W.}, \bibinfo{author}{Lei, L.}, \bibinfo{author}{Wang,
  Y.Z.}, \bibinfo{author}{Wang, B.}, \bibinfo{author}{Wang, Y.Y.},
  \bibinfo{author}{Chen, C.}, \bibinfo{author}{Shen, Z.Q.},
  \bibinfo{author}{Cai, Y.F.}, \bibinfo{author}{Fan, Y.Z.},
  \bibinfo{year}{2024}.
\newblock \bibinfo{title}{{Rapidly growing primordial black holes as seeds of
  the massive high-redshift JWST Galaxies}}.
\newblock \bibinfo{journal}{Sci. China Phys. Mech. Astron.}
  \bibinfo{volume}{67}, \bibinfo{pages}{109512}.
\newblock \DOIprefix\doi{10.1007/s11433-024-2433-3},
  \href{http://arxiv.org/abs/2303.09391}{{\tt arXiv:2303.09391}}.
\bibitem[{Zhang et~al.(2025)Zhang, Li, Du, Zhou, Gao, Zhang and
  Zhang}]{Zhang:2025dwu}
\bibinfo{author}{Zhang, Y.M.}, \bibinfo{author}{Li, T.N.}, \bibinfo{author}{Du,
  G.H.}, \bibinfo{author}{Zhou, S.H.}, \bibinfo{author}{Gao, L.Y.},
  \bibinfo{author}{Zhang, J.F.}, \bibinfo{author}{Zhang, X.},
  \bibinfo{year}{2025}.
\newblock \bibinfo{title}{{Alleviating the $H_0$ tension through new
  interacting dark energy model in light of DESI DR2}}
  \href{http://arxiv.org/abs/2510.12627}{{\tt arXiv:2510.12627}}.
\bibitem[{Zhou et~al.(2025)Zhou, Li, Du, Jiang, Zhang and Zhang}]{Zhou:2025nkb}
\bibinfo{author}{Zhou, S.H.}, \bibinfo{author}{Li, T.N.}, \bibinfo{author}{Du,
  G.H.}, \bibinfo{author}{Jiang, J.Q.}, \bibinfo{author}{Zhang, J.F.},
  \bibinfo{author}{Zhang, X.}, \bibinfo{year}{2025}.
\newblock \bibinfo{title}{{Measuring neutrino masses with joint JWST and DESI
  DR2 data}}.
\newblock \bibinfo{journal}{Phys. Rev. D} \bibinfo{volume}{112},
  \bibinfo{pages}{123532}.
\newblock \DOIprefix\doi{10.1103/mtdg-hbqt},
  \href{http://arxiv.org/abs/2509.10836}{{\tt arXiv:2509.10836}}.
\bibitem[{Ziegler et~al.(2025)Ziegler, Freese, Lozano and
  Montefalcone}]{Ziegler:2025plz}
\bibinfo{author}{Ziegler, J.J.}, \bibinfo{author}{Freese, K.},
  \bibinfo{author}{Lozano, J.}, \bibinfo{author}{Montefalcone, G.},
  \bibinfo{year}{2025}.
\newblock \bibinfo{title}{{Explaining the ''too massive'' high-redshift
  galaxies in JWST data: numerical study of three effects and a simple
  relation}} \href{http://arxiv.org/abs/2507.21409}{{\tt arXiv:2507.21409}}.
\bibitem[{Zuckerman and Anchordoqui(2022)}]{Zuckerman:2021kgm}
\bibinfo{author}{Zuckerman, E.}, \bibinfo{author}{Anchordoqui, L.A.},
  \bibinfo{year}{2022}.
\newblock \bibinfo{title}{{Spatial curvature sensitivity to local H0 from the
  Cepheid distance ladder}}.
\newblock \bibinfo{journal}{JHEAp} \bibinfo{volume}{33},
  \bibinfo{pages}{10--13}.
\newblock \DOIprefix\doi{10.1016/j.jheap.2021.10.002},
  \href{http://arxiv.org/abs/2110.05346}{{\tt arXiv:2110.05346}}.

\end{thebibliography}
\bibliographystyle{elsarticle-harv}

\end{document}